\documentclass{biophys-new}
\usepackage[utf8]{inputenc}
\usepackage{graphicx}
\usepackage[colorlinks,allcolors=cyan!70!black]{hyperref}
\usepackage{color}
\usepackage{enumerate}
\usepackage{siunitx}
\usepackage{array}

\usepackage[resetlabels]{multibib}
\newcites{Sup}{Supporting References}
\newcounter{supplementary}

\usepackage{amssymb} %required for /eqcirc

\usepackage{lipsum}

\title{Search and capture efficiency of dynamic microtubules for centrosome relocation during IS formation}
\runningtitle{Capture efficiency of MTs in T cell} %% For page header

\author[1,*]{Apurba Sarkar}
\author[2,*]{Heiko Rieger}
\author[1,*]{Raja Paul}

\runningauthor{Sarkar et al.}

\affil[1]{School of Mathematical \& Computational Sciences, Indian Association for the Cultivation of Science, Kolkata, West Bengal, India}

\affil[2]{Department of Theoretical Physics and Center for Biophysics, Saarland University, Saarbrücken, Germany }

\corrauthor[*]{sspas2@iacs.res.in, h.rieger@mx.uni-saarland.de, or raja.paul@iacs.res.in}
\papertype{Article}

\begin{document}

\begin{frontmatter}
\bigskip
\begin{abstract}
Upon contact with antigen presenting cells (APCs) cytotoxic T lymphocytes 
(T cells) establish a highly organized contact zone denoted as immunological synapse (IS). The formation of the IS implies relocation of the microtubule organizing center (MTOC) towards the contact zone, which necessitates a proper connection between MTOC and IS via dynamic microtubules (MTs). The efficiency of the MTs finding the IS within relevant time scale is, however, still illusive. We investigate how MTs search the three-dimensional constrained cellular volume for the IS and bind upon encounter to dynein anchored at the IS cortex. The search efficiency is estimated by calculating the time required for the MTs to reach the dynein-enriched region of the IS. In this study, we develop simple mathematical and numerical models incorporating relevant components of a cell and propose an optimal search strategy. Using the mathematical model, we have quantified the average search time for a wide range of model parameters and proposed an optimized set of values leading to the minimum capture time. Our results show that search times are minimal when the IS formed at the nearest or at the farthest sites on the cell surface, with respect to the perinuclear MTOC. The search time increases monotonically away from these two specific sites and are maximal at an intermediate position near the equator of the cell. We observed that search time strongly depends on the number of searching MTs and distance of the MTOC from the nuclear surface.

\end{abstract}

\end{frontmatter}

\section*{Introduction}
Cytotoxic T lymphocytes (T cells) play a crucial role in adaptive immunity by defending against virus infected and tumorigenic cells. Directional killing of an antigen presenting cell (APC) by T cells is completed in multiple steps including: (a) binding of the T cell receptor (TCR) to the cognate antigen presented by the major histocompatibility complex (MHC) on the surface of the APC. The interaction leads to the formation of a specialized T cell/APC junction denoted as the immunological synapse (IS) consisting of several supramolecular activation clusters (SMACs)~\cite{Huang2005,Andre1990,Dustin2010,Monks1998}, (b) establishment of the connection between the T cell's centrosome or microtubule organizing center (MTOC) and the synapse by the dynamic microtubules~\cite{Hammer2014, Hui2017}, (c) translocation of the MTOC to or near the target contact site achieved by the development of tension on microtubules~\cite{Schatten2011,Hammer2014,Kuhn_and_Peonie2002}, and (d) the subsequent secretion of the cytotoxic content of lytic vesicles at the IS via exocytosis, which kills the target cell~\cite{Kupfer1984,Yannelli1986,Pasternack1986,Krzewski_and_Coligan2012}.

Immediately after the establishment of the IS the T cell's microtubule organizing center (MTOC) moves to a position that is just underneath the plasma membrane at the center of the IS~\cite{Geiger1982,Stinchcombe2006,Kupfer1984}. One important consequence of MTOC relocation in T cells is that the MT minus end directed transport of vesicles containing cytotoxic material can be directed towards and terminated immediately adjacent to the bound APC for subsequent secretion via exocytosis~\cite{Kupfer1984,Yannelli1986,Pasternack1986,Krzewski_and_Coligan2012}. In quest of understanding MTOC translocation towards the synapse, suggested by \citet{Geiger1982}, modulated polarization microscopy (MPM) facilitated the visualization of the cytoskeleton and MTOC in living cells~\cite{Kuhn_and_Peonie2002,Kuhn2001}. MPM studies revealed that MTOC repositioning is associated with a development of tension on the microtubules resulting in a pulling on the MTOC towards the IS. Prior to the initiation of the MTOC relocation, the microtubules are seen to be nucleated from the MTOC, curve spontaneously following the shape of the cell boundary contacting the immunological synapse~\cite{Kuhn_and_Peonie2002}. It is suggested that actin polymerizing in a small patch at the center of the IS may facilitate microtubules to connect through IQGAP or CIP4 proteins~\cite{Fukata2002,Watanabe2004,Banerjee2007,Stinchcombe2006,Lansbergen2006,Kuroda1996}. During the development of the IS actin clears out from the center of the synapse. As actin spreads out in the form of an expanding ring, tension develops on the MTs as the ring enlarges~\cite{Stinchcombe2006,Bunnel2001}. The microtubule-associated molecular motor protein dynein assists this tension development on the MTs and it has been suggested that dyneins drive the MTOC towards the IS via cortical sliding~\cite{Kuhn_and_Peonie2002,Combs2006,Stinchcombe2014}.

\begin{figure*}[hbt!]
\centering
\includegraphics[width=0.78\linewidth]{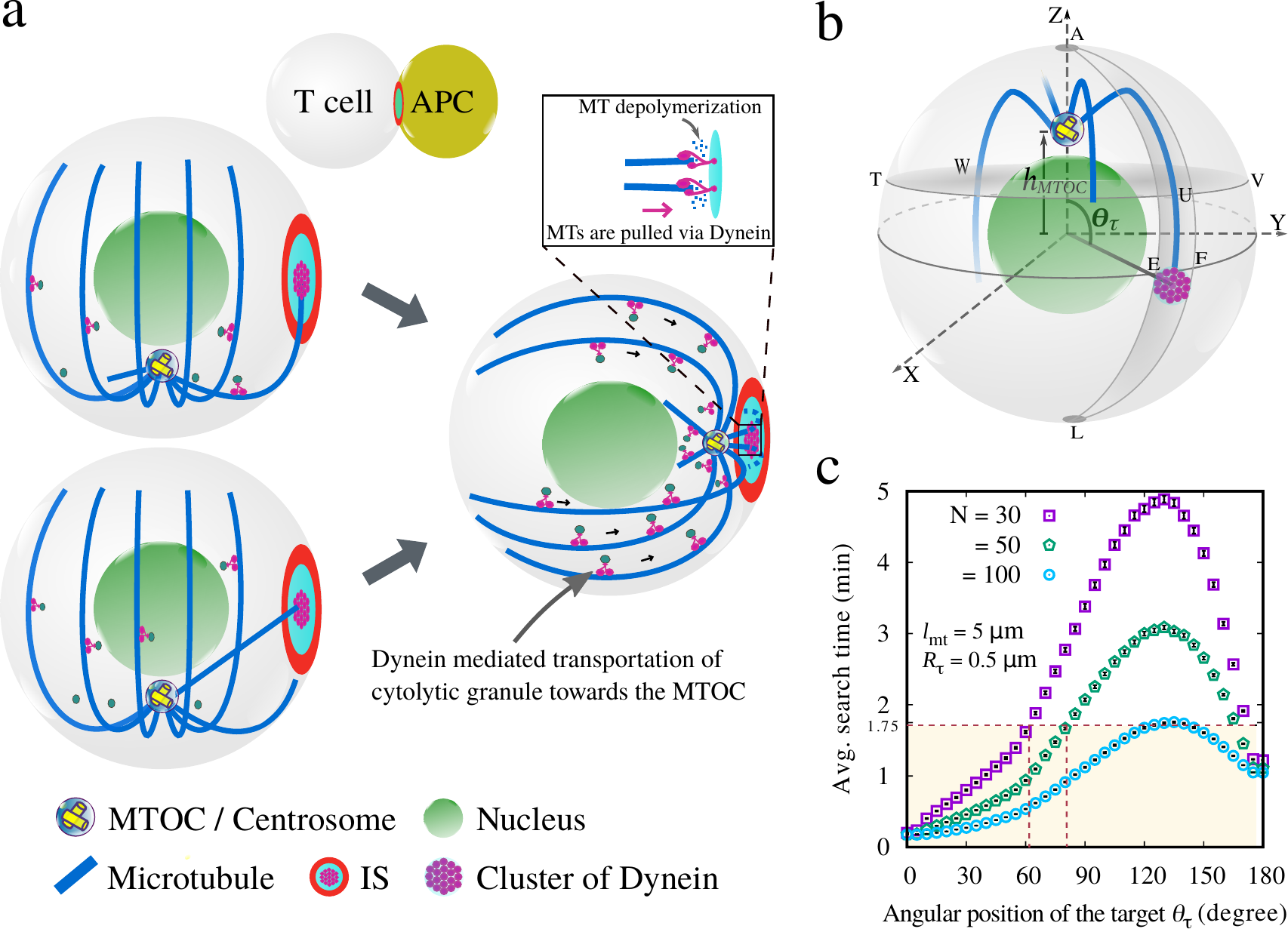}
\caption{{\bf{Schematics of the microtubule (MT) mediated end-on capture of the target localized at the center of the Immunological Synapse (IS) in T cell initiating the MTOC relocation}}. (a) Sketch of MTOC relocation driven by the `MT mediated end on capture shrinkage' mechanism \cite{Hammer2014}. T cell is represented by the outer sphere, the nucleus by the inner sphere, MTOC/Centrosome is yellow, MTs blue, IS red, and target (a cluster of dynein) is the central disc within IS. Relocation of the MTOC near the IS occurs following the capture of MTs by the dynein motors localized in a patch at the IS center and subsequent development of the tension on the MTs by dynein and MT depolymerization acting in concert (see the Inset). MTs can capture the target in two different ways: directly and/or by gliding along the cell periphery.
(b) Sketch of the theoretical model demonstrating MTs growing from the MTOC in random directions and capturing the target upon encounter. The MTOC is placed along the z-axis above the nucleus in a distance $h_{MTOC}$ from the cell center. The target is located on the cell periphery at an arbitrary polar angle $\theta_{\tau}$ (where $0^{\circ}\leq\theta_{\tau}\leq 180^{\circ}$) measured from the positive z-axis. For illustration, the target in this figure is positioned at a polar angle $\sim 90^{\circ}$. The target located above the imaginary horizontal plane $TUVW$ can be captured by the MTs directly and indirectly, whereas the target located below the plane is captured only indirectly. 
%Two semi-great circles $AEL$ and $AFL$ are tangential to the target and form the spherical lune $AELFA$. 
(c) Simulation results showing average search time as a function of the target polar angle $\theta_{\tau}$ with 30, 50 and 100 searching MTs. Notice that for $100$ searching MTs, the search time is within $1.75 \; min$, whereas for $30$ and $50$ searching MTs the target is successfully captured within this time scale if the target is located below the polar angle $~62^{\circ}$ and $~82^{\circ}$ respectively. Average search times maximize around $\theta_{\tau}=132^{\circ}$. The model parameters are: cell radius $R_C=5\;\SI{}{\micro\metre}$, $h_{MTOC}=$ nuclear radius $R_N=2.5 \;\SI{}{\micro\metre}$, average MT length $l_{mt}=5 \;\SI{}{\micro\metre}$, target radius $R_{\tau}=0.5\;\SI{}{\micro\metre}$, and the MT dynamic instability parameters: growth velocity $v_g=14.3 \;\SI{}{\micro\metre}/\text{min}$, shrink velocity $v_s=16 \;\SI{}{\micro\metre}/\text{min}$, catastrophe frequency $f_c=v_g/l_{mt}$ and rescue frequency $f_r=0$. The standard error of mean is smaller than the corresponding point sizes. To see this figure in color, go online.}
\label{fig:figure1}
\end{figure*}

A quantitative theoretical model for MTOC relocation based on the `cortical 
sliding' mechanism has been suggested by \citet{kim2009}. According to the proposed model the T cell is represented by a sphere with a radius of $5-10\;\SI{}{\micro\metre}$ that forces the individual MTs to run underneath and along the plasma membrane. MTs that pass over the T cell/APC interface are pulled by the dynein motors anchored at the IS cortex causing the MTs to slide past the dynein while drawing the MTOC towards the IS. For a range of conditions, \citet{kim2009} find that the cortical pulling mechanism is capable of reorienting the MTOC near the cell interface and the misorientations of the MTOC were found to increase if the IS formed at a position near the location that is symmetrically opposite to the location of T cell's MTOC. The inclusion of dynamic instability of the MTs in the model was found to have significant effects in improving the problem of misorientations. 

More recently \citet{Hammer2014} presented experimental evidence against the `cortical sliding' mechanism of MTOC relocation operating at the IS periphery and performed a series of experiments suggesting that the centrosome repositioning is driven by dynein by a `MT end-on capture-shrinkage' mechanism operating at the center of the IS. Using an optical trap APCs were placed such that the initial contact point with the T cell appears exactly opposite to the location of T cells MTOC. Soon after the initiation of a successful contact, several MTs in T cell are observed to project from the MTOC approaching towards the IS center by spontaneously bend along the cell outlet (as observed in Fig.5A of \citet{Hammer2014}), consistent with the previously observed MT structure in T cells (Fig. 6a of \citet{Kuhn_and_Peonie2002}). Once the microtubules plus end reach the IS center, dynein acts on it in an end-on fashion. The dynein power stroke then reels in the captured MTs that depolymerize (``shrink'') when pulled against the IS cortex, which results in a movement of the MTOC towards the IS. A schematic visualization of this `microtubule end-on capture-shrinkage' mechanism of MTOC repositioning mediated by dynein is depicted in FIG.~\ref{fig:figure1}a.

In contrast to the `cortical sliding' mechanism, where all MTs tangential to the IS can potentially bind to dynein, the `MT end-on capture-shrinkage' mechanism necessitates the plus ends of MTs to hit a small area in the center of the IS and bind to dynein within. Here we propose a stochastic `search and capture' model that assumes astral MTs grow and shrink rapidly in all directions from the MTOC, probing the inner surface of the plasma membrane until they capture a small dynein patch at the center of the IS. The remarkable concept of `search \& capture' was first proposed in the context of mitotic spindle assembly after the discovery of dynamic instability of MTs~\cite{Mitchison_and_Kirschner1984}. Dynamic MTs nucleate from the MTOC probing the cellular volume isotropically and arrest the chromosome upon encounter with the kinetochore, a proteinous complex assembled at the centromeric region of the chromosomes~\cite{Pavin2014, Hill1985}. Following several theoretical investigations~\cite{Hill1985,Leibler1994,Wollman2005,Paul2009,Gopalakrishnan2011,Mulder2012} it was postulated that for an efficient search (a) MT's rescue frequency must be zero (i.e., MTs do not waste time searching in a direction missing the target)~\cite{Leibler1994}, (b) catastrophe frequency is such that MTs grow on an average up to the target (i.e., MTs do not experience any early catastrophe while growing in the direction of the target)~\cite{Wollman2005, Leibler1994}. In addition, several proposed mechanisms that can accelerate the search process include spatially biased MT growth towards the chromosomes~\citep{Wollman2005, Paul2009, Rafael1999}, spatial arrangement of chromosome~\cite{Paul2009, Magidson2011}, kinetochore diffusion~\cite{Leibler1994, Paul2009}, interaction of MTs nucleated from kinetochore or/and chromosomes with the centrosomal MTs~\cite{Pavin2014, Paul2009, Kitamura2010, Karsenti1984, Sikirzhytski2014}, Augmin promoted nucleation of MTs from the preexisting MTs and their interaction with centrosomal MTs~\cite{Sabine2013, Carlos2015}, change in kinetochore architecture upon encounter MTs~\cite{Magidson2015}, and recently observed `pivoting and search' mechanism where MTs pivot around the spindle pole in a random manner exploring the space laterally~\cite{Pavin2014, Kalinina2013, Blackwell2017}. 

Although the term `search \& capture' was first coined in the context of chromosome search during mitosis, the concept is very generic and fundamental for numerous cellular processes. MTs are the key regulator of the `search \& capture' process and capture the target contact sites for successive cargo transport or serving as ropes for the molecular motors on which they can sit and pull the MTs facilitating a number of cellular functions like nuclear migration in budding yeast~\cite{Adames2000}, spindle positioning in {\it C.~elegans} embryos~\cite{Grill2003}, centrosome positioning in interphase cells~\cite{Som2019,Zhu2010}, orientation of cortical MT array in plant cells~\cite{Ambrose2011}, and nuclear-positioning in fission yeast facilitated by the formation of parallel microtubule bundles~\cite{Tran2001, Pavin2014}. Moreover, in certain yeast species, such as {\it C.~neoformans}, MT mediated capture and subsequent aggregation of the MTOCs organizes the pre-mitotic assembly of spindle pole body (SPB)~\cite{Sutradhar2015}. Likewise, reassembly of Golgi via the concentrated effort of centrosomal MTs and the Golgi driven MTs also thought to rely on a `search-capture' hypothesis~\cite{Vinogradova2012}.

In order to initialize the reorientation of the T cell’s MTOC towards the IS the efficiency of the `search-capture' process is crucial. Any perturbations in the MT dynamics may cause severe delay in the MTOC relocation process~\cite{Hammer2014}. To estimate the efficiency of the search strategy, we calculate the time required by the MTs to bind to the dynein enriched target at the central region of the IS (henceforth, synonymous with target). The faster the MTs find the target, the faster the MTOC can reorient and deliver various cargoes to the target site efficiently. We formulate a mathematical model for this process and show that for biologically relevant parameters of the process the MT capture at the IS, and concomitantly the initiation of the MTOC relocation, is robustly achieved on time-scales comparable to experimental measurements~\cite{Hammer2014}. Further, we developed a computational model using stochastic searching MTs, which confer additional support to the analysis and provide significant insights about the system. Utilizing the model, we could predict optimized values of the parameters corresponding to the minimal search time.

\section*{Methods}

\subsection*{Computational Model}
T cells under the conditions as in \cite{Hammer2014} have a spherical shape, with a radius of ca. $5-7\;\SI{}{\micro\metre}$, and have a large spherical nucleus leaving only a few $\SI{}{\micro\metre}$ thick shell for cytoplasm and cytoskeleton, in particular for the MTOC and MTs. Therefore we consider the cell and the nucleus as two concentric spheres centered at the origin of radii $R_{C}$ and $R_{N}$ respectively. The MTOC is placed along the z-axis above the north pole of the nucleus at $h_{MTOC}$ (the north pole of the nucleus is the point on the periphery of the nuclear envelope with $z=R_{N}$, see FIG.~\ref{fig:figure1}b). The dynein-enriched target zone in the center of the IS is modeled as a circular stationary patch embedded on the cell periphery depicted in FIG.~\ref{fig:figure1}b. MTs nucleate from the MTOC, grow and shrink rapidly while probing the space isotropically and capture the target upon encounter. Each MT is considered as a rod of negligible thickness with the plus end growing at a constant velocity ($v_g$) until a catastrophe occurring at a frequency $f_c$. Following a catastrophic event, the plus end of the MT shrinks with a constant velocity $v_s$. Upon rescue with a finite frequency $f_r$, MT grows again in the same direction, whereas the new cycle starts in a random direction if $f_r=0$.

Statistically, the distance covered by the MT tip, without undergoing a catastrophe is crucial to find the target placed on the cell periphery indicating that the average search time is a function of the average MT length $l_{mt}$ and dynamical parameters. Earlier studies on `search and capture' of chromosomes~\cite{Leibler1994,Wollman2005,Paul2009} suggested that an efficient search would require the MT to explore the space randomly in all possible directions. Accordingly, upon completion of an unsuccessful attempt to capture the target, MTs should not be rescued. An optimal search, therefore, primarily requires setting the rescue frequency to zero ($f_r=0$) and allowing the MTs to grow, on an average, up to the length $l_{mt}$ by regulating $f_c$ at a fixed $v_g$ through the relation $l_{mt}=v_g/f_c$ (see the \ref{s:supplementary}).

We study the microtubule dynamics by a stochastic model incorporating the four parameters ( $v_g,~v_s,~f_c,~f_r$) described earlier. The direction of the MT nucleated from the MTOC is described by a polar angle $\theta$ (where $\theta \in [0,\pi]$), and an azimuthal angle $\phi$ (where $\phi \in [0,2\pi]$) in the standard spherical polar frame of reference. Depending on the direction of MT nucleation two distinct scenarios arise: i) MT hit the nucleus and undergoes instantaneous catastrophe, leading to complete depolymerization and {\it de novo} nucleation of MT in a random direction and ii) MT encounter with the cell periphery spontaneously curves along the cell surface preferring the direction of minimum curvature. Prior to reaching the cell periphery, the MT remains straight. The MT is stabilized when the plus end makes a contact with the target and the target is said to be captured. Our model hypothesis of spontaneous gliding of MTs along the cell periphery is based on the observed MT contours, where individual MTs projected from the MTOC appeared to glide along the cell periphery approaching the target~\cite{Hammer2014, Kuhn_and_Peonie2002}. Additionally, we also considered the effect of cell boundary induced catastrophe of the MTs i.e., upon interacting with the cell wall, the MTs either undergo catastrophe or glide along the cell periphery, depending on the MT nucleation direction. Unless until explicitly specified elsewhere, the results are reported without considering any perturbation of MT dynamics due to the interaction between cell wall and MT plus tip. Additional technical details of the computational model are presented in the \ref{s:supplementary}.

Unless otherwise specified, we simulated with cell radius $R_C=5\;\SI{}{\micro\metre}$, nuclear radius $R_N=2.5 \;\SI{}{\micro\metre}$, target radius $R_{\tau}=0.5\;\SI{}{\micro\metre}$, and MT dynamic instability parameters: growth velocity $v_g=14.3 \;\SI{}{\micro\metre}/\text{min}$, shrink velocity $v_s=16 \;\SI{}{\micro\metre}/\text{min}$, catastrophe frequency $f_c=v_g/l_{mt}$ and rescue frequency $f_r=0$. The reference parameters and additional parameters around the reference value are recorded in Table~\ref{tab:table} in the \ref{s:supplementary}.

\section*{Results and Discussion}
\subsection*{Unbiased search by dynamic microtubules efficiently capture the target within experimentally relevant time scale}

The central question of the stochastic search-and-capture process described by our model is whether such a random process can lead to the capture of one or more of the MTs at the IS in a few minutes. Before we present the results of our detailed mathematical model (\ref{s:supplementary}; FIG.~\ref{fig:figureS1}), we estimate briefly the expected time scales based on a simplified picture. 
We assume that the T cell is spherical with radius $R_C=5\;\SI{}{\micro\metre}$ and, for simplicity, the MTOC is located somewhere immediately underneath the cell boundary. We identify the position of the MTOC with the north pole of the sphere and assume that the center of the target with radius $R_{\tau}=0.5\;\SI{}{\micro\metre}$ is located on the surface of the sphere at a polar angle $\theta_{\tau}$ with respect to the north pole. T cells are small, so we can consider the simplistic scenario that there is not much space between the nucleus and the plasma membrane and we can assume that MTs grow from the MTOC along the cell surface in a random direction. They grow on the average to a length $l_{mt} \approx \pi R_C \approx 15 \;\SI{}{\micro\metre}$ (i.e. from the north to the south pole) with a growth velocity $v_g\approx 15\;\SI{}{\micro\metre}/min$~\cite{Leibler1994} and then shrink with approximately the same speed $v_s$ unless they hit the target where the plus end of the MT is immediately captured. In case the target is missed, the average time for a MT to grow and shrink again is $T_u=l_{mt}/v_g+l_{mt}/v_s \approx 2\;min$; on completion of the cycle MT starts to grow in a new random direction. Thus $2\;min$ is the average time scale for an unsuccessful attempt of one MT to hit the target. The probability, $P_{direction}$, for a successful growth towards the target is the ratio between the diameter of the target ($2R_{\tau}=1 \;\SI{}{\micro\metre}$) and the perimeter of the circle at the polar angle $\theta_{\tau}$, $L=2\pi R_C \sin \theta_{\tau}$, provided $\theta_{\tau}$ is large. Note that, this imaginary circle inscribed at latitude $\theta_{\tau}$ is parallel to the equatorial circle. Since $L$ is the largest for the equatorial positioning of the target ($\theta_{\tau}=\pi/2$), $P_{direction}$ will be larger than $R_{\tau}/\pi R_C\approx 0.032$ elsewhere. The average number of unsuccessful trials (i.e. MTs grow and shrink without hitting the target) is $N_u=1/P_{direction} = \pi R_C/R_{\tau} \approx 31$, takes about $T_{av}=N_uT_u = 62$ minutes. Consequently a single dynamic MT needs, on the average, less than $62$ minutes to hit the target; likewise, for $N$ dynamic MTs, the average time reduces to $T_{av}(N)= max\{T_{av}/N, R_C \theta_{\tau}/v_g\}$; thus, for several tens of MTs the average capture time is just the time a MT needs to grow up to the target at polar angle $\theta_{\tau}$, i.e. less than $2$ minutes. This estimate agrees well with the experimentally observed time between the formation of the IS and the start of the movement of the MTOC~\cite{Hammer2014}.

In the following, we support this estimate by extensive simulations of a detailed quantitative model described earlier. We place the IS at various locations on the periphery of the cell to estimate the average search time as a function of the target position. The MTOC is placed along the z-axis above the nuclear surface at a distance $h_{MTOC}$ measured from the center of the cell/nucleus. MTs nucleate from the MTOC and search for the target uniformly in all possible directions. In the present model system, a single MT can capture the target in two different ways: a) directly and b) by gliding along the cell surface. 

\subsection*{Capture time is maximal for targets located near equatorial positions}

We choose the experimentally relevant cell radius $5\; \SI{}{\micro\metre}$ ~\cite{Hammer2014} and nucleus radius $2.5\; \SI{}{\micro\metre}$ \cite{Peglow2013,Maccari2016}. We set the average MT length ($l_{mt}$), to be determined by the dynamic instability parameters of the MT (See eq.~\ref{eq:l_mt} and eq.~\ref{eq:l_g} in the \ref{s:supplementary}), to $l_{mt}=5\; \SI{}{\micro\metre}$ and we use the optimal zero rescue frequency (FIG.~\ref{fig:figureS2}, \ref{s:supplementary}). 
In FIG.~\ref{fig:figure1}c we show the average search time as a function of the target polar angle. The maximum average capture time $\sim 1.75\; min$ for $100$ searching MTs is close to the experimental observation~\cite{Hammer2014} and supported by the result of the scaling arguments discussed in the last section. When the number of MTs is reduced to $50$ and $30$ the target is successfully captured within the same time limit provided the target polar angles does not exceed $\sim 82^{\circ}$ and $\sim 62^{\circ}$ respectively. For $N=50,30$ MTs the maximum capture time is $\sim 3\; min$ and $\sim 4.9 \; min$, respectively (FIG.~\ref{fig:figure1}c).
FIG.~\ref{fig:figure1}c shows that the mean search time as a function of target polar angle monotonically increases to a maximum and then falls off rapidly. The search time is minimum when the IS is formed at the proximal pole on the cell surface with respect to the MTOC (i.e., $\theta_{\tau}=0^{\circ}$). Interestingly, average search time increase only marginally when the target is located at the distal pole (i.e., $\theta_{\tau}=180^{\circ}$). Search time maximizes at an intermediate position of the target away from the poles. 

In FIG.~\ref{fig:figure2}a-c, we show the capture time due to a single MT as a function of the angular position $\theta_{\tau}$ of the target for different cell and nucleus radii ($R_C$ and $R_N$), target radii ($R_\tau$) and average MT lengths ($l_{mt}$)). We find that the search time decreases rapidly with increasing target size and that the variation is maximal when the target is localized around the equatorial periphery of the cell (i.e., $\theta_{\tau}=90^{\circ}$). FIG.~\ref{fig:figure2}b shows that the capture time is significantly reduced in the smaller T cell. 

Our simulation and analytically predicted data (see \ref{s:supplementary} for detailed analysis) independently predict the rise in the search time up to a certain angular position of the target reaching a maximum and then falling rapidly back to a value close to the initial one. A noticeable feature of the average search time profile is the kink at a specific polar angle of the target (indicated by arrows in FIG.~\ref{fig:figure2}a). The feature appears when straight MT filaments directed towards the target are hindered by the nucleus and incidentally lose minimum visibility of the target required for the direct capture (i.e. target transit from above the imaginary tangential plane $TUVW$ (FIG.~\ref{fig:figure1}b) to below).
%; $\angle AOD'$ to $\angle AOD''$ as shown in FIG. S1 in Supporting Material). 
Clearly, as long as the target is located above the imaginary tangential plane $TUVW$, it is fully visible to the MTs implying that the MTs can directly capture the target (with probability $P_{\;{direct,\; \tau\Uparrow}}$, c.f. \ref{s:supplementary}). Once the target starts crossing the plane $TUVW$ and becomes partially visible to the MTs, the direct capture probability ($P_{\;{direct,\; \tau\eqcirc}}$, c.f. \ref{s:supplementary}) vanishes rapidly leading to a sudden increase in the average capture time.

For longer MTs ($l_{mt}>> R_C$) the search time becomes maximum at a polar angle $\theta_{\tau} \sim 90^{\circ}$ of the target (FIG.~\ref{fig:figure2}a). Here the probability that the MT will be directed towards the target ($P_{direction}$) is minimal. For shorter MTs the probability that the MT will survive (i.e. no catastrophe), $P_{no \; cat}$, is small whenever the polar angle of the target is large since the MT has to grow a long distance to find the target. Thus, the effect of $P_{no \; cat}$ becomes more significant and the maximum of the average search time shifts towards larger polar angle $\theta_{\tau}$ of the target (FIG.~\ref{fig:figure2}b).

\subsection*{MTOC positioning within the cell significantly influences the capture time}

\begin{figure*}[hbt!]
\centering
\includegraphics[width=1\linewidth]{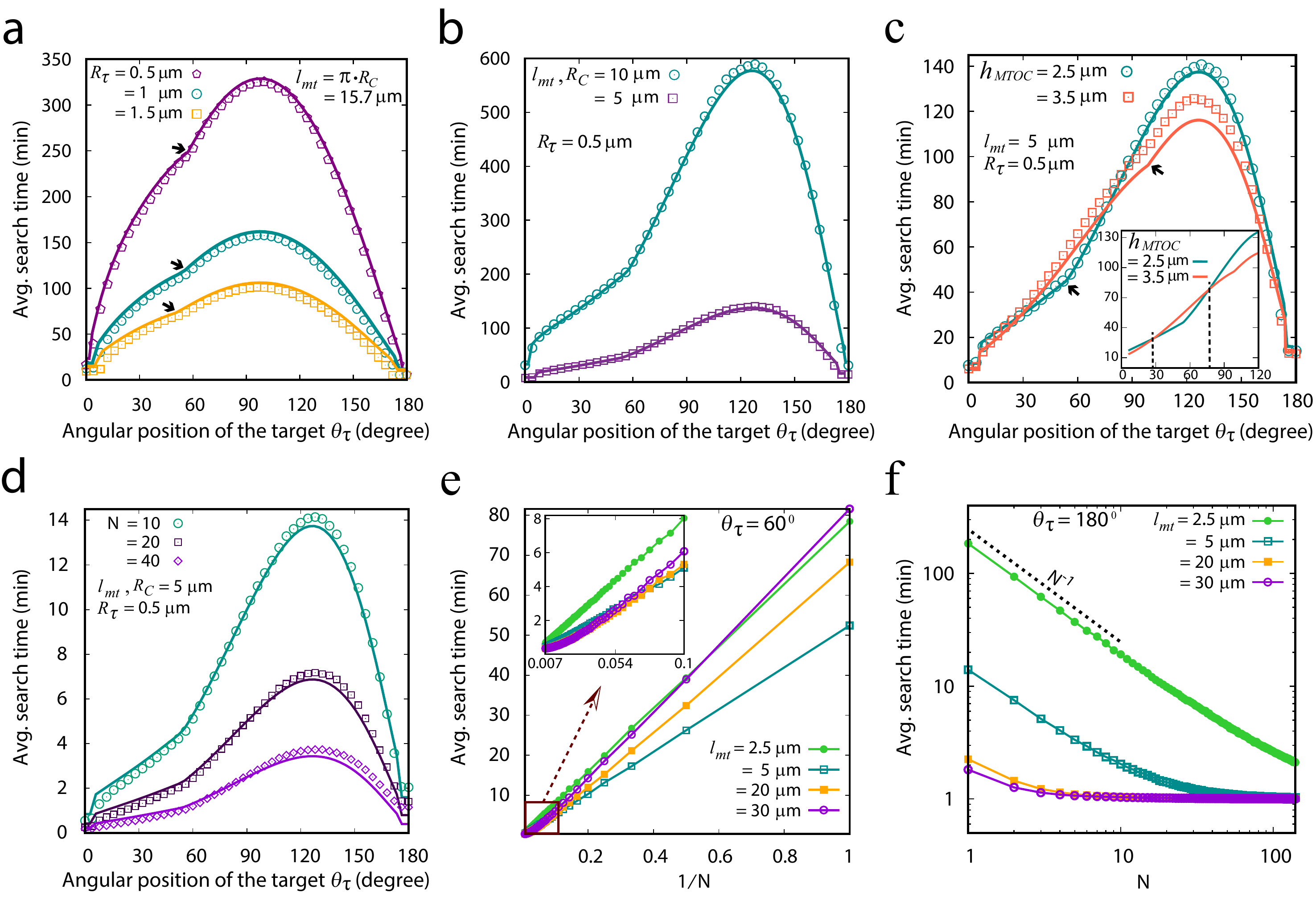}
\caption {{\bf{(a-c) Search time for a single MT varies with the angular position of the target $\theta_{\tau}$ for different parameters.}} The solid lines are from the analytical prediction whereas the simulated data are shown by points (a-c). (a) Search time for cell radius ${R_C=10 \;\SI{}{\micro\metre}}$, ${h_{MTOC}=R_{N}=5 \;\SI{}{\micro\metre}}$, and for varying target radius $R_{\tau}=0.5\; \SI{}{\micro\metre}$, $1\; \SI{}{\micro\metre}$ and $1.5\; \SI{}{\micro\metre}$. The black arrow indicates a sudden increase in the average search time, see text. (b) Search time for two different cell sizes: $R_C=10 \; \SI{}{\micro\metre}, \; h_{MTOC}=R_{N}=5 \; \SI{}{\micro\metre}$ and $R_C=5 \; \SI{}{\micro\metre}, \;h_{MTOC}=R_{N}=2.5 \; \SI{}{\micro\metre}$. Avg. MT length $l_{mt}=R_C$ and target size $R_{\tau}=0.5 \; \SI{}{\micro\metre}$. (c) Search time for different MTOC positions: $R_{\tau}=0.5 \; \SI{}{\micro\metre}$, $l_{mt}=R_C=5 \;\SI{}{\micro\metre}$, and $R_N=2.5\;\SI{}{\micro\metre}$, search times are shown for $h_{MTOC}=2.5\; \SI{}{\micro\metre}$ and $h_{MTOC}=3.5\; \SI{}{\micro\metre}$. Besides, having characteristic similarities with (a) and (b), we find two crossovers in the average search time around $\theta_{\tau}=30^{\circ} \; \text{and} \; 75^{\circ}$ (shown in the Inset). {\bf{(d-f) Search time decreases with the number of microtubules.}} Here we have fixed ${R_C=5}\;\SI{}{\micro\metre}$, ${h_{MTOC}=R_N=2.5\;\SI{}{\micro\metre}}$, ${R_{\tau}=0.5}\;\SI{}{\micro\metre}$ and $f_c=v_g/l_{mt}$. (d) Search time as a function of the angular position of the target $\theta_{\tau}$ with number of searching MTs $N=$ $10$, $20$ and $40$. The solid lines are obtained from the analytical prediction and the points represent the simulated value. (e) Search time is plotted against the inverse of the number of searching MTs (i.e. $1/N$) showing the time until capture decreases linearly with $1/N$ as the number of MTs increased. However, in the limit $N>>1$ average search time deviates from the linear behavior (shown in the Inset). (f) Average search time as a function of number of searching MTs ($N$) for different average length $l_{mt}$, in log-log scale. The target is placed at $\theta_{\tau}=180^{\circ}$. Notice, the average search time follows the inverse law (with $N$) indicated by the slope of the dotted reference line. To see this figure in color, go online.
%Following the predicted linear decrease, the search time tends to saturate %beyond a certain number of MTs depending on the average MT length $l_{mt}$. %Larger the average MT length, faster is the saturation.
}
\label{fig:figure2}
\end{figure*}

The proximity of the MTOC to the cell periphery also influences significantly the capture time of the IS. FIG.~\ref{fig:figure2}c demonstrates the variation of the average capture time as a function of the target polar angle $\theta_{\tau}$ for two different MTOC locations: in the proximity and away from the nuclear surface, quantified by the MTOC distances $h_{MTOC}$. The larger the $h_{MTOC}$ is, the closer the MTOC is to the membrane; thus targets in the proximity (small target polar angles) and targets at the opposite side of the nucleus (large target polar angles, for which searching in the MTOC proximity is ineffective), can be found earlier. FIG.~\ref{fig:figure2}c confirm this expectation and also shows that for intermediate angles a MTOC position closer to the nucleus is more advantageous. 

In order to understand MTOC position dependence on the capture time more quantitatively, we resort to the analytical calculation presented in the \ref{s:supplementary}. Essentially, the net capture probability is a sum of the probabilities due to the MTs capable of capturing the IS directly ($P_{direct}$) and the MTs having no direct access to the target but can search by gliding along the cell surface (i.e., $P_{glide}$ ). Comparison of the two probabilities provides a clear picture of the underlying process and described in the \ref{s:supplementary} (see FIG.~\ref{fig:figureS3}).
We notice that the simulation data deviate from our analytical predictions
for the larger $h_{MTOC}=3.5 \;\SI{}{\micro\metre}$, i.e., when MTOC is relatively closer to the plasma membrane or away from the nuclear surface. A large separation between the MTOC and the nuclear surface allows the MTs to explore the space between them. Although the search remains unsuccessful due the absence of IS in this region, the futile attempts by the MTs increases the average search time for capturing the target. The mathematical model presented here ignores this local search and hence deviate from the numerical data whenever MTOC is located away from the nuclear surface.

\subsection*{Capture time depends on the number of searching microtubules}

Search time shortens if more than one MT participate in the search process (see FIG.~\ref{fig:figure2}d-f). FIG.~\ref{fig:figure2}d demonstrates the average search time as a function of the target polar angle for $N=$ 10, 20 and 40 searching MTs. We find that the search time decreases linearly as $1/N$ (FIG.~\ref{fig:figure2}e) which is again supported by our analytical considerations (see the \ref{s:supplementary}). 
%Nevertheless, the analytical and the numerical curves marginally deviate from each other if the number of MT is large as shown in FIG.~\ref{fig:figure2}d for $N=40$. This discrepancy further emphasizes the departure of the average search time from $1/N$ behavior if $N$ is large (see the Inset of FIG.~\ref{fig:figure2}e). 

We further analyze this scenario by plotting the average search time as a function of the number of MTs for different target location ($\theta_{\tau}$). The search time tends to saturate beyond a certain number of MTs that depends on the average MT length. The saturation occurs earlier with longer average MT length $l_{mt}$ and with fewer MTs when the target is located at the distal pole of the cell (i.e., $\theta_{\tau}=180^{\circ}$) (FIG.~\ref{fig:figure2}f).
According to our analytical consideration elaborated in the \ref{s:supplementary} the probability of a successful capture by $N$ number of MTs is $N$ times the probability of a successful search by a single MT, $P_c$. When $NP_c$ approaches $1$ further increase of the number of MTs does not increase the capture probability anymore and the search time as a function of $N$ saturates. For instance, in a cell with radius $R_C=5 \;\SI{}{\micro\metre}$, perinuclear MTOC ($h_{MTOC}=R_N=2.5\;\SI{}{\micro\metre}$), average MT length $l_{mt}=30\; \SI{}{\micro\metre}$ and target radius $R_{\tau}=0.5 \;\SI{}{\micro\metre}$ located at the distant pole ($\theta_{\tau}=180^{\circ}$), two searching MTs ($N=2$) yield $NP_c \sim 2 \times 0.6 =1.2 \geq1$, ($P_c$ is determined using eq.~\ref{eq:p_capture_s}, \ref{s:supplementary}) ensuing the capture process purely deterministic.

\subsection*{Optimized MT length minimize the average capture time for a small number of MTs}

\begin{figure*}[hbt!]
\centering
\includegraphics[width=1\linewidth]{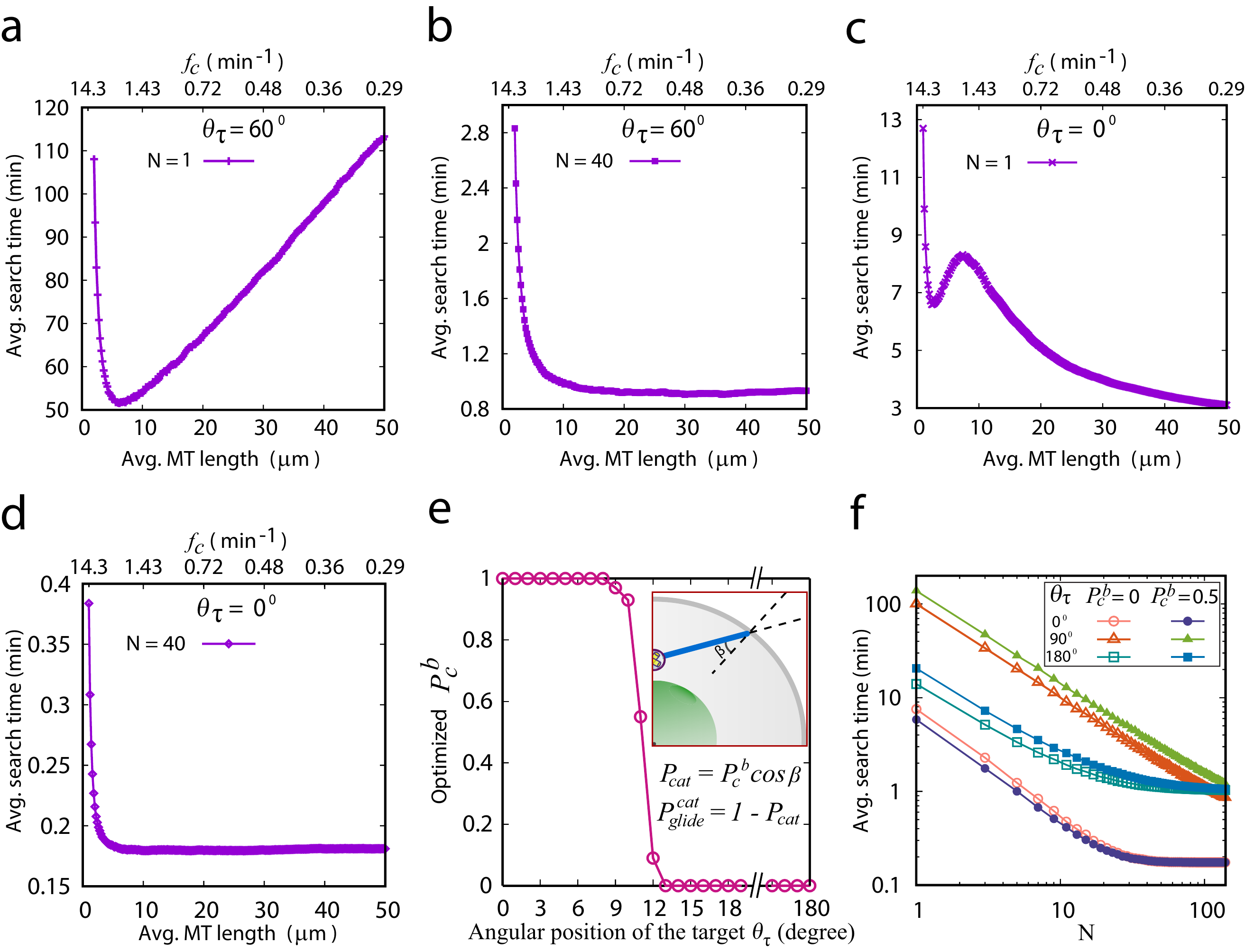}
\caption {{\bf{Search time as a function of the average MT length (a-d), and the effect of boundary induced catastrophe of the MTs (e-f)}}. Target is fixed at $\theta_{\tau}=60^{\circ}$ and $h_{MTOC}=2.5 \;\SI{}{\micro\metre}$. The upper $x$-axis indicate the MT catastrophe frequency $f_c$ defining the average MT length. 
(a) Search time as function of average MT length $l_{mt}$ for $N=1$,
target is fixed at $\theta_{\tau}=60^{\circ}$.
The minimum is at $l_{mt}\sim 6\;\SI{}{\micro\metre}$.
(b) Same as (a) but for $N=40$. The minimum vanished. 
(c) Same as (a) but with target fixed at $\theta_{\tau}=0^{\circ}$. 
The local minimum is at $l_{mt}\sim 2.5\;\SI{}{\micro\metre}$.
(d) Same as in (c) but for $N=40$. The minimum vanished. 
(e) Optimal probability parameter for boundary induced MT catastrophe as a function of the target location ($\theta_{\tau}$). 
%As long as the target remains in touch with the proximal pole of the cell, the minimum in the search time is obtained for $P_c^b=1$. As the target position shifts away from the north pole, the optimized value of $P_c^b$ falls suddenly to $0$. 
(f) Search time as a function of the number of searching MTs for three different target positions: $\theta_{\tau}=0^{\circ}$, $90^{\circ}$, and $180^{\circ}$ and for two different probability parameters for boundary induced MT catastrophe $P_c^b$: $0$, and $0.5$. To see this figure in color, go online.
%Note that for large number of microtubules, the average search time does not differ much from each other. For the target positioned at $\theta_{\tau}=0^{\circ}$ and $180^{\circ}$, the capture time becomes almost the same for two different $P_c^b$'s if the MT numbers are large. Further, the difference in the average search times for $\theta_{\tau}=90^{\circ}$ is insignificant and are $\leq 2 min$ when the number of MTs is large.
} 
\label{fig:figure3}
\end{figure*}

In order to explore the behavior of the search time with the average MT length ($l_{mt}$), the target is placed at a fixed polar angle $\theta_{\tau}=60^{\circ}$. The average search time is plotted as a function of the average MT length for different number of MTs in FIG.~\ref{fig:figure3}a-b. Our data shows that a single MT participating in the search process gives rise to a sharp minimum in the average search time for average MT length $\sim 6 \;\SI{}{\micro\metre}$ (i.e., $f_c \sim 2.38 \; min^{-1} $). This feature is characteristically very similar to the ``Search and Capture'' scenarios studied in the context of spindle assembly during mammalian mitosis ~\cite{Leibler1994,Wollman2005,Paul2009}. If the catastrophe frequency ($f_c$) is small, MTs waste time searching in the wrong directions; on the contrary, if the catastrophe frequency is large, MTs undergo premature catastrophe before reaching the target. Therefore, for an optimal search, $f_c$ (or $l_{mt}$) must be adjusted such that the MT grows just enough to reach the target. As the number of MTs increases, the minimum in the average search time becomes shallower for longer MTs (data not shown here) and finally vanishes for $N>>1$ (FIG.~\ref{fig:figure3}b). In this limit, the effective search probability $NP_c$ that at least one MT (out of $N$ MTs) will be able to capture the target, becomes very large and hence the optimization-feature is lost.

As expected, for the target localized near the south pole of the cell (i.e., $\theta_{\tau} \sim 180^{\circ}$), MTs nucleated at any azimuthal angle $\phi$ will have a finite probability to capture the target. However, for a short MT, this probability is too small resulting the capture time to diverge. On the contrary, longer MTs can always find the target; thus, the sharp minimum in the average search time vanishes even when a single MT is executing the search (data not shown here). If the target is located near the north pole of the cell (i.e., $\theta_{\tau} \sim 0^{\circ}$) and a single MT carries out the search, two possible ways the MT can progress: the MT captures the target directly or glides along the cell periphery and capture the target after a complete circle. For this specific position of the target, shorter MTs has a higher probability for direct capture, however the capture probability diminishes if the direct capture fails and subsequently MTs glide along the cell periphery. Hence we find a minimum in the average capture time if the average MT length $l_{mt}$ is close to the distance between the MTOC and the target (i.e., $2.5\;\SI{}{\micro\metre}$). Further increase of the $l_{mt}$ increases the average capture time. For large $l_{mt}$, probability of capturing the target along the cell periphery dominates, leading to a further decrease in the average capture time beyond a certain MT length (FIG.~\ref{fig:figure3}c). In the limit the number of MTs is large ($N>>1$), we see a monotonic decrease in the search time (FIG.~\ref{fig:figure3}d).

\subsection*{Effect of boundary induced catastrophe of MTs}
While in T cells, MTs emanating from the centrosome appear to follow the curvature of the plasma membrane towards the IS (c.f. \cite{Hammer2014}), it has been observed in other cell lines that MTs impinging upon the plasma membrane either undergo catastrophe or continue to grow along the cell periphery~\cite{Picone2010, Laan2012, Pavin2014}. Several studies show that the distribution of the MTs along the cell periphery is determined by the cell shape and the chances of the MTs to bend or catastrophe upon interaction with the cell periphery is guided by the angle of incidence at the cell cortex~\cite{Picone2010, Laan2012, Pavin2014}. Therefore, we analyze in this section the effect of boundary induced catastrophe upon the search process.

It appears plausible that MTs hitting the cell boundary nearly tangential at large angles of incidence (i.e. small angle between the cell periphery and the MT) are prone to glide along the cell periphery and those hitting at small angles undergo rapid catastrophe. The catastrophe can also be enhanced by an opposing force at the MT plus end, or by motor proteins such as kinesin-8 that does not depend on the angle of interaction~\cite{Foethke2009, Varga2009}. 
In the following, we consider only a probabilistic catastrophe induced by the cell boundary and depending on the MT's angle of incidence. Plausibly the catastrophe probability $P_{cat}$ should depend on the angle ($\beta$) between the MT and the local normal vector of the plasma membrane. We therefore assume $P_{cat} = P_c^{b}\cos\beta$ and gliding probability as $P^{\;cat}_{glide} = 1 - P_{cat}$, where $P_c^{b}$ is the catastrophe probability parameter varied between 0 and 1 (see \ref{s:supplementary}). Consequently, upon interaction with the cell periphery MTs undergo a catastrophe with zero probability for tangential incidence ($\beta=\pi/2$) and with probability $P_c^{b}$ for normal incidence (i.e., $\beta=0$) (see inset of FIG.~\ref{fig:figure3}e). Note that, in the absence of boundary induced catastrophe, MTs spontaneously glide along the cell periphery as considered earlier in this study. 

In FIG.~\ref{fig:figure3}e, we analyze the effect of the boundary induced catastrophe of individual MTs on the average search time for different target location. The optimized value of $P_c^b$ which corresponds to the minimum of the average search time is plotted as a function of target polar angle ($\theta_{\tau}$) for 10 searching MTs. We find that as long as the target remains in touch with the proximal pole of the cell with respect to the MTOC (north pole), minimum in the average search time is obtained for $P_c^b=1$. For this specific target positioning, MTs has a lesser chance to capture the target by gliding along the cell surface as the MTs having smaller average length can rarely capture the target by traversing a complete circle around the cell periphery. The shorter MTs that grows steadily and capture the target directly, dominate the capture process. In this case, $P_c^b=1$ would prevent most of the MTs to glide along the cell periphery resulting the emergence of a large number of shorter MTs which ultimately decrease the average search time by capture the target directly. Strikingly, the target located slightly away from the proximal pole (north pole), optimized $P_c^b$ decreases rapidly and eventually for all large polar angles of the target ($\theta_{\tau}$) the minimum in the average search time is obtained with $P_c^b=0$. For such target positioning, MTs that hit the cell periphery at polar angles smaller than the target location, glide along the cell periphery towards the target dominating the capture process. 

Next, in FIG.~\ref{fig:figure3}f, we plot the average search time as a function of the number of searching MTs for $P_c^b=0.5$ and for target positioned at $\theta_{\tau}=0^{\circ}$, $\theta_{\tau}=90^{\circ}$ and $\theta_{\tau}=180^{\circ}$ respectively. We compare these with the data obtained for $P_c^b=0$ (i.e. no catastrophe upon interaction with the cell membrane). 
We find that for small numbers of MTs the search times for catastrophe probability parameter $P_c^b=0.5$ differ marginally from $P_c^b=0$, whereas for large number of MTs the difference vanishes.

\subsection*{Proper tuning of the MT parameters and the number of searching MTs lead to faster searching}
In this section, we analyze how possible combinations of the relevant parameters affect quantitatively the efficiency of the capture process. First we calculate the search time as a function of the average MT length $l_{mt}$ and the number of searching MTs $N$ for three different positions of the target on the cell periphery viz, $\theta_{\tau} = 0^{\circ}$ (north pole), $90^{\circ}$ (equatorial plane) and $180^{\circ}$ (south pole) (FIG.~\ref{fig:figureS4}a-c, \ref{s:supplementary}). 
%The results are obtained from a parameter scan of average MT length $l_{mt}$ from $2\;\SI{}{\micro\metre}$ to $35\;\SI{}{\micro\metre}$ and number of searching MTs $N$ from $1$ to $50$. 
For $\theta_{\tau} \sim 0^{\circ}$, target is found early (in $\leq 2\; \text{min}$) even with short MTs (FIG.~\ref{fig:figureS4}a). For the target located at $\theta_{\tau}=90^{\circ}$, the search takes longer if the number of MTs is small and a large number of MTs ( $N\gtrsim 28$ ) is required to find the target in $\leq 2\; \text{min}$ (FIG.~\ref{fig:figureS4}b). For a wide range of $N$ and $l_{mt}$ the average search time can be less than $2\; \text{min}$ if the target located at south pole of the cell (i.e., $\theta_{\tau}=180^{\circ}$) (FIG.~\ref{fig:figureS4}c). Furthermore, the search time is reduced to $\sim 1.03 \; \text{min}$ and $\sim 0.83 \; \text{min}$ for $\theta_{\tau}=180^{\circ}$ with 10 searching MTs in a cell with $R_C=5 \;\SI{}{\micro\metre}$ and $4 \;\SI{}{\micro\metre}$~\cite{Maccari2016} respectively (FIG.~\ref{fig:figureS4}d).

Next, we analyze the variation of the search time with the MT growth velocity ($v_g$) at constant catastrophe frequencies (FIG.~\ref{fig:figureS5}a-c, \ref{s:supplementary}). The prevalent trend of the data indicate that the search time decreases rapidly with growth velocity and tends to saturate at large $v_g$. Capture time is reduced even further at smaller catastrophe frequencies with large number of MTs (FIG.~\ref{fig:figureS5}a, \ref{s:supplementary}). Intuitively, one can correlate this with rapidly growing MTs search the space relatively faster than sluggish MTs and the smaller catastrophe frequencies stabilize the MTs growing in the direction of the target. However, the search carried out with smaller number of MTs (e.g. $N=10$) shows deviation from the monotonic decrease in the search time; a minimum in the average search time is observed as a function of $v_g$ at relatively lower catastrophe frequencies (FIG.~\ref{fig:figureS5}b, \ref{s:supplementary}). Clearly, with a few searching MTs, large $v_g$ and small catastrophe frequency ($f_c$) tend to increase the average MT length and the MTs elapse more time searching in the wrong direction. When the number of MTs become too large, the probability that at least one MT will move towards the target converge to unity eliminating the optimization-feature (FIG.~\ref{fig:figureS5}a, \ref{s:supplementary}). However, if the target is located at the distal pole (south pole) of the cell (i.e., $\theta_{\tau}=180^ {\circ}$), the minimum in the average search time as a function of $v_g$ is not observed even with a small number of searching MTs (FIG.~\ref{fig:figureS5}c, \ref{s:supplementary}). The specific configuration facilitates the MT to pass through the target location; therefore, small catastrophe frequencies and large growth velocities always lead to smaller search time. The effect of the MT shrink velocity ($v_s$) on the average search time (FIG.~\ref{fig:figureS5}d, \ref{s:supplementary}) is not substantial for large number of MTs, as $v_s$ does not affect the average MT length significantly. Nevertheless, the effect become significant for a small number of searching MTs, because higher the shrink velocity, faster the MT can move back to the MTOC and grow afresh in another random direction looking for the target.

Another important parameter that can regulate the search time is the size of the nucleus. Depending on the target location on the cell periphery, average search time decreases non-monotonously with the increasing nuclear radius. Data suggests that if the target location is not directly accessible to the searching MTs due to steric hindrance from the nucleus (e.g. at $\theta_{\tau}=90^{\circ}$ in FIG.~\ref{fig:figureS5}e), the monotonous behavior is characteristically altered. The monotonous decrease in the average search time is due to the dominant effect of direct capture over the indirect capture which is evident from the feature of the average search time for the target polar angle $\theta_{\tau}=0^{\circ}$. Since the target remains always visible to the MTs at this position, direct capture leads to the monotonous variation of the average search time for all values of the nuclear radius (FIG.~\ref{fig:figureS5}e). Clearly, the change occurs when the radius of the nucleus crosses beyond a certain value for which the direct capture probability starts to diminish (see plot for $\theta_{\tau}=60^{\circ}$ in FIG.~\ref{fig:figureS5}e). In consistent with the expectation, the average search time is found to decrease with the target size (FIG.~\ref{fig:figureS5}f).

\section*{Conclusions}
Faithful reorientation of the T cell's MTOC towards the interface with the target cell is a pre-requisite for the polarized secretion of the lytic granules containing performin and granozymes at the target site triggering the target cell lysis~\cite{Schatten2011,Hammer2014,Kuhn_and_Peonie2002,Kupfer1984,Kupfer1986,Yannelli1986,Pasternack1986,Krzewski_and_Coligan2012}. Central to this, two mechanisms are thought to be involved in regulating the centrosome relocation. In the cortical sliding mechanism, dynein motors anchored at the periphery of the IS cortex reel in the microtubules emanating from the centrosome causing them to slide past the IS while drawing the centrosome towards the IS~\cite{Kuhn_and_Peonie2002,Combs2006,Stinchcombe2014,kim2009}. Recently~\citet{Hammer2014} presented evidence favoring a `microtubule end-on capture shrinkage' mechanism of MTOC repositioning in which dyneins localize at the IS center interact with the microtubule's plus end in an end-on fashion so as to couple the microtubule's depolymerization with the movement of the MTOC towards the IS. According to the capture shrinkage mechanism~\cite{Hammer2014}, the force generators (e.g., dyneins) are thought to localize at the center of the IS but not at the IS periphery as perceived earlier in cortical sliding mechanism. Invagination of the T cell membrane at the center of the IS towards the MTOC, observed in `frustrated' T cell/APC conjugates where the MTOC is stuck behind the nucleus, clearly argued that the force generation mechanism is focused at the IS center. Additionally, dynamical imaging of the MTs during normal repositioning showed a microtubule end-on capture shrinkage operating at the IS center. 

%Profound blockage of the MTOC repositioning by inhibition of both the dynein and MT depolymerization ensures the dependency of the mechanisms both on dynein activity and the MT depolymerization force~\cite{Hammer2014}, consistent with two other in vitro~\cite{Laan2012} and in vivo observation~\cite{Nguyen-Ngoc2007}. 

Our study addressed the question of effectiveness of this dynein mediated `capture-shrinkage mechanism' in the light of a `search-capture' model where individual microtubules, nucleated from MTOC, undergo dynamic instability until they are captured by dynein anchored at the IS cortex. We combined mathematical and numerical analyses to estimate the average time taken by the aster of MTs to secure an end-on attachment with dynein. Exponentially distributed time until capturing the target, obtained numerically is conceived as an input conjecture for the mathematical model (FIG.~\ref{fig:figureS6}, \ref{s:supplementary}). We find that the capture of the target is essentially a combination of a direct and an indirect processes: direct capture occurs when MTs hit the target without prior interaction with the cell or nucleus – this process is prevalent for target located in the cell-hemisphere containing the MTOC; indirect capture arises due to MTs that miss the target directly but glide along the cell surface seeking the target – this process is dominant in both the cell-hemispheres and is the only mechanism of capture in the hemisphere lacking the MTOC. 

We observed, in general, the search time largely depends on the relative size of the cell and the nucleus; e.g., search process becomes efficient upon reduction of the T cell diameter. Average search time for various location of the IS on the cell periphery suggests that a single microtubule would rapidly capture the target located near the proximal or the distal poles of the cell. Away from the polar regions, capture time increases and maximize at an intermediate angle determined by the system parameters. The monotonic increasing in the capture time transit through a sudden jump away from the proximal pole occurring when the chance of directly capturing the target diminishes. 

Although the target is rapidly captured at the distal pole, such positioning of the target might lead to a tug-of-war like scenario during MTOC relocation process. According to the observation of \citet{kim2009} and others~\cite{Serrador1999}, the MTOC sometime is stuck for long time behind the nucleus for this specific target positioning. Due to the symmetry of the MT array nucleated from the MTOC, the target can be reached by the MTs from all directions at the same time resulting a vanishing dynein mediated net pull applied on the MTOC. The pulling forces towards the IS are so symmetrically distributed that to create an imbalance in the forces, requiring to translocate the MTOC towards the IS, a long delay may occur. Interestingly \citet{Hammer2014} showed that in such `frustrated' condition the membrane of the IS center often invaginates in order to reach the stuck MTOC demonstrating that IS membrane can also move to the MTOC instead of MTOC comes to the IS.

During IS formation, often the centrosomes appear to be dissociated from the Nucleus both in migrating and in resting T cells ~\cite{Lui-Roberts2012}, while in most other cell types the centrosomes remain closely associated with the nuclear membrane through nesprins, a family of transmembrane proteins residing on the nuclear surface~\cite{Malone2003, Schneider2011}. \citet{Lui-Roberts2012} showed that T cells in which centrosomes were irreversibly `glued' to the nucleus by expressing GFP-BICD2-NT-nesprin-3, were still able to kill the target, following successful migration of their centrosome towards the IS. Naturally, the question arises why the position of the centrosomes varies in T cell. \citet{Lui-Roberts2012} speculated that the dynamic centrosome positioning might help the migrating cells to jiggle around seeking for the target cell. However, why the centrosomes in resting cells adopt different positions during IS formation has remained enigmatic. One possibility could be that the cell attempts to alter the search strategy that might be linked with the location of the target - as our study suggests that distance of the MTOC from the nuclear surface significantly alters the capture time. Depending on the MTOC position above the nucleus, efficiency of the capture process varies for various angular position of the target along the cell surface. Interestingly, the average search times for the MTOC located proximal to the nucleus and away (close to the cell membrane), cross each other at two distinct angular positions of the target (see FIG.~\ref{fig:figure2}c). Moreover, the MT array from the distant MTOC found to be more efficient in capturing the IS at small polar angles. For intermediate angular positions of the IS, MTs from the MTOC proximal to the nucleus are more effective while MTs from the distant MTOC capture the large-angle targets relatively early. Analyzing the dependency of the average search time on the MTOC position, we argue that T cell's MTOC may adopt different location largely to optimize the time required to capture the target.

According to our analysis, the average search time decreases with the number of searching MTs and saturates beyond a certain number of MTs that depends on the position of the IS and the average MT length. Since all MTs search in parallel, efficiency of the search process increases with the number of MTs. When a single MT carries out the search, an optimized value of the average MT length (or the catastrophe frequency) appears which corresponds to the minimum in the average search time. Interestingly, the optimization feature as a function of the average MT length diminishes if the number of MTs is very large. However, if the IS formed at the distal pole of the cell, the sharp minimum in the average capture time does not appear even in the context of a single searching MT. Since the average MT length is controlled by the growth velocity and catastrophe frequency, a minimum in the average search time as a function of the growth velocity appears at smaller catastrophe frequencies with a fewer searchers. A similar optimization feature at smaller catastrophe frequencies is not found with large number of MTs; instead, a higher growth velocity together with smaller catastrophe frequency leads to a rapid capture. Capture time drops monotonically and does not pass through a minima for the target located at the distal pole of the cell, even if the number of searching MTs is small. A similar type of feature is also observed in the context of kinetochore capture in fission yeast where the efficiency of the capture process is primarily determined by the average MT length~\cite{Blackwell2017}. Since the catastrophe frequency and the growth velocity primarily control the average MT length, these parameters largely influence the capture process. A tabular list of the different capture time scenarios depending on the number of searching MTs and the target location is shown in Table 1.

\begin{figure*}
\centering
\includegraphics[width=0.95\linewidth]{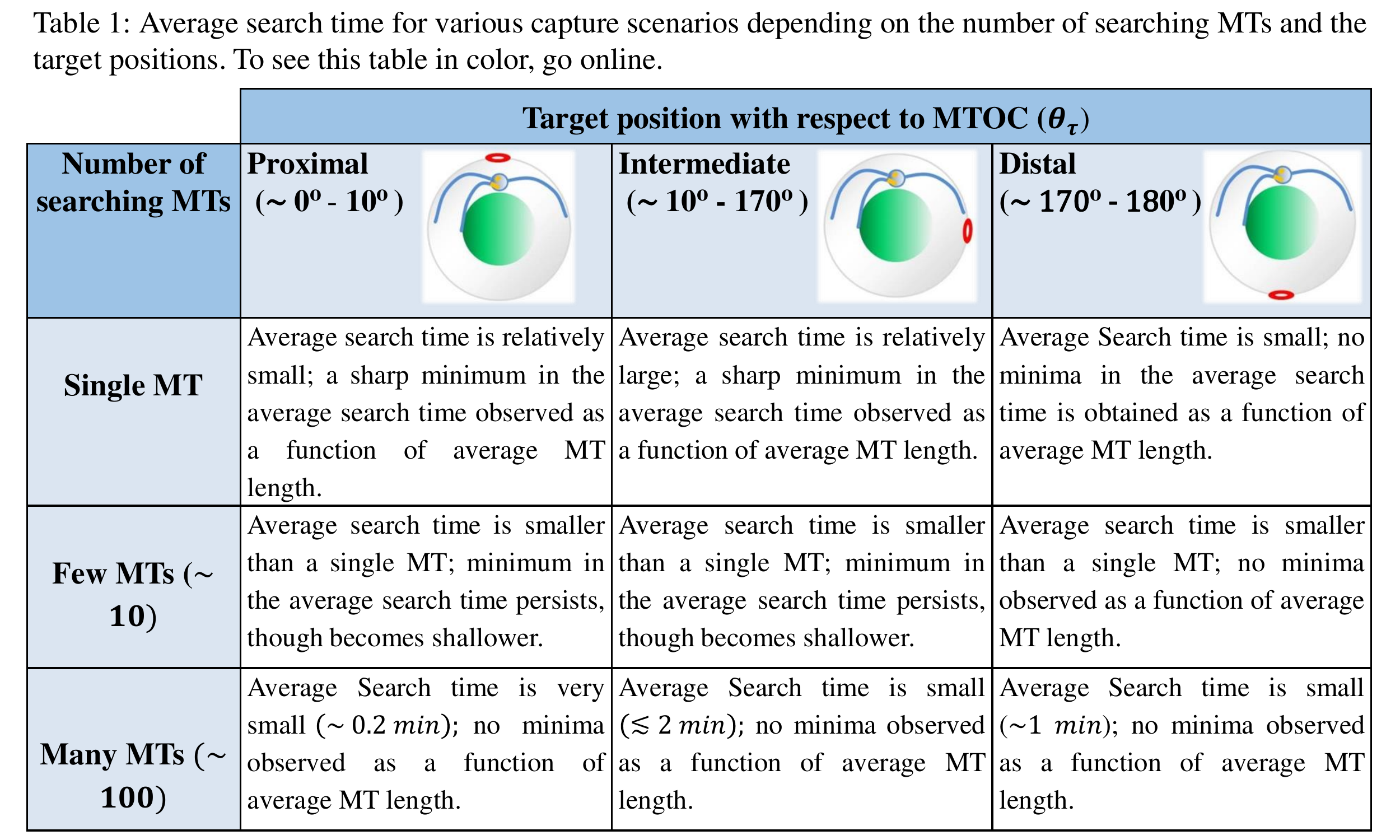}
\captionsetup{labelformat=empty}
%\caption {}
\label{fig:figure6}
\end{figure*}

Based on the observation, we argue that optimization in the average search time is a robust mechanism that leads to the rapid capture of the target and therefore a prerequisite with a fewer searchers. However, the optimization is not essential if the system contains a large number of MTs. Consequently, the set of fewer MTs also needs to be more dynamic to capture the target rapidly, whereas the search becomes faster with large number of MTs when they are stable. Analysis of the average search time suggests that larger targets reduce the capture time. Search time also changes with the size of the nucleus. With increasing nuclear radius, average search time decreases non-monotonically for targets with no direct capture possibility; targets visible to the MTs prior to hitting the cell periphery, direct capture probability dominates over the indirect capture resulting a monotonous decrease in the average search time with nuclear size. 

It is very likely that the cell and the nucleus could be different from a sphere; therefore, an extension of the current approach could include the analysis of capture efficiency considering aspherical architecture of cell and nucleus. In fact, it would be interesting to consider a spheroidal cell with the MTOC placed along its major/minor axis. Efficiency of the MTs capturing the target would depend on the specific positioning of the MTOC inside the cell. Similarly, the nucleus can also take a spheroidal shape. Since the nucleus acts as a steric hindrance for the MTs growing straightforwardly towards the target, the capture time would clearly depend on the angular position of the MTOC and the target. Besides, off-centered positioning of nucleus in the current model could also be relevant and more realistic. Nucleus positioned away from the center of the cell can regulate the capture time by increasing/decreasing the effective distance between the target and the perinuclear MTOC. 

Although we have developed the model to understand the optimal search strategy in T cell, the model itself is more general. Most importantly, our model can be used to understand search processes where direct as well as indirect search guided by the topology of the gliding surface on a curved manifold are significant. In fact, in the pre-mitotic assembly of central pole body (SPB) in yeast ({\it C. neoformans}) via MT mediated capture and aggregation of MTOCs, the current model can be extended to estimate the time required to form an SPB. In future, it would also be interesting to generalize the model to predict the mitotic assembly time, with emphasis on the recently explored MT features. Moreover, our current model can be used directly or can be modified in a context dependent manner to elucidate the efficiency of search strategies that crucially regulate many cellular functions in various organisms.

Certain aspects of our model motivate further experiments. It appears that the MT must glide along the plasma membrane in order to obtain an MT cytoskeleton organization referred in~\cite{Hammer2014}, a systematic study observing the dynamics of MTs growing along the plasma membrane in T Cells remain obscure.
We analyzed potential boundary induced MT catastrophe in a phenomenological way by introducing a hitting angle dependent catastrophe probability and found that boundary induced MT catastrophe has even an advantageous effect for small target angles and only mild effects for large target angles. 
For a better microscopic understanding, the gliding of MTs along the plasma membrane while one end being clamped at the MTOC raises two major concerns: (a) whether the MTs have sufficient resistance to bend without disintegrating the MT lattice. Previous studies suggest that the MTs have mechanical properties analogous to plexiglass and rigid plastics~\cite{Gittes1993}, and the yield strength of MT is similar to that of polymethylmethacrylate (i.e., 40-70 MPa)~\cite{Botvinick2004}, and with 25 nm diameter they are sufficient enough to bend along the plasma membrane bearing the tensile load applied on it~\cite{kim2009}. Strikingly, recent studies by~\citet{Schaedel2015} showed that MTs that damaged over extensive load might recover their initial stiffness by incorporating tubulin dimer into their lattice. (b) Whether the MT catastrophe upon hitting the plasma membrane is biochemically suppressed or whether for instance vimentin filaments induce the increased curvature of MTs in the T cell. The MT plus-end tracking proteins or +TIPs are well known for its ability to accumulate at the MT plus end and could modulate the MT dynamics. The most conserved +TIPs are the end-binding proteins (EBs) which may either interact directly with the MT plus end or in combination with other +TIPs (e.g., cytoplasmic dynein, dynactin, CLIP-170, $p150^{Glued}$ etc.)~\cite{Wu2006}. An observation made in the mammalian cells by \citet{Komarova2009} showed that EBs (particularly EB1 and EB2) have little effect on MTs rescue but actively suppress the MT catastrophe. The observation of \citet{Komarova2009} is in the line with the observed anti-catastrophe activity promoted by the proper homologs of mammalian EBs in fission yeast~\cite{Busch2004} and in Xenopus extracts~\cite{Tirnauer2002}. In addition, the intermediate filaments vimentin (VIF), regarded as a long live copy of the MT filaments, could also regulate the MT dynamics via interacting with the microtubule through several proteins~\cite{Prahlad1998,Huber2015}. It is observed that the motile cells like fibroblasts and lymphocytes have higher VIF levels~\cite{Mendez2010} and recently ~\citet{Gan2016} showed that VIF could increase the persistence of the MTs to enhance the directed cell migration. Likewise, other cytoskeleton proteins, such as stable F-actin structures, could also guide MTs organization. Similar experiments in T cell could provide a clear picture of the underlying mechanism of MTs growth along the Plasma Membrane in T cell. It would be crucial also to identify the characteristics of dynein's cortical activities at the IS center~\cite{Hammer2014} and the molecular basis of the bond formation between the MT plus end and the dynein.

\section*{Author Contributions}

R.P. and H.R. designed the research, A.S. carried out all simulations and analyzed the data, R.P., A.S. and H.R. wrote the article.

\section*{Acknowledgments}
R.P. thanks Grant No. EMR/2017/001346 of SERB, DST, India for the computational facility. This work was financially supported by the German Research Foundation (DFG) within the Collaborative Research Center SFB 1027 (H.R.) and R.P. thanks the SFB 1027 for supporting his visit to the Saarland University for discussion and finalizing the project. A.S.'s fellowship was supported by the University Grants Commission (UGC), India.

\section*{Supporting Citations}
References~\cite{Verde1992, Dogterom1993, Sokolnikoff1966, Todhunter1886, Ross1972} appear in the \ref{s:supplementary}.

\bibliography{ManuscriptBibTex}

\begin{thebibliography}{77}
\providecommand{\url}[1]{\texttt{#1}}
\providecommand{\urlprefix}{ }

\bibitem[Huang et~al.(2005)Huang, Norton, Precht, Martindale, Burkhardt, and
  Wange]{Huang2005}
Huang, Y., D.~Norton, P.~Precht, J.~Martindale, J.~Burkhardt, and R.~Wange,
  2005.
\newblock Deficiency of ADAP/Fyb/SLAP-130 Destabilizes SKAP55 in Jurkat T
  Cells.
\newblock \emph{The Journal of Biological Chemistry} 280:23576--23583.

\bibitem[Andre et~al.(1990)Andre, Benoliel, Capo, Foa, Buferne, Boyer,
  Schmitt-Verhulst, and Bongrand]{Andre1990}
Andre, P., A.~Benoliel, C.~Capo, C.~Foa, M.~Buferne, C.~Boyer,
  A.~Schmitt-Verhulst, and P.~Bongrand, 1990.
\newblock Use of conjugates made between a cytolytic T cell clone and target
  cells to study the redistribution of membrane molecules in cell contact
  areas.
\newblock \emph{Journal of Cell Science} 97:335--347.

\bibitem[Dustin et~al.(2010)Dustin, Chakraborty, and Shaw]{Dustin2010}
Dustin, M., A.~Chakraborty, and A.~Shaw, 2010.
\newblock Understanding the Structure and Function of the Immunological
  Synapse.
\newblock \emph{Cold Spring Harb Perspect Biol. 2(10):a002311} 2.

\bibitem[Monks et~al.(1998)Monks, Freiberg, Kupfer, Sciaky, and
  Kupfer]{Monks1998}
Monks, C., B.~Freiberg, H.~Kupfer, N.~Sciaky, and A.~Kupfer, 1998.
\newblock Three-dimensional segregation of supramolecular activation clusters
  in Tcells.
\newblock \emph{Nature} 395:82--86.

\bibitem[Yi et~al.(2014)Yi, Wu, Chung, Chen, Kapoor, and Hammer]{Hammer2014}
Yi, J., X.~Wu, A.~H. Chung, J.~K. Chen, T.~M. Kapoor, and J.~A. Hammer, 2014.
\newblock Centrosome repositioning in T cells is biphasic and driven by
  microtubule end-on capture-shrinkage.
\newblock \emph{J. Cell Biol.} 202:779–792.

\bibitem[Hui and Upadhyaya(2017)]{Hui2017}
Hui, K.~L., and A.~Upadhyaya, 2017.
\newblock Dynamic microtubules regulate cellular contractility during T-cell
  activation.
\newblock \emph{Proc Natl Acad Sci U S A.} 114:E4175--E4183.

\bibitem[Schatten(2011)]{Schatten2011}
Schatten, H., 2011.
\newblock The Centrosome: Cell and Molecular Mechanisms of Functions and
  Dysfunctions in Disease.
\newblock Humana Press, Department of Veterinary Pathobiology, University of
  Missouri-Columbia, Columbia MI, USA.

\bibitem[Kuhn and Poenie(2002)]{Kuhn_and_Peonie2002}
Kuhn, R.~J., and M.~Poenie, 2002.
\newblock Dynamic Polarization of the Microtubule Cytoskeleton during
  CTL-Mediated Killing.
\newblock \emph{Immunity} 16:111--121.

\bibitem[Kupfer and Dennert(1984)]{Kupfer1984}
Kupfer, A., and G.~Dennert, 1984.
\newblock Reorientation of the microtubule-organizing center and the Golgi
  apparatus in cloned cytotoxic lymphocytes triggered by binding to lysable
  target cells.
\newblock \emph{J Immunol.} 133:2762--2766.

\bibitem[Yannelli et~al.(1986)Yannelli, Sullivan, Mandell, and
  Engelhard]{Yannelli1986}
Yannelli, J., J.~Sullivan, G.~Mandell, and V.~Engelhard, 1986.
\newblock Reorientation and fusion of cytotoxic T lymphocyte granules after
  interaction with target cells as determined by high resolution
  cinemicrography.
\newblock \emph{J Immunol.} 136:377--382.

\bibitem[Pasternack et~al.(1986)Pasternack, Verret, Liu, and
  Eisen]{Pasternack1986}
Pasternack, M.~S., C.~R. Verret, M.~A. Liu, and H.~N. Eisen, 1986.
\newblock Serine esterase in cytolytic T lymphocytes.
\newblock \emph{Nature} 322:740--743.

\bibitem[Krzewski and Coligan(2012)]{Krzewski_and_Coligan2012}
Krzewski, K., and J.~Coligan, 2012.
\newblock Human NK cell lytic granules and regulation of their exocytosis.
\newblock \emph{Front Immunol.} 3.

\bibitem[Geiger et~al.(1982)Geiger, Rosen, and Berke]{Geiger1982}
Geiger, B., D.~Rosen, and G.~Berke, 1982.
\newblock Spatial relationships of microtubule-organizing centers and the
  contact area of cytotoxic T lymphocytes and target cells.
\newblock \emph{J. Cell Biol.} 95:137--143.

\bibitem[Stinchcombe et~al.(2006)Stinchcombe, Majorovits, Bossi, Fuller, and
  Griffiths]{Stinchcombe2006}
Stinchcombe, J.~C., E.~Majorovits, G.~Bossi, S.~Fuller, and G.~M. Griffiths,
  2006.
\newblock Centrosome polarization delivers secretory granules to the
  immunological synapse.
\newblock \emph{Nature} 443:462–465.

\bibitem[Kuhn et~al.(2001)Kuhn, Wu, and Poenie]{Kuhn2001}
Kuhn, J., Z.~Wu, and M.~Poenie, 2001.
\newblock Modulated polarization microscopy: a promising new approach to
  visualizing cytoskeletal dynamics in living cells.
\newblock \emph{Biophys J.} 80:972--985.

\bibitem[Fukata et~al.(2002)Fukata, Watanabe, Noritake, Nakagawa, Yamaga,
  Kuroda, Matsuura, Iwamatsu, Perez, and Kaibuchi]{Fukata2002}
Fukata, M., T.~Watanabe, J.~Noritake, M.~Nakagawa, M.~Yamaga, S.~Kuroda,
  Y.~Matsuura, A.~Iwamatsu, F.~Perez, and K.~Kaibuchi, 2002.
\newblock Rac1 and Cdc42 capture microtubules through IQGAP1 and CLIP-170.
\newblock \emph{Cell} 109:873--885.

\bibitem[Watanabe et~al.(2004)Watanabe, Wang, Noritake, Sato, Fukata, Takefuji,
  Nakagawa, Izumi, Akiyama, and Kaibuchi]{Watanabe2004}
Watanabe, T., S.~Wang, J.~Noritake, K.~Sato, M.~Fukata, M.~Takefuji,
  M.~Nakagawa, N.~Izumi, T.~Akiyama, and K.~Kaibuchi, 2004.
\newblock Interaction with IQGAP1 Links APC to Rac1, Cdc42, and Actin Filaments
  during Cell Polarization and Migration.
\newblock \emph{Dev Cell.} 7:871–883.

\bibitem[Banerjee et~al.(2007)Banerjee, Pandey, Zheng, Suhoski, Monaco-Shawver,
  and Orange]{Banerjee2007}
Banerjee, P., R.~Pandey, R.~Zheng, M.~Suhoski, L.~Monaco-Shawver, and
  J.~Orange, 2007.
\newblock Cdc42-interacting protein-4 functionally links actin and microtubule
  networks at the cytolytic NK cell immunological synapse.
\newblock \emph{J Exp Med.} 204:2305--2320.

\bibitem[Lansbergen and Akhmanova(2006)]{Lansbergen2006}
Lansbergen, G., and A.~Akhmanova, 2006.
\newblock Microtubule plus end: a hub of cellular activities.
\newblock \emph{Traffic.} 7:499--507.

\bibitem[Kuroda et~al.(1996)Kuroda, Fukata, Kobayashi, Nakafuku, Nomura,
  Iwamatsu, and Kaibuchi]{Kuroda1996}
Kuroda, S., M.~Fukata, K.~Kobayashi, M.~Nakafuku, N.~Nomura, A.~Iwamatsu, and
  K.~Kaibuchi, 1996.
\newblock Identification of IQGAP as a Putative Target for the Small GTPases,
  Cdc42 and Rac1.
\newblock \emph{The Journal of Biological Chemistry} 271:23363--23367.

\bibitem[Bunnell et~al.(2001)Bunnell, Kapoor, Trible, Zhang, and
  Samelson]{Bunnel2001}
Bunnell, S.~C., V.~Kapoor, R.~P. Trible, W.~Zhang, and L.~E. Samelson, 2001.
\newblock Dynamic Actin Polymerization Drives T Cell Receptor–Induced
  Spreading: A Role for the Signal Transduction Adaptor LAT.
\newblock \emph{Immunity} 14:315--329.

\bibitem[Combs et~al.(2006)Combs, Kim, Tan, Ligon, Holzbaur, Kuhn, and
  Poenie]{Combs2006}
Combs, J., S.~Kim, S.~Tan, L.~Ligon, E.~Holzbaur, J.~Kuhn, and M.~Poenie, 2006.
\newblock Recruitment of dynein to the Jurkat immunological synapse.
\newblock \emph{Proc Natl Acad Sci U S A.} 103:14883--8.

\bibitem[Stinchcombe and Griffiths(2014)]{Stinchcombe2014}
Stinchcombe, J., and G.~Griffiths, 2014.
\newblock Communication, the centrosome and the immunological synapse.
\newblock \emph{Phil. Trans. R. Soc. B} 369.

\bibitem[Kim and Maly(2009)]{kim2009}
Kim, M., and I.~V. Maly, 2009.
\newblock Deterministic Mechanical Model of T-Killer Cell Polarization
  Reproduces the Wandering of Aim between Simultaneously Engaged Targets.
\newblock \emph{PLoS Comput Biol} 5.

\bibitem[Mitchison and Kirschner(1984)]{Mitchison_and_Kirschner1984}
Mitchison, T., and M.~Kirschner, 1984.
\newblock Dynamic instability of microtubule growth.
\newblock \emph{Nature} 312:237--242.

\bibitem[Pavin and Tolić-Nørrelykke(2014)]{Pavin2014}
Pavin, N., and I.~M. Tolić-Nørrelykke, 2014.
\newblock Swinging a sword: how microtubules search for their targets.
\newblock \emph{Syst Synth Biol.} 8:1103.

\bibitem[Hill(1985)]{Hill1985}
Hill, T.~L., 1985.
\newblock Theoretical problems related to the attachment of microtubules to
  kinetochores.
\newblock \emph{Proc Natl Acad Sci U S A} 82:4404–4408.

\bibitem[Holy and Leibler(1994)]{Leibler1994}
Holy, T., and S.~Leibler, 1994.
\newblock Dynamic instability of microtubules as an efficient way to search in
  space.
\newblock \emph{Proc Natl Acad Sci U S A.} 91:5682–5685.

\bibitem[Wollman et~al.(2005)Wollman, Cytrynbaum, Jones, Meyer, Scholey, and
  Mogilner]{Wollman2005}
Wollman, R., E.~Cytrynbaum, J.~Jones, T.~Meyer, J.~Scholey, and A.~Mogilner,
  2005.
\newblock Efficient chromosome capture requires a bias in the
  `search-and-capture’ process during mitotic spindle assembly.
\newblock \emph{Curr Biol.} 15:828--832.

\bibitem[Paul et~al.(2009)Paul, Wollman, Silkworth, Nardi, Cimini, and
  Mogilner]{Paul2009}
Paul, R., R.~Wollman, W.~T. Silkworth, I.~K. Nardi, D.~Cimini, and A.~Mogilner,
  2009.
\newblock Computer simulations predict that chromosome movements and rotations
  accelerate mitotic spindle assembly without compromising accuracy.
\newblock \emph{Proc Natl Acad Sci U S A} 106:15708--15713.

\bibitem[Gopalakrishnan and Govindan(2011)]{Gopalakrishnan2011}
Gopalakrishnan, M., and B.~S. Govindan, 2011.
\newblock A First-Passage-Time Theory for Search and Capture of Chromosomes by
  Microtubules in Mitosis.
\newblock \emph{Bulletin of Mathematical Biology} 73:2483--2506.

\bibitem[Mulder(2011)]{Mulder2012}
Mulder, B.~M., 2011.
\newblock Microtubules interacting with a boundary: mean length and mean
  first-passage times.
\newblock \emph{Phys. Rev. E} 86:011902.

\bibitem[Carazo-Salas et~al.(1999)Carazo-Salas, Guarguaglini, Gruss, Segref,
  Karsenti, and Mattaj]{Rafael1999}
Carazo-Salas, R.~E., G.~Guarguaglini, O.~J. Gruss, A.~Segref, E.~Karsenti, and
  I.~W. Mattaj, 1999.
\newblock Generation of GTP-bound Ran by RCC1 is required for chromatin-induced
  mitotic spindle formation.
\newblock \emph{Nature} 400.

\bibitem[Magidson et~al.(2011)Magidson, O'Connell, Lončarek, Paul, Mogilner,
  and Khodjakov]{Magidson2011}
Magidson, V., C.~B. O'Connell, J.~Lončarek, R.~Paul, A.~Mogilner, and
  A.~Khodjakov, 2011.
\newblock The Spatial Arrangement of Chromosomes during Prometaphase
  Facilitates Spindle Assembly.
\newblock \emph{Cell} 146:555--567.

\bibitem[Kitamura et~al.(2010)Kitamura, Tanaka, Komoto, Kitamura, Antony, and
  Tanaka]{Kitamura2010}
Kitamura, E., K.~Tanaka, S.~Komoto, Y.~Kitamura, C.~Antony, and T.~U. Tanaka,
  2010.
\newblock Kinetochores Generate Microtubules with Distal Plus Ends: Their Roles
  and Limited Lifetime in Mitosis.
\newblock \emph{Dev Cell.} 18:248--259.

\bibitem[Karsenti et~al.(1984)Karsenti, Newport, and Kirschner]{Karsenti1984}
Karsenti, E., J.~Newport, and M.~Kirschner, 1984.
\newblock Respective roles of centrosomes and chromatin in the conversion of
  microtubule arrays from interphase to metaphase.
\newblock \emph{J Cell Biol} 99:47s--54s.

\bibitem[Sikirzhytski et~al.(2014)Sikirzhytski, Magidson, Steinman, He, Berre,
  Tikhonenko, Ault, McEwen, Chen, Sui, Piel, Kapoor, and
  Khodjakov]{Sikirzhytski2014}
Sikirzhytski, V., V.~Magidson, J.~B. Steinman, J.~He, M.~L. Berre,
  I.~Tikhonenko, J.~G. Ault, B.~F. McEwen, J.~K. Chen, H.~Sui, M.~Piel, T.~M.
  Kapoor, and A.~Khodjakov, 2014.
\newblock Direct kinetochore–spindle pole connections are not required for
  chromosome segregation.
\newblock \emph{J Cell Biol.} 206:231--243.

\bibitem[Petry et~al.(2013)Petry, Groen, Ishihara, Mitchison, and
  Vale]{Sabine2013}
Petry, S., A.~C. Groen, K.~Ishihara, T.~J. Mitchison, and R.~D. Vale, 2013.
\newblock Branching microtubule nucleation in Xenopus egg extracts mediated by
  augmin and TPX2.
\newblock \emph{Cell} 152:768--777.

\bibitem[Sánchez-Huertas and Lüders(2015)]{Carlos2015}
Sánchez-Huertas, C., and J.~Lüders, 2015.
\newblock The Augmin Connection in the Geometry of Microtubule Networks.
\newblock \emph{Curr Biol.} 25:R294--R299.

\bibitem[Magidson et~al.(2015)Magidson, Paul, Yang, Ault, O'Connell,
  Tikhonenko, McEwen, Mogilner, and Khodjakov]{Magidson2015}
Magidson, V., R.~Paul, N.~Yang, J.~G. Ault, C.~B. O'Connell, I.~Tikhonenko,
  B.~F. McEwen, A.~Mogilner, and A.~Khodjakov, 2015.
\newblock Adaptive changes in the kinetochore architecture facilitate proper
  spindle assembly.
\newblock \emph{Nature Cell Biology} 17.

\bibitem[Kalinina et~al.(2013)Kalinina, Nandi, Delivani, Chacón, Klemm,
  Ramunno-Johnson, Krull, Lindner, Pavin, and Tolić-Nørrelykke]{Kalinina2013}
Kalinina, I., A.~Nandi, P.~Delivani, M.~R. Chacón, A.~H. Klemm,
  D.~Ramunno-Johnson, A.~Krull, B.~Lindner, N.~Pavin, and I.~M.
  Tolić-Nørrelykke, 2013.
\newblock Pivoting of microtubules around the spindle pole accelerates
  kinetochore capture.
\newblock \emph{Nature Cell Biology} 15:82--87.

\bibitem[Blackwell et~al.(2017)Blackwell, Sweezy-Schindler, Edelmaier, Gergely,
  Flynn, Montes, Crapo, Doostan, McIntosh, Glaser, and
  Betterton]{Blackwell2017}
Blackwell, R., O.~Sweezy-Schindler, C.~Edelmaier, Z.~R. Gergely, P.~J. Flynn,
  S.~Montes, A.~Crapo, A.~Doostan, R.~McIntosh, M.~A. Glaser, and M.~D.
  Betterton, 2017.
\newblock Contributions of Microtubule Dynamic Instability and Rotational
  Diffusion to Kinetochore Capture.
\newblock \emph{Biophysical Journal} 112:552--563.

\bibitem[Adames and Cooper(2000)]{Adames2000}
Adames, N.~R., and J.~A. Cooper, 2000.
\newblock Microtubule Interactions with the Cell Cortex Causing Nuclear
  Movements in Saccharomyces cerevisiae.
\newblock \emph{Molecular Biology of the Cell} 149:863--874.

\bibitem[Grill et~al.(2003)Grill, Howard, Schäffer, Stelzer, and
  Hyman]{Grill2003}
Grill, S.~W., J.~Howard, E.~Schäffer, E.~H.~K. Stelzer, and A.~A. Hyman, 2003.
\newblock The Distribution of Active Force Generators Controls Mitotic Spindle
  Position.
\newblock \emph{Science} 301:518--521.

\bibitem[Som et~al.(2019)Som, Chatterjee, and Paul]{Som2019}
Som, S., S.~Chatterjee, and R.~Paul, 2019.
\newblock Mechanistic three-dimensoinal model to study centrosome positioning
  in the interphase cell.
\newblock \emph{Phys. Rev. E} 99:012409.

\bibitem[Zhu et~al.(2010)Zhu, Burakov, Rodionov, and Mogilner]{Zhu2010}
Zhu, J., A.~Burakov, V.~Rodionov, and A.~Mogilner, 2010.
\newblock Finding the Cell Center by a Balance of Dynein and Myosin Pulling and
  Microtubule Pushing: A Computational Study.
\newblock \emph{Molecular Biology of the Cell} 21:4418--4427.

\bibitem[Ambrose et~al.(2011)Ambrose, Allard, Cytrynbaum, and
  Wasteneys]{Ambrose2011}
Ambrose, C., J.~F. Allard, E.~N. Cytrynbaum, and G.~O. Wasteneys, 2011.
\newblock A CLASP-modulated cell edge barrier mechanism drives cell-wide
  cortical microtubule organization in Arabidopsis.
\newblock \emph{Nature Communications} 2:518--521.

\bibitem[Tran et~al.(2001)Tran, Marsh, Doye, Inoué, and Chang]{Tran2001}
Tran, P., L.~Marsh, V.~Doye, S.~Inoué, and F.~Chang, 2001.
\newblock A Mechanism for Nuclear Positioning in Fission Yeast Based on
  Microtubule Pushing.
\newblock \emph{J Cell Biol.} 153:397--412.

\bibitem[Sutradhar et~al.(2015)Sutradhar, Sridhar, Sreekumar, Bhattacharyya,
  Ghosh, Paul, and Sanyal]{Sutradhar2015}
Sutradhar, S., S.~Sridhar, L.~Sreekumar, D.~Bhattacharyya, S.~K. Ghosh,
  R.~Paul, and K.~Sanyal, 2015.
\newblock A comprehensive model to predict mitotic division in budding yeasts.
\newblock \emph{Molecular Biology of the Cell} 26.

\bibitem[Vinogradova et~al.(2012)Vinogradova, Paul, Grimaldi, Lončarek,
  Miller, Yampolsky, Magidson, Khodjakov, Mogilner, and
  Kaverina]{Vinogradova2012}
Vinogradova, T., R.~Paul, A.~D. Grimaldi, J.~Lončarek, P.~M. Miller,
  D.~Yampolsky, V.~Magidson, A.~Khodjakov, A.~Mogilner, and I.~Kaverina, 2012.
\newblock Concerted effort of centrosomal and Golgi-derived microtubules is
  required for proper Golgi complex assembly but not for maintenance.
\newblock \emph{Mol Biol Cell.} 23:820--833.

\bibitem[Peglow et~al.(2013)Peglow, Niemeyer, Hoth, and Rieger]{Peglow2013}
Peglow, M., B.~A. Niemeyer, M.~Hoth, and H.~Rieger, 2013.
\newblock A calcium-redox feedback loop controls human monocyte immune
  responses: The role of ORAI Ca2+ channels.
\newblock \emph{New Journal of Physics} 15:055022.

\bibitem[Maccari et~al.(2016)Maccari, Zhao, Peglow, Schwarz, Hornak, Pasche,
  Quintana, Hoth, Qu, and Rieger]{Maccari2016}
Maccari, I., R.~Zhao, M.~Peglow, K.~Schwarz, I.~Hornak, M.~Pasche, A.~Quintana,
  M.~Hoth, B.~Qu, and H.~Rieger, 2016.
\newblock Cytoskeleton rotation relocates mitochondria to the immunological
  synapse and increases calcium signals.
\newblock \emph{Cell Calcium.} 60:309--321.

\bibitem[Picone et~al.(2010)Picone, Ren, Ivanovitch, Clarke, McKendry, and
  Baum]{Picone2010}
Picone, R., X.~Ren, K.~D. Ivanovitch, J.~D.~W. Clarke, R.~A. McKendry, and
  B.~Baum, 2010.
\newblock A Polarised Population of Dynamic Microtubules Mediates Homeostatic
  Length Control in Animal Cells.
\newblock \emph{PLoS Biol.} 8:e1000542.

\bibitem[Laan et~al.(2012)Laan, Pavin, Husson, Romet-Lemonne, van Duijn,
  López, Vale, Jülicher, Reck-Peterson, and Dogterom]{Laan2012}
Laan, L., N.~Pavin, J.~Husson, G.~Romet-Lemonne, M.~van Duijn, M.~López,
  R.~Vale, F.~Jülicher, S.~Reck-Peterson, and M.~Dogterom, 2012.
\newblock Cortical Dynein Controls Microtubule Dynamics to Generate Pulling
  Forces that Reliably Position Microtubule Asters.
\newblock \emph{Cell} 148:502--514.

\bibitem[Foethke et~al.(2009)Foethke, Makushok, Brunner, and
  Nédélec]{Foethke2009}
Foethke, D., T.~Makushok, D.~Brunner, and F.~Nédélec, 2009.
\newblock Force- and length-dependent catastrophe activities explain interphase
  microtubule organization in fission yeast.
\newblock \emph{Molecular Systems Biology} 5.

\bibitem[Varga et~al.(2009)Varga, Leduc, Bormuth, Diez, and Howard]{Varga2009}
Varga, V., C.~Leduc, V.~Bormuth, S.~Diez, and J.~Howard, 2009.
\newblock Kinesin-8 motors act cooperatively to mediate length-dependent
  microtubule depolymerization.
\newblock \emph{Cell} 138:1174--1183.

\bibitem[Kupfer et~al.(1986)Kupfer, Singer, and Dennert]{Kupfer1986}
Kupfer, A., S.~Singer, and G.~Dennert, 1986.
\newblock On the mechanism of unidirectional killing in mixtures of two
  cytotoxic T lymphocytes. Unidirectional polarization of cytoplasmic
  organelles and the membrane-associated cytoskeleton in the effector cell.
\newblock \emph{J Exp Med.} 163:489–498.

\bibitem[Serrador et~al.(1999)Serrador, Nieto, and
  Sánchez-Madrid]{Serrador1999}
Serrador, J.~M., M.~Nieto, and F.~Sánchez-Madrid, 1999.
\newblock Cytoskeletal rearrangement during migration and activation of T
  lymphocytes.
\newblock \emph{Trends in Cell Biology} 9:228--233.

\bibitem[Lui-Roberts et~al.(2012)Lui-Roberts, Stinchcombe, Ritter, Akhmanova,
  Karakesisoglou, and Griffiths]{Lui-Roberts2012}
Lui-Roberts, W. W.~Y., J.~C. Stinchcombe, A.~T. Ritter, A.~Akhmanova,
  I.~Karakesisoglou, and G.~M. Griffiths, 2012.
\newblock Cytotoxic T lymphocyte effector function is independent of
  nucleus–centrosome dissociation.
\newblock \emph{Eur J Immunol.} 42:2132–2141.

\bibitem[Malone et~al.(2003)Malone, Misner, Bot, Tsai, Campbell, Ahringer, and
  White]{Malone2003}
Malone, C.~J., L.~Misner, N.~L. Bot, M.-C. Tsai, J.~M. Campbell, J.~Ahringer,
  and J.~G. White, 2003.
\newblock The C. elegans Hook Protein, ZYG-12, Mediates the Essential
  Attachment between the Centrosome and Nucleus.
\newblock \emph{Cell} 115:825--836.

\bibitem[Schneider et~al.(2011)Schneider, Lu, Neumann, Brachner, Gotzmann,
  Noegel, and Karakesisoglou]{Schneider2011}
Schneider, M., W.~Lu, S.~Neumann, A.~Brachner, J.~Gotzmann, A.~A. Noegel, and
  I.~Karakesisoglou, 2011.
\newblock Molecular mechanisms of centrosome and cytoskeleton anchorage at the
  nuclear envelope.
\newblock \emph{Cellular and Molecular Life Sciences} 68:1593–1610.

\bibitem[Gittes et~al.(1993)Gittes, Mickey, and Howard]{Gittes1993}
Gittes, F., B.~Mickey, and J.~Howard, 1993.
\newblock Flexural rigidity of microtubules and actin filaments measured from
  thermal fluctuations in shape.
\newblock \emph{Nat Mater.} 120:923.

\bibitem[Botvinick et~al.(2004)Botvinick, Venugopalan, Shah, Liaw, and
  Berns]{Botvinick2004}
Botvinick, E., V.~Venugopalan, J.~Shah, L.~Liaw, and M.~Berns, 2004.
\newblock Controlled ablation of microtubules using a picosecond laser.
\newblock \emph{Biophys J.} 87:4203--4212.

\bibitem[Schaedel et~al.(2015)Schaedel, John, Gaillard, Nachury, Blanchoin, and
  Théry]{Schaedel2015}
Schaedel, L., K.~John, J.~Gaillard, M.~V. Nachury, L.~Blanchoin, and M.~Théry,
  2015.
\newblock Microtubules self-repair in response to mechanical stress.
\newblock \emph{Nat Mater.} 14:1156--1163.

\bibitem[Wu et~al.(2006)Wu, Xiang, and Hammer]{Wu2006}
Wu, X., X.~Xiang, and J.~A. Hammer, 2006.
\newblock Motor proteins at the microtubule plus-end.
\newblock \emph{Trends Cell Biol.} 16:135--143.

\bibitem[Komarova et~al.(2009)Komarova, Groot, Grigoriev, Gouveia, Munteanu,
  Schober, Honnappa, Buey, Hoogenraad, Dogterom, Borisy, Steinmetz, and
  Akhmanova]{Komarova2009}
Komarova, Y., C.~O.~D. Groot, I.~Grigoriev, S.~M. Gouveia, E.~L. Munteanu,
  J.~M. Schober, S.~Honnappa, R.~M. Buey, C.~C. Hoogenraad, M.~Dogterom, G.~G.
  Borisy, M.~O. Steinmetz, and A.~Akhmanova, 2009.
\newblock Mammalian end binding proteins control persistent microtubule growth.
\newblock \emph{J Cell Biol.} 184:691--706.

\bibitem[Busch and Brunner(2004)]{Busch2004}
Busch, K.~E., and D.~Brunner, 2004.
\newblock The microtubule plus end-tracking proteins mal3p and tip1p cooperate
  for cell-end targeting of interphase microtubules.
\newblock \emph{Curr Biol.} 14:548--559.

\bibitem[Tirnauer et~al.(2002)Tirnauer, Grego, Salmon, and
  Mitchison]{Tirnauer2002}
Tirnauer, J.~S., S.~Grego, E.~Salmon, and T.~J. Mitchison, 2002.
\newblock EB1-microtubule interactions in Xenopus egg extracts: role of EB1 in
  microtubule stabilization and mechanisms of targeting to microtubules.
\newblock \emph{Mol Biol Cell.} 13:3614--3626.

\bibitem[Prahlad et~al.(1998)Prahlad, Yoon, Moir, Vale, and
  Goldman]{Prahlad1998}
Prahlad, V., M.~Yoon, R.~D. Moir, R.~D. Vale, and R.~D. Goldman, 1998.
\newblock Rapid Movements of Vimentin on Microtubule Tracks: Kinesin-dependent
  Assembly of Intermediate Filament Networks.
\newblock \emph{jcb} 143:159.

\bibitem[Huber et~al.(2015)Huber, Boire, López, and Koenderink]{Huber2015}
Huber, F., A.~Boire, M.~P. López, and G.~H. Koenderink, 2015.
\newblock Cytoskeletal crosstalk: when three different personalities team up.
\newblock \emph{Current Opinion in Cell Biology} 32:39--47.

\bibitem[Mendez et~al.(2010)Mendez, Kojima, and Goldman]{Mendez2010}
Mendez, M.~G., S.-I. Kojima, and R.~D. Goldman, 2010.
\newblock Vimentin induces changes in cell shape, motility, and adhesion during
  the epithelial to mesenchymal transition.
\newblock \emph{FASEB J.} 24:1838--1851.

\bibitem[Gan et~al.(2016)Gan, Ding, Burckhardt, Lowery, Zaritsky, Sitterley,
  Mota, Costigliola, Starker, Voytas, Tytell, Goldman, and Danuser]{Gan2016}
Gan, Z., L.~Ding, C.~J. Burckhardt, J.~Lowery, A.~Zaritsky, K.~Sitterley,
  A.~Mota, N.~Costigliola, C.~G. Starker, D.~F. Voytas, J.~Tytell, R.~D.
  Goldman, and G.~Danuser, 2016.
\newblock Vimentin Intermediate Filaments Template Microtubule Networks to
  Enhance Persistence in Cell Polarity and Directed Migration.
\newblock \emph{Cell Systems} 3:252--263.

\bibitem[Verde et~al.(1992)Verde, Dogterom, Stelzer, Karsenti, and
  Leibler]{Verde1992}
Verde, F., M.~Dogterom, E.~Stelzer, E.~Karsenti, and S.~Leibler, 1992.
\newblock Control of microtubule dynamics and length by cyclin A- and cyclin
  B-dependent kinases in Xenopus egg extracts.
\newblock \emph{J. Cell Biol.} 118:1097--1108.

\bibitem[Dogterom and Leibler(1993)]{Dogterom1993}
Dogterom, M., and S.~Leibler, 1993.
\newblock Physical aspects of the growth and regulation of microtubule
  structures.
\newblock \emph{Phys Rev Lett.} 70:1347--1350.

\bibitem[Sokolnikoff and Redheffer(1966)]{Sokolnikoff1966}
Sokolnikoff, I., and R.~Redheffer, 1966.
\newblock Mathematics of Physics and Modern Engineering.
\newblock New York: McGraw-Hill.

\bibitem[Todhunter(1886)]{Todhunter1886}
Todhunter, I., 1886.
\newblock Spherical Trigonometry: For the Use of Colleges and Schools.
\newblock London: MACMILLAN AND CO.

\bibitem[Ross(1972)]{Ross1972}
Ross, S., 1972.
\newblock Introduction to Probability Models.
\newblock New York: Academic Press, Department of Industrial Engineering and
  Operations Research, University of California, Berkeley, California.

\end{thebibliography}


\begin{thebibliography}{14}
\providecommand{\url}[1]{\texttt{#1}}
\providecommand{\urlprefix}{ }

\bibitem[Verde et~al.(1992)Verde, Dogterom, Stelzer, Karsenti, and
  Leibler]{Verde}
Verde, F., M.~Dogterom, E.~Stelzer, E.~Karsenti, and S.~Leibler, 1992.
\newblock Control of microtubule dynamics and length by cyclin A- and cyclin
  B-dependent kinases in Xenopus egg extracts.
\newblock \emph{J. Cell Biol.} 118:1097--1108.

\bibitem[Dogterom and Leibler(1993)]{Dogterom}
Dogterom, M., and S.~Leibler, 1993.
\newblock Physical aspects of the growth and regulation of microtubule
  structures.
\newblock \emph{Phys Rev Lett.} 70:1347--1350.

\bibitem[Holy and Leibler(1994)]{Leibler}
Holy, T., and S.~Leibler, 1994.
\newblock Dynamic instability of microtubules as an efficient way to search in
  space.
\newblock \emph{Proc Natl Acad Sci U S A.} 91:5682–5685.

\bibitem[Wollman et~al.(2005)Wollman, Cytrynbaum, Jones, Meyer, Scholey, and
  Mogilner]{Wollman}
Wollman, R., E.~Cytrynbaum, J.~Jones, T.~Meyer, J.~Scholey, and A.~Mogilner,
  2005.
\newblock Efficient chromosome capture requires a bias in the
  `search-and-capture’ process during mitotic spindle assembly.
\newblock \emph{Curr Biol.} 15:828--832.

\bibitem[Paul et~al.(2009)Paul, Wollman, Silkworth, Nardi, Cimini, and
  Mogilner]{Paul}
Paul, R., R.~Wollman, W.~T. Silkworth, I.~K. Nardi, D.~Cimini, and A.~Mogilner,
  2009.
\newblock Computer simulations predict that chromosome movements and rotations
  accelerate mitotic spindle assembly without compromising accuracy.
\newblock \emph{Proc Natl Acad Sci U S A} 106:15708--15713.

\bibitem[Sokolnikoff and Redheffer(1966)]{Sokolnikoff}
Sokolnikoff, I., and R.~Redheffer, 1966.
\newblock Mathematics of Physics and Modern Engineering.
\newblock New York: McGraw-Hill.

\bibitem[Todhunter(1886)]{Todhunter}
Todhunter, I., 1886.
\newblock Spherical Trigonometry: For the Use of Colleges and Schools.
\newblock London: MACMILLAN AND CO.

\bibitem[Ross(1972)]{Ross}
Ross, S., 1972.
\newblock Introduction to Probability Models.
\newblock New York: Academic Press, Department of Industrial Engineering and
  Operations Research, University of California, Berkeley, California.

\bibitem[Yi et~al.(2014)Yi, Wu, Chung, Chen, Kapoor, and Hammer]{Hammer}
Yi, J., X.~Wu, A.~H. Chung, J.~K. Chen, T.~M. Kapoor, and J.~A. Hammer, 2014.
\newblock Centrosome repositioning in T cells is biphasic and driven by
  microtubule end-on capture-shrinkage.
\newblock \emph{J. Cell Biol.} 202:779–792.

\bibitem[Picone et~al.(2010)Picone, Ren, Ivanovitch, Clarke, McKendry, and
  Baum]{Picone}
Picone, R., X.~Ren, K.~D. Ivanovitch, J.~D.~W. Clarke, R.~A. McKendry, and
  B.~Baum, 2010.
\newblock A Polarised Population of Dynamic Microtubules Mediates Homeostatic
  Length Control in Animal Cells.
\newblock \emph{PLoS Biol.} 8:e1000542.

\bibitem[Laan et~al.(2012)Laan, Pavin, Husson, Romet-Lemonne, van Duijn,
  López, Vale, Jülicher, Reck-Peterson, and Dogterom]{Laan}
Laan, L., N.~Pavin, J.~Husson, G.~Romet-Lemonne, M.~van Duijn, M.~López,
  R.~Vale, F.~Jülicher, S.~Reck-Peterson, and M.~Dogterom, 2012.
\newblock Cortical Dynein Controls Microtubule Dynamics to Generate Pulling
  Forces that Reliably Position Microtubule Asters.
\newblock \emph{Cell} 148:502--514.

\bibitem[Pavin and Tolić-Nørrelykke(2014)]{Pavin}
Pavin, N., and I.~M. Tolić-Nørrelykke, 2014.
\newblock Swinging a sword: how microtubules search for their targets.
\newblock \emph{Syst Synth Biol.} 8:1103.

\bibitem[Peglow et~al.(2013)Peglow, Niemeyer, Hoth, and Rieger]{Peglow}
Peglow, M., B.~A. Niemeyer, M.~Hoth, and H.~Rieger, 2013.
\newblock A calcium-redox feedback loop controls human monocyte immune
  responses: The role of ORAI Ca2+ channels.
\newblock \emph{New Journal of Physics} 15:055022.

\bibitem[Maccari et~al.(2016)Maccari, Zhao, Peglow, Schwarz, Hornak, Pasche,
  Quintana, Hoth, Qu, and Rieger]{Maccari}
Maccari, I., R.~Zhao, M.~Peglow, K.~Schwarz, I.~Hornak, M.~Pasche, A.~Quintana,
  M.~Hoth, B.~Qu, and H.~Rieger, 2016.
\newblock Cytoskeleton rotation relocates mitochondria to the immunological
  synapse and increases calcium signals.
\newblock \emph{Cell Calcium.} 60:309--321.

\end{thebibliography}

\clearpage

\title{Supporting Material: Search and capture efficiency of dynamic microtubules for centrosome relocation during IS formation}
%\runningtitle{Capture efficiency of MTs in T cell} %% For page header

%\author[1,*]{Apurba Sarkar}
%\author[2,*]{Heiko Rieger}
%\author[1,*]{Raja Paul}
%\author[]{Apurba Sarkar$^{\text{1,*}}$, Heiko Rieger$^{\text{2,*}}$, and Raja Paul$^{\text{1,*}}$}
%\author[1,*]{Apurba Sarkar}
%\author[2,*]{Heiko Rieger}
%\author[1,*]{Raja Paul}

\runningauthor{Sarkar et al.}
%\runningauthor{Apurba Sarkar, Raja Paul, and Heiko Rieger} %% For page header

%\affil[1]{School of Mathematical \& Computational Sciences, Indian Association for the Cultivation of Science, Kolkata, West Bengal, India}

%\affil[2]{Department of Theoretical Physics and Center for Biophysics, Saarland University, Saarbrücken, Germany }

%\corrauthor[*]{sspas2@iacs.res.in, h.rieger@mx.uni-saarland.de, or raja.paul@iacs.res.in}

%\affil{School of Mathematical \& Computational Sciences, Indian Association for the Cultivation of Science, Kolkata, India}

%\affil[2]{Department of Theoretical Physics and Center for Biophysics, Saarland University, Saarbrücken, Germany }

%\corrauthor[*]{raja.paul@iacs.res.in}

%.....................................%

\begin{frontmatter}
\end{frontmatter}

\refstepcounter{supplementary}
\label{s:supplementary}

\setcounter{equation}{0}
\setcounter{figure}{0}
\setcounter{table}{0}
\makeatletter
\renewcommand{\theequation}{S\arabic{equation}}
\renewcommand{\thefigure}{S\arabic{figure}}
\renewcommand{\thetable}{S\arabic{table}}

\vspace{7 mm}
\section{SUPPORTING METHODS}
Here we describe the modeling of the numerical simulation and the mathematical analyses that concur with the statements proposed in the main text. The model parameters are recorded in the table (Table~\ref{tab:table}).

\subsection{Computational Model}
MTs assemble via dynamic instability regulated by the four MT dynamic instability parameters (i.e., $v_g$, $v_s$, $f_c$, and $f_r$). Average MT length $l_{mt}$ can be expresses as~\citeSup{Verde,Dogterom},
\begin{align}
l_{mt} &= \frac{v_gv_s}{f_cv_s-f_rv_g} \label{eq:l_mt} \\
&=\frac{v_g}{f_c} \;\;\;\;\;\;\;\;\;[\; \text{when} \;\;f_r=0\;] \label{eq:l_g}
\end{align}

Since $l_{mt}$ can be interpreted as the average distance covered by the MT tip without undergoing a catastrophe, the average search time for a target placed on the cell periphery, is a function of $l_{mt}$ and dynamical parameters of the MT. Earlier studies on ``Search and Capture'' of chromosomes~\citeSup{Leibler,Wollman,Paul} suggested that an efficient search would require the MT to explore the space randomly in all possible directions. Accordingly upon completion of an unsuccessful attempt to capture the target, MTs should not be rescued and hence $f_r=0$. 

Microtubule dynamics is simulated using a Monte Carlo algorithm incorporating these four parameters. At each computational time step ($\Delta t=0.01 \; \text{sec}$) the switching of the MT state from growing to shortening determined by the probability $[1-\exp(-f_c\Delta t)]$. Direction of the MT nucleated from the MTOC is specified by the polar angle $\theta$ (where $\theta \in [0,  \pi]$), and azimuthal angle $\phi$ (where $\phi \in [0,  2\pi]$). Direction of the MT nucleation raises possibilities that the plus-tip i) undergoes instant catastrophe impinging upon the nucleus, leading to complete depolymerization and new MT nucleation in a random direction and ii) encounters the cell periphery and spontaneously curves along the cell surface such that curvature of the MT is minimum. Before reaching the cell periphery, the MT remains straight and the MT plus-tip coordinate is specified by the length of the MT $L(t)$ at that instant of time, polar angle $\theta$ and azimuthal angle $\phi$, and in this case the only variable is $L(t)$; $\theta$ and $\phi$ remains constant at which the MT is nucleated from the MTOC. Following encounter with the inner cell surface, MT grows along the periphery and subsequently the polar angle of the MT plus-tip varies; thus, $\theta$ changes with the length of the MT ($L(t)$) and the radius of the cell ($R_C$). Whenever MT crosses a pole (two poles are the points on the cell periphery with $z=R_C$ and $z=-R_C$ respectively), the azimuthal angle $\phi$ changes to $\phi \pm \pi $ depending on $\phi \leq \pi $ or $\phi \ge \pi$ prior to the event. MT is stabilized when the plus end makes a contact with the target and the target is said to be captured.

\subsection{Mathematical Formalism}

We propose a mathematical framework estimating the average time taken by a single MT to find a stationary target residing on the surface of the cell. The procedure is based on the conjecture that the time until capture of the target by a stochastic searcher is distributed exponentially. Furthermore, we propose a general expression of the average search time for arbitrary number of MTs.

Generally, a successful search event is preceded by several unsuccessful searches. A new MT grows upon complete shrinkage of the unsuccessful MT; the number of unsuccessful search continues until a successful search when MT finds the target by chance. Consider the probability that a single stationary target is captured by a single searching MT at time $T$ denoted by $Pr(t{=}T)$, where the random variable $t$ represents the search time. Therefore, the probability that the target is eventually captured at a time less than or equal to $T$ is the Cumulative Density Function (CDF) $F(T)=Pr(t\leq T)$ of the corresponding Probability Density Function (PDF) $Pr(t{=}T)$. 

According to the law of total probability, $\text{Pr}(t\leq T)$ can be written as~\citeSup{Wollman,Sokolnikoff}:
\begin{equation} \label{eq:Pr}
{Pr}(t\leq T)=\sum_{n=0}^{\infty}{Pr}(t\leq T|n).{Pr}(n)
\end{equation}
where the probability $Pr(n)$ of $n$ unsuccessful search events before a successful MT-target attachment is a geometric random variable~\citeSup{Sokolnikoff}: $Pr(n)$=$P_c(1-P_c)^n$, where $P_c$ is the probability of a successful search (capture probability). $Pr(t\leq T|n)$ is the probability that the total time taken by $n$ unsuccessful searches is less than $T$ and at ($n+1$)$^{st}$ time-step the MT captures the target, given by a gamma distribution\citeSup{Wollman,Sokolnikoff}. 

Now, if the number of unsuccessful search $n>>1$, the probability of a successful search $P_c<<1$~\citeSup{Wollman} and in this limit the expression for $Pr(t\leq T)$ becomes
\begin{equation} \label{eq:F_T}
F(T)=Pr(t\leq T) \approx 1-e^{-\frac{P_c}{T_u}T}.
\end{equation}
The corresponding Probability Density Function (PDF) would be 
\begin{equation} \label{eq:f_T}
Pr(t{=}T) =\dfrac{dF(T)}{dT}=\tfrac{P_c}{T_u}e^{-\frac{P_c}{T_u}T}.
\end{equation}
The mean value of this exponentially distributed random variable $t=T$ is the approximate average capture time:
\begin{align}
T_{avg.} &=\frac{T_u}{P_{c}}. \label{eq:T^1_avg} 
\end{align}
where $T_{avg.}$ is the average search time of one MT searching for the target. The average duration taken by a microtubule for an unsuccessful search $T_u$ is the average time to grow, ($l_{mt}/v_g$), plus the corresponding time to shrink back to the MTOC, ($l_{mt}/v_s$):
\begin{align}
\qquad\qquad\qquad\qquad\;\;\;\;\;\;\;\;T_u &=\frac{l_{mt}}{v_{g}}+\frac{l_{mt}}{v_s}=\frac{v_g+v_s}{v_s.f_{c}}. \label{eq:T_u}\\
&\qquad\qquad\;\;\;\;\;\;\;\;\;\;\; \left(\text{here},\; l_{mt}=v_g/f_{c}\; (\text{see eq.~\ref{eq:l_g}})\right) \nonumber
\end{align}
Therefore, the approximate average search time for the single searching MT becomes
\begin{align}
T_{avg.} &=\left[\frac{v_g+v_s}{v_s.f_{c}}\right].\;P_{c}^{\;-1} \label{eq:T_avg} 
\end{align} 
and, the approximate average search time for $N$ searching MTs is given by 
\begin{equation}
T^N_{avg.} =\frac{T_{avg.}}{N} \label{eq:T_avg_main} 
\end{equation} 
which is $\frac{1}{N}$ times the avg. search time for a single MT (see the section given below for detailed analysis).

The capture probability ($P_c$) can be calculated using the conjecture of two probabilities: the probability that the MT emanates in the direction so that it can reach the target ($P_{direction}$) and the probability that the MT survives before the target is reached i.e. the MT does not undergo any catastrophe until a successful MT-target attachment ($P_{no \;cat}$). The capture probability ($P_c$) is a function of the position of the target on the cell periphery and the estimation of $P_c$ for different target position (FIG.~\ref{fig:figureS1}) is analyzed in the section given below.

\subsubsection*{\textbf{Calculation of Capture Probability ($\mathbf{P_{c}}$) : }}
\subsubsection*{\textbf{$I$. One MT and one target}}
%\clearpage 
\renewcommand{\thefigure}{S\arabic{figure}}
\begin{figure}[ht!]
\centering
\includegraphics[width=1.0\textwidth]{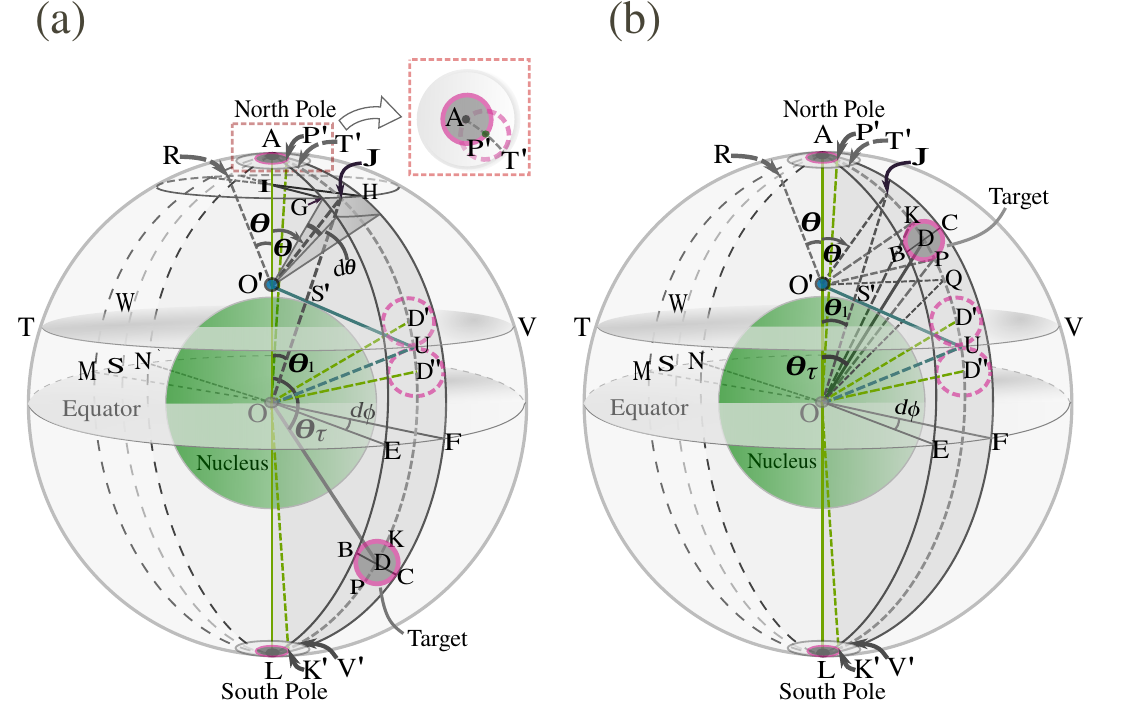}
\caption {A schematic diagram of the geometry of the theoretical model. \textbf{(a)} MTs are nucleated from the MTOC, $O^{\prime}$ within polar angle $\theta$ and $\theta+d\theta$ and azimuthal angle $\phi$ and $\phi+d\phi$ so that the MT tip falls within the shaded solid cone subtended at $O^{\prime}$. $O^{\prime}U$ is a line segment tangential to the nuclear surface at the point $S^{\prime}$ impinging on the cell surface at $U$ and $TUVW$ is a cross-section plane passing through the point $U$. MTs nucleated with a polar angle greater than $AO^{\prime}U$ will hit the nuclear envelope and shrink instantly. The target is located at a random point $D$ on the cell surface below the plane $TUVW$ and $\theta_{\tau}$ is the polar angle of the target. $AEL$ and $AFL$ are two semi-great circles passing through points $B$ and $C$ tangential to the target and form a spherical lune $AELFA$. Geometrical construct of a target in the north pole region of the cell is shown in the Inset; a similar geometry would also emerge for the target placed in the south pole of the cell. \textbf{(b)} Schematic of the theoretical model with the target located in the upper half of the $TUVW$ plane. To see this figure in color, go online.}
\label{fig:figureS1}
\end{figure}

FIG.~\ref{fig:figureS1} illustrates our mathematical model. MTs nucleated isotropically from the MTOC, can be specified by the polar angle $\theta \in [0, \pi]$ and the azimuthal angle $\phi \in [0, 2\pi]$. If the MT is nucleated with $\theta > \angle AO^{\prime}U$, it will retract upon interaction with the nuclear envelope; thus, MTs with $0 \leq \theta \leq \angle AO^{\prime}U$ will be able to participate in the capture process. Henceforth, $\angle AO^{\prime}U$ is denoted by $\theta_{\text{eff}}$.

Consider the MTOC positioned at $z=h_{MTOC} \ge R_N$ and radius of the target is $R_{\tau}$. Applying the rules of trigonometry and considering $OS^{\prime}\perp O^{\prime}U$ (since $O^{\prime}U$ is a tangent at the point $S^{\prime}$), we can write:
\begin{equation}
%\begin{align}
\theta_{\text{eff}} =\angle AO^{\prime}U=\pi - \sin^{-1}\left(\frac{R_{N}}{h_{MTOC}}\right) \label{eq:AO'U}
\end{equation}
and
\begin{equation}
\angle AOU =\frac{\pi}{2}-\sin^{-1}\left(\frac{R_{N}}{h_{MTOC}}\right)+\cos^{-1}\left(\frac{R_{N}}{R_{C}}\right) \label{eq:AOU}
%\end{align}
\end{equation}

Here, $\theta_{\text{eff}}$ becomes $\frac{\pi}{2}$ if $h_{MTOC}=R_N$ (i.e. if the MTOC is embedded at the nuclear surface). The target may remain very close to the poles; proximity to the north pole marked by the polar angle ranging from 0 to $\angle AOP^{\prime}=\frac{R_{\tau}}{R_{C}}$ (considering $AP^{\prime}\approx R_{\tau}$ as $R_C>>R_{\tau}$) (see Inset of FIG.~\ref{fig:figureS1}a). A similar scenario emerges for the south pole if the target is within the polar angle $\angle AOK^{\prime}=\pi-\frac{R_{\tau}}{R_C}$ (considering $LK^{\prime}=AP^{\prime} \approx R_{\tau}$) and $\pi$.

%\clearpage
%We consider that the microtubule can nucleate in all direction with polar angle $\theta$ between 0 to $\pi$ and azimuthal angle $\phi$ in between 0 to $2\pi$.

Since the nucleated MT has probability ${P_{c}}$ to capture the target, it can be decomposed into the product of (i) the Probability that the MT emanates in the direction to reach the target (${P_{direction}}$) and (ii) the Probability that the MT does not undergo any catastrophe until a successful MT-target attachment (${P_{no \; cat}}$).

In our study, the capture probability is the combination due to direct and indirect capture processes: In direct capture, MT can capture the target directly without interacting cell membrane with a probability $P_{direct}$ and indirect capture arises if the MT fails to hit the target directly but glide along the cell surface toward the target. We denote the probability contribution due to the gliding MTs as $P_{glide}$.

In addition, depending on the target location at different positions on the cell periphery, gliding MT can capture the target in multiple ways. Considering these possibilities, we adopt the following nomenclature for the probabilities due to the gliding MT (i.e., $P_{glide}$): (i) $P_{glide}^{\;above}$ denotes the probability of capture when the MT hit the cell membrane above the target position (i.e. polar angle of the point of incidence of the MT on the cell membrane is smaller than $\theta_{\tau}$) and glide along the cell cortex toward the target, (ii) $P_{glide}^{\;below}$ denotes the probability of capture if the MT hit the cell membrane with polar angle greater than $\theta_{\tau}$ and glide a long way through the distal and proximal (south and north) poles toward the target, and (iii) $P_{glide}^{\;reverse}$ corresponding to the third possibility when MT glide in the reverse direction through the lune $AMLNA$ (FIG.~\ref{fig:figureS1}, opposite to the lune $AELFA$ associated with the target) toward the target.

\subsubsection*{\textbf{{Capturing the target at various locations on the cell surface}}}

Calculation of the capture time has been carried out for three different regions of target position on the cell periphery:
\begin{itemize}
\item Target is away from the polar region $\left(\frac{R_{\tau}}{R_C} \leq \theta_{\tau} \leq (\pi-\frac{R_{\tau}}{R_C})\right)$,
\item Target is in the north polar region $\left(0\leq \theta_{\tau} \leq \frac{R_{\tau}}{R_C}\right)$,
\item Target is in the south polar region $\left((\pi-\frac{R_{\tau}}{R_C}) \leq \theta_{\tau} \leq \pi\right):$
\end{itemize}

%We can see from the FIG.~\ref{fig:figure2} that if the target is in between angle $\angle AOT$ of the cell, then the target is in the north polar region. 

\subsubsection*{\textbf{{Target is away from the polar region $(\frac{R_{\tau}}{R_C} \leq \theta_{\tau} \leq (\pi-\frac{R_{\tau}}{R_C})):$}}} We consider three distinct segments of the cell periphery where the target can be located.

\begin{enumerate}[(i)]
\item { \bf The Target is located below the horizontal plane $TUVW$:}
Referring to FIG.~\ref{fig:figureS1}a, in this region, the relevant polar angle of the target ($\theta_{\tau}$) would range from $\angle AOD^{{\prime}{\prime}}$ to $\angle AOK^{{\prime}}$ ($\angle AOK^{{\prime}}=(\pi-\frac{R_{\tau}}{R_C})$), where, 
\begin{align}
\angle AOD^{{\prime}{\prime}}&=\angle AOU + (UD^{{\prime}{\prime}}/R_C)=\frac{\pi}{2}-\sin^{-1}\left(\frac{R_{N}}{h_{MTOC}}\right)+\cos^{-1}\left(\frac{R_{N}}{R_{C}}\right) + \frac{R_{\tau}}{R_C}. \label{eq:AOD''}\\ 
&\qquad\qquad\;\;\;\;\;\;\;\;\;\;\;\;\;\;\;\;\;\;\;\;\;\;(\text{using}\; \text{eq.~\ref{eq:AOU}}, \;\text{and}\; UD^{{\prime}{\prime}} \approx R_{\tau} \;\;\text{since}\;\; R_C>>R_{\tau}) \nonumber
\end{align}
\item {\bf The Target is located above the plane $TUVW$:}
In this region, the allowed polar angle of the target ($\theta_{\tau}$) would be from $\angle AOP^{{\prime}}$ ($\angle AOP^{{\prime}}=\frac{R_{\tau}}{R_C}$) to $\angle AOD^{{\prime}}$ .
where, 
\begin{align}
\angle AOD^{{\prime}}&=\angle AOU - (UD^{{\prime}}/R_C)=\frac{\pi}{2}-\sin^{-1}\left(\frac{R_{N}}{h_{MTOC}}\right)+\cos^{-1}\left(\frac{R_{N}}{R_{C}}\right)- \frac{R_{\tau}}{R_C}. \label{eq:AOD'}\\ 
&\qquad\qquad\;\;\;\;\;\;\;\;\;\;\;\;\;\;\;\;\;\;\;\;\;\;(\text{using}\; \text{eq.~\ref{eq:AOU}}, \;\text{and} \; UD^{{\prime}} \approx R_{\tau} \;\;\text{since}\;\; R_C>>R_{\tau}) \nonumber
\end{align}

\item {\bf The target is in contact with the plane $TUVW$:}

According to FIG.~\ref{fig:figureS1}, plane $TUVW$ always intersects the target if the polar angle of the target falls between $\angle AOD^{{\prime}}$ and $\angle AOD^{{\prime}{\prime}}$, where $\angle AOD^{{\prime}{\prime}}$ and $\angle AOD^{{\prime}}$ are given by eq.~\ref{eq:AOD''} and eq.~\ref{eq:AOD'} respectively.
\end{enumerate}

Let us now consider each of these scenarios independently.

%%%%%%%%%%%%%
\subsubsection*{\textbf{(i) Target is located below the horizontal plane $TUVW$ $\left(\;\left(\frac{\pi}{2}-\sin^{-1}(\frac{R_{N}}{h_{MTOC}})+\cos^{-1}(\frac{R_{N}}{R_{C}}) + \frac{R_{\tau}}{R_C}\right) \; \;\leq \;\theta_{\tau}\; \leq \; \;(\pi-\frac{R_{\tau}}{R_C})\right)$:}}

\begin{enumerate}[(a)]
\item Calculating $P_{direction}$:\\
Probabilities that the MT is nucleated between polar angles $\theta$ and $\theta +d\theta$ (note the maximum effective polar angle for a MT is $\theta_{\text{eff}}$ is $\angle AO^{\prime}U$) and azimuthal angles $\phi$ and $\phi+d\phi$ are given by $p_{\theta}$ and $p_{\phi}$ respectively. Clearly, if the MT is nucleated in a direction so that the tip falls within the arc $GH$ (depends on the diameter of the target, see in FIG.~\ref{fig:figureS1}a), it will move along the lune AELFA and be able to capture the target. Therefore, the probability that the MT tip would fall within the arc $GH$ is same as the probability that it will be in the lune AELFA and is equivalent to the probability $p_{\phi}$, where,

\begin{align}
p_{\phi}&=\frac{arc\; GH}{ \text{Circumference of the small circle passing through} \; GH} . \nonumber \\
&=\frac{\angle GIH\;.\;radius \;HI}{2\pi.\;radius \;HI} . \nonumber \\
&=\frac{\angle GIH}{2\pi} . \label{eq:p_phi}
\end{align}

Therefore, the directional probability that the MT will capture the target is 
\begin{align}
p_{direction} &=p_{\theta}\;\;.\;\;p_{\phi} \nonumber\\
&=\frac{d\theta}{\theta_{\text{eff}}} \;. \;\frac{\angle GIH}{2\pi} \label{eq:p_direction}
\end{align}
where $\theta_{\text{eff}}$ is given by eq.~\ref{eq:AO'U}.

According to FIG.~\ref{fig:figureS1}a, $IH$, $IG$, $OF$, $OE$ are perpendicular to $OA$, because the planes $GIH$ and $EOF$ are perpendicular to $OA$; therefore $IG$ is parallel to $OE$, and $IH$ is parallel to $OF$. Hence $\angle GIH = \angle EOF$. Further, $\angle EOF$ is equal to the lune angle $A$ (i.e. the angle between two planes passing through the arc $AE$ and $AF$) and can be evaluated following simple rules of spherical trigonometry.

{\bf{Derivation of the lune angle A:}} Three great circular arcs are required to form a spherical triangle. The great circles are the largest possible circle that can be drawn in a given sphere, the center of the great circle coincides with the center of the sphere, and the diameter of the great circle is the same as the diameter of the sphere. The two great circles intersect at two antipodal points on the surface of the sphere, provided the circles do not overlap. The crescent shape formed by two such great circles on the surface of the sphere is called a lune. A third great circle will cut the lune into two spherical triangles.

In FIG.~\ref{fig:figureS1}, we have considered two great circles ($AELNA$ and $AFLMA$) intersecting at the north and south poles and tangential to the target at points $B$ and $C$. For each position of the target on the surface of the cell, a lune can be constructed by considering similar great circles.

Consider the great circular arcs $AEL$ and $AFL$ passing through the edges $B$ and $C$ of the target forming the crescent lune $AELFA$. If we follow the remaining halves of the great circular arcs $ANL$ and $AML$ in rear side of the sphere, their intersection creates a second lune $AMLNA$ identical with the first one. Here, we consider $BC$ as an arc of the third great circle so that arcs $AB$, $BC$, and $AC$ forms an Isosceles spherical triangle $ABC$ with two equal sides $AB$ and $AC$. We can compare the length of arc $BC$ with the diameter of the spherical target (i.e., $2R_{\tau}$); since the radius of the target is very small compared to the radius of the cell, we write 
\begin{equation} \label{eq:diameter}
BC \approx 2R_{\tau} . 
\end{equation} 

Now, arc $AD = R_C \theta_{\tau} $ and arc $AD \perp$ arc $BC$. Applying Pythagorean law of spherical trigonometry~\citeSup{Todhunter} to spherical triangles $ADB$ or $ADC$ we find,
\begin{align}
\cos(AB/R_C)= \cos(AC/R_C) &= \cos(AD/R_C)\cos(BD/R_C). \label{eq:cos_AB_Rc}  \\
 &\qquad\;\;\;\;\;\;\;\;\;\;\;\;\;\;\;\;\;\; (\text{here,} \; BD=DC) \nonumber   
\end{align}
Therefore,
\begin{align}
AB=AC&= R_C.\cos^{-1}[\cos(AD/R_C).\cos(BD/R_C)] \nonumber \\
&=R_C\cos^{-1}[cos(\theta_{\tau}).\cos(R_{\tau}/R_C)] .\label{eq:AB} \\
&\qquad\;\;\;\;\;\;\;\;\;\;\;\;\;\;\;\;\;\; \;\; \;(\text{using}\; AD= R_C \theta_{\tau}, \;\text{and}\; BD=BC/2=R_{\tau}) \nonumber
\end{align}
Applying the cosine rule in spherical trigonometry\citeSup{Todhunter} to the spherical triangle $ABC$,
\begin{equation} \label{eq:cos_A}
\cos (A) .\sin(AB/R_C). \sin(AC/R_C) = \cos (BC/R_C) - \cos(AC/R_C).\cos(AB/R_C)
\end{equation}

Using eqs.~\ref{eq:diameter},~\ref{eq:AB} and rearranging eq.~\ref{eq:cos_A} we find an expression of the lune angle
\begin{equation} \label{eq:A}
A= \cos^{-1}\left[\frac{\cos(2R_{\tau}/R_C)-\cos^2\theta_{\tau} \; \cos^2(R_{\tau}/R_C)}{1-\cos^2\theta_{\tau}\; \cos^2(R_{\tau}/R_C)}\right] .
\end{equation}

Thus, eq.~\ref{eq:p_direction} becomes
\begin{equation} \label{eq:p_direction_final}
P_{direction}= \frac{d\theta}{2\pi.\theta_{\text{eff}}}\;\;.\;\;\cos^{-1}\left[\frac{\cos(2R_{\tau}/R_C)-\cos^2\theta_{\tau} \; \cos^2(R_{\tau}/R_C)}{1-\cos^2\theta_{\tau}\; \cos^2(R_{\tau}/R_C)}\right].
\end{equation}

where $\theta_{\text{eff}}$ is given by eq.~\ref{eq:AO'U}.

%% calculate P_nocat

\item Calculating ${P_{no \; cat}}$:\\
For a successful capture, the microtubule has to survive up to the distance ($O^{\prime} J+arc \;JK)$ without catastrophe (see FIG.~\ref{fig:figureS1}a). The probability of not undergoing catastrophe before arriving at point $K$ is approximately an exponential probability density function \citeSup{Ross,Wollman}. 

Therefore,
\begin{equation} \label{eq:p_no cat}
P_{no \;cat}=\exp[-(O^{\prime} J+ arc \;JK).f_c/{v_g}].
\end{equation}

Applying the ``Laws of Cosine'' to triangle $OO^{\prime}J$ we find,
\begin{equation} \label{eq:o_dah_j}
O^{\prime}J=\sqrt{R_C^2+h_{MTOC}^2-2R_C\;h_{MTOC}\cos\theta_1} .
\end{equation}
where $\theta_1$ is the angle between $OO^{\prime}$ and $OJ$.

Also,
\begin{align}
arc \;\;JK&=R_C(\theta_{\tau}-\theta_1)-arc \;KD. \nonumber\\
&= R_C(\theta_{\tau}-\theta_1)-R_{\tau}. \label{eq:JK}\\
&\qquad\qquad\;\;\;\;\;\;\;\;\;\;\;\;\;\;\;\;\;\;\;\;\;\;(\text{here},\; KD \approx R_{\tau} \;\;\text{since}\;\; R_C>>R_{\tau}) \nonumber
\end{align}

Polar angle of the MT tip with respect to the cell center (i.e., $\theta_1$) and with respect to the MTOC (i.e., $\theta$ ) can be expressed together by applying the ``Laws of Sine'' to the triangle $OO^{\prime}J$,
\begin{align}
\theta_1&=\theta-\sin^{-1}\left(\frac{h_{MTOC}\sin\theta}{R_C}\right). \label{eq:theta_1}\\
and, \nonumber \\
\theta&= \cot^{-1}\left(\cot\theta_1-\frac{h_{MTOC}}{R_C}\mathrm{cosec}\;\theta_1\right). \label{eq:theta}
\end{align}

Now substituting $O^{\prime}J$ (eq.~\ref{eq:o_dah_j}) and $arc \;\;JK$ (eq.~\ref{eq:JK}) in eq.~\ref{eq:p_no cat} and using eq.~\ref{eq:theta_1}, we find the explicit form of 
\begin{multline} 
P_{no\;cat}=\exp\left[-\left(\sqrt{R_C^2+h_{MTOC}^2-2R_C\;h_{MTOC}\cos\left(\theta-\sin^{-1}(h_{MTOC}\sin\theta/R_C)\right)} \right. \right. \\
+\left. \left. R_C\left(\theta_{\tau}-\theta+\sin^{-1}(h_{MTOC}\sin\theta/R_C)\right)-R_{\tau}\right).\frac{f_c}{v_g}\right] . \label{eq:p_no_cat_final}
\end{multline}
\end{enumerate}

Therefore, the probability that the MT will reach the target at point $K$ (FIG.~\ref{fig:figureS1}a) would be the product of $P_{direction}$ and $P_{no \;cat}$:
\begin{multline} 
P_{direction}\times P_{no \;cat}=\frac{1}{2\pi.\theta_{\text{eff}}}\;\cos^{-1}\left[\frac{\cos(2R_{\tau}/R_C)-\cos^2\theta_{\tau} \; \cos^2(R_{\tau}/R_C)}{1-\cos^2\theta_{\tau}\; \cos^2(R_{\tau}/R_C)}\right] \\
\times \;\exp\left[-\left(\sqrt{R_C^2+h_{MTOC}^2-2R_C\;h_{MTOC}\cos\left(\theta-\sin^{-1}(h_{MTOC}\sin\theta/R_C)\right)}\right. \right. \\
+ \left. \left. \;R_C\left(\theta_{\tau}-\theta+\sin^{-1}(h_{MTOC}\sin\theta/R_C)\right)-R_{\tau}\right).\frac{f_c}{v_g}\right]d\theta . \label{eq:p_cap}
\end{multline}

where $\theta_{\text{eff}}$ is given by eq.~\ref{eq:AO'U}. Since MTs can nucleate at any polar angle between $0$ and $\theta_{\text{eff}}$ (i.e.$\angle AO^{\prime}U$), for which it has a chance to reach the target, eq.~\ref{eq:p_cap} must be integrated over $\theta$. Therefore, capture probability

\begin{multline} \label{eq:p_capture_prime}
%P_{c}^{\prime}
P_{{glide,\; \tau \Downarrow}}^{\;{above}}=\frac{1}{2\pi.\theta_{\text{eff}}}\;\cos^{-1}\left[\frac{\cos(2R_{\tau}/R_C)-\cos^2\theta_{\tau} \; \cos^2(R_{\tau}/R_C)}{1-\cos^2\theta_{\tau}\; \cos^2(R_{\tau}/R_C)}\right] \\
\;
\times \;\int_0^{\theta_{\text{eff}}}\exp\left[-\left(\sqrt{R_C^2+h_{MTOC}^2-2R_C\;h_{MTOC}\cos\left(\theta-\sin^{-1}(h_{MTOC}\sin\theta/R_C)\right)}\right. \right. \\
+ \;\left. \left. R_C\left(\theta_{\tau}-\theta+\sin^{-1}(h_{MTOC}\sin\theta/R_C)\right)-R_{\tau}\right).\frac{f_c}{v_g}\right]d\theta . 
\end{multline}
where $\theta_{\text{eff}}$ is given by eq.~\ref{eq:AO'U}. Here $P_{{glide,\; \tau\Downarrow}}^{\;{above}}$ is the probability of capture the target following the shorter path when the target is located below the plane $TUVW$ (FIG.~\ref{fig:figureS1}).

%%%%%%5 2nd P_c
Note that, we did not need to integrate over the azimuthal angle $\phi$ since the azimuthal angle of the MT remains constant. However, there is a possibility that the MT would have a chance to capture the target if the azimuthal angle is $\phi + \pi$. Consider the MT nucleates in a direction (along $O^{\prime} R$) to fall in the lune AMLNA (FIG.~\ref{fig:figureS1}a). In this case, the MT plus tip will reach the target if it does not undergo any catastrophe before reaching the point $P$ through the path $RSLP$ ($arc \; RP$).

In this case, the directional probability ($P_{direction}$) would be the same as eq.~\ref{eq:p_direction_final} as the lune AMLNA is identical in shape and area with the first lune AELFA; however the probability that the MT would survive the distance ($O^{\prime}R+arc \; RP$) would be
\begin{align}
P^{\prime}_{no\;cat}&=\exp[-(O^{\prime} R+ arc \;RP).f_c/{v_g}] .\label{eq:p_no_cat'}
\end{align}

From FIG.~\ref{fig:figureS1}, we find 
\begin{align}
\;\;\;O^{\prime}R=O^{\prime}J=\sqrt{R_C^2+h_{MTOC}^2-2R_C\;h_{MTOC}\cos\theta_1} . \label{eq:O'R} 
\end{align} 
and,
\begin{align}
arc \;RP&= 2\pi R_C -R_C(\theta_{\tau}+\theta_1)-arc \;PD. \nonumber \\
&=2\pi R_C -R_C(\theta_{\tau}+\theta_1)-R_{\tau} . \label{eq:RP} \\
&\qquad\qquad\;\;\;\;\;\;\;\;\;\;\;\;\;\;\;\;\;\;\;\;\;\;(\text{here},\; PD \approx R_{\tau} \;\;\text{since}\;\; R_C>>R_{\tau}) \nonumber
\end{align}

Substituting eqs.~\ref{eq:RP},~\ref{eq:O'R} in eq.~\ref{eq:p_no_cat'} and using eq.~\ref{eq:theta_1}, we find 
\begin{multline} 
P_{no \;cat}^{\prime} =\exp\left[-\left(\sqrt{R_C^2+h_{MTOC}^2-2R_C\;h_{MTOC}\cos\left(\theta-\sin^{-1}(h_{MTOC}\sin\theta/R_C)\right)}\right. \right. \\
+\left. \left.2 \pi R_C-R_C\left(\theta_{\tau}+\theta-\sin^{-1}(h_{MTOC}\sin\theta/R_C)\right)-R_{\tau}\right).\frac{f_c}{v_g}\right] . \label{eq:p_no_cat_final'}
\end{multline}

%P_{c}^{{\prime}{\prime}}
Therefore, the capture probability corresponding to the lune AMLNA would be 
\begin{multline} \label{eq:p_capture_doubleprime}
{P^{\;{reverse}}_{\;{glide}}}=\frac{1}{2\pi.\theta_{\text{eff}}}\;\cos^{-1}\left[\frac{\cos(2R_{\tau}/R_C)-\cos^2\theta_{\tau} \; \cos^2(R_{\tau}/R_C)}{1-\cos^2\theta_{\tau}\; \cos^2(R_{\tau}/R_C)}\right] . \\
\;\times \;\int_0^{\theta_{\text{eff}}}{}\exp\left[-\left(\sqrt{R_C^2+h_{MTOC}^2-2R_C\;h_{MTOC}\cos\left(\theta-\sin^{-1}(h_{MTOC}\sin\theta/R_C)\right)} \right. \right.\\
+ \;\left. \left. 2\pi R_{C} -R_C\left(\theta_{\tau}+\theta-\sin^{-1}(h_{MTOC}\sin\theta/R_C)\right)-R_{\tau}\right).\frac{f_c}{v_g}\right]d\theta . 
\end{multline}
where $\theta_{\text{eff}}$ is given by eq.~\ref{eq:AO'U}.

Therefore, the total probability the capture the target would be 
\begin{equation} \label{eq:p_capture}
P_{c}={P^{\;{above}}_{\;{glide,\; \tau\Downarrow}}}+{P^{\;{reverse}}_{\;{glide}}}
\end{equation}
where ${P^{\;{above}}_{\;{glide,\; \tau\Downarrow}}}$ and ${P^{\;{reverse}}_{\;{glide}}}$ are given by eq.~\ref{eq:p_capture_prime} and eq.~\ref{eq:p_capture_doubleprime} respectively.

Therefore from eq.~\ref{eq:T_avg}, the approximate average search time is given by
\begin{align}
T_{avg.} = \left[\frac{v_g+v_s}{v_s.f_{c}}\right] \times \left[{P^{\;{above}}_{\;{glide,\; \tau\Downarrow}}}+{P^{\;{reverse}}_{\;{glide}}}\right]^{-1} \label{eq:T_avg_final}
\end{align}
where ${P^{\;{above}}_{\;{glide,\; \tau\Downarrow}}}$ and ${P^{\;{reverse}}_{\;{glide}}}$ are given by eq.~\ref{eq:p_capture_prime} and eq.~\ref{eq:p_capture_doubleprime} respectively.

%%%%%%%%

\subsubsection*{\textbf {(ii) The Target is located above the plane $TUVW$ $\left(\frac{R_{\tau}}{R_C} \; \;\leq \theta_{\tau} \; \; \leq \left(\frac{\pi}{2}-\sin^{-1}(\frac{R_{N}}{h_{MTOC}})+\cos^{-1}(\frac{R_{N}}{R_{C}}) - \frac{R_{\tau}}{R_C}\right)\; \right) $:}}

From FIG.~\ref{fig:figureS1}b, we see that there are four possible ways by which the MT can capture the target:

{(a)} If the MT tip falls in the lune $AELFA$ (FIG.~\ref{fig:figureS1}b) but above the target position, the growing MT will move along the cell periphery and hit the target:
Here, the probability that the MT will grow in the direction of the target (${P_{direction}}$), would be the same as reported earlier in eq.~\ref{eq:p_direction_final} and the probability that the MT would not undergo any catastrophe before the target is reached is the probability that it would survive the distance ($O^{\prime}J + arc \;JK $)(FIG.~\ref{fig:figureS1}b) which is again same as eq.~\ref{eq:p_no cat}:
%Therefore, the probability that the MT will reach the point J (FIG. 1) would be the product of P direction and P no cat and is given by eq.~\ref{eq:p_cap} : 

Therefore, in this case, the capture probability would be given by eq.~\ref{eq:p_capture_prime} but the limit of integration for the polar angle would be from $0$ to $\angle AO^{\prime}K$. For this range of polar angle, MT will hit the cell periphery and move along the surface to capture the target.
Using eq.~\ref{eq:theta}, we can find the $\angle AO^{\prime}K$ in terms of the angle $\angle AOK$ as
\begin{align}
\angle AO^{\prime}K&= \cot^{-1}\left(\cot\angle AOK-\frac{h_{MTOC}}{R_C}\;\mathrm{cosec}\;\angle AOK\right). \nonumber \\
\angle AO^{\prime}K&= \cot^{-1}\left(\cot(\theta_{\tau}-\frac{R_{\tau}}{R_C})-\frac{h_{MTOC}}{R_C}\;\mathrm{cosec}\;(\theta_{\tau}-\frac{R_{\tau}}{R_C})\right) . \label{eq:AO'K}\\
&\qquad\;\;\;\;\;\;\;\;\;\;\;\;\;\;\;\;\;\;\;\;\;\; \;\; \;\left(\text{using}\; \angle AOK=\theta_{\tau}-(KD/R_C)=\theta_{\tau}-\frac{R_{\tau}}{R_C}\right) \nonumber
\end{align}

Therefore, the probability that the MT will capture the target can be written as, 
\begin{multline} \label{eq:p_capture_a}
{P^{\;{above}}_{\;{glide,\; \tau \Uparrow}}}=\frac{1}{2\pi.\theta_{\text{eff}}}\;\cos^{-1}\left[\frac{\cos(2R_{\tau}/R_C)-\cos^2\theta_{\tau} \; \cos^2(R_{\tau}/R_C)}{1-\cos^2\theta_{\tau}\; \cos^2(R_{\tau}/R_C)}\right] \\
\times\;\int_0^{\angle AO^{\prime}K}\exp\left[-\left(\sqrt{R_C^2+h_{MTOC}^2-2R_C\;h_{MTOC}\cos\left(\theta-\sin^{-1}(h_{MTOC}\sin\theta/R_C)\right)} \right. \right.\\
+\left. \left. \; R_C\left(\theta_{\tau}-\theta+\sin^{-1}(h_{MTOC}\sin\theta/R_C)\right)-R_{\tau}\right).\frac{f_c}{v_g}\right]d\theta . 
\end{multline}

where $\theta_{\text{eff}}$ and $\angle AO^{\prime}K$ are given by eq.~\ref{eq:AO'U} and eq.~\ref{eq:AO'K} respectively and the symbol `$\tau\Uparrow$' is used to denote that the target is located above the plane $TUVW$(FIG.~\ref{fig:figureS1}b).

{(b)} The second possibility is that the MT captures the target directly:

Referring to FIG.~\ref{fig:figureS1}b, If the MT nucleates in between the angle $AO^{\prime}K$ and $AO^{\prime}P$, the growing MT will grow steadily and hit the target. In this case, the probability that the MT will reach the target ($P_{\;{direct,\; \tau\Uparrow}}$) is given by the product of the probability that MT will grow in the direction of the target (${P_{direction}}$) that is same as eq.~\ref{eq:p_direction_final} and the probability that it would not undergo any catastrophe before the target is reached:

Therefore, $P_{\;{direct,\; \tau\Uparrow}}$ can be written as,
\begin{multline} \label{eq:p_capture_b}
P_{\;{direct,\; \tau\Uparrow}}=\frac{1}{2\pi.\theta_{\text{eff}}}\;\cos^{-1}\left[\frac{\cos(2R_{\tau}/R_C)-\cos^2\theta_{\tau} \; \cos^2(R_{\tau}/R_C)}{1-\cos^2\theta_{\tau}\; \cos^2(R_{\tau}/R_C)}\right] \\
\;
\times \;\int_{\angle AO^{\prime}K}^{\angle AO^{\prime}P}\exp\left[-\left(\sqrt{R_C^2+h_{MTOC}^2-2R_C\;h_{MTOC}\cos\left(\theta-\sin^{-1}(h_{MTOC}\sin\theta/R_C)\right)} 
\right).\frac{f_c}{v_g}\right]d\theta.
\end{multline}

where we have considered $\sqrt{R_C^2+h_{MTOC}^2-2R_C\;h_{MTOC}\cos\left(\theta-\sin^{-1}(h_{MTOC}\sin\theta/R_C)\right)}$ to be the distance from the MTOC to the target up to which MT has to survive for a successful capture, measured in terms of the polar angle $\theta$ of the MT. Here $\theta_{\text{eff}}$ and $\angle AO^{\prime}K$ are given by eq.~\ref{eq:AO'U} and eq.~\ref{eq:AO'K} respectively. 
Also, $\angle AO^{\prime}P$ can be written in terms of the polar angle $\theta_{\tau}$ of the target using eq.~\ref{eq:theta}:
Therefore,
\begin{align}
\angle AO^{\prime}P&= \cot^{-1}\left(\cot(\theta_{\tau}+\frac{R_{\tau}}{R_C})-\frac{h_{MTOC}}{R_C}\;\mathrm{cosec}\;(\theta_{\tau}+\frac{R_{\tau}}{R_C})\right) .\label{eq:AO'P}
\end{align}

{(c)} If the MT tip falls in the lune $AELFA$ (FIG.~\ref{fig:figureS1}b) but below the target: 

In FIG.~\ref{fig:figureS1}b, let the MT hits the cell periphery at point $Q$, therefore it has to grow all the way through the path $QULSAK$ for a successful capture. Here, (${P_{direction}}$) would be the same as eq.~\ref{eq:p_direction_final} and the probability that the MT would not undergo any catastrophe before covering the distance ($O^{\prime}Q+arc \; QLSAK$) would be
\begin{multline} 
P_{no \;cat}^{\;c} =exp\left[-\left(\sqrt{R_C^2+h_{MTOC}^2-2R_C\;h_{MTOC}\cos\left(\theta-\sin^{-1}(h_{MTOC}\sin\theta/R_C)\right)}\right. \right.\\
+\left. \left. 2 \pi R_C-R_C\left(\theta-\sin^{-1}(h_{MTOC}\sin\theta/R_C)-\theta_{\tau}\right)-R_{\tau}\right).\frac{f_c}{v_g}\right] . \label{eq:p_no_cat_final^c}
\end{multline}

where $O^{\prime}Q=\sqrt{R_C^2+h_{MTOC}^2-2R_C\;h_{MTOC}\cos\left(\theta-\sin^{-1}(h_{MTOC}\sin\theta/R_C)\right)}$ , and

$arc \; QLSAK=2 \pi R_C-R_C\left(\theta-\sin^{-1}(h_{MTOC}\sin\theta/R_C)-\theta_{\tau}\right)-R_{\tau}$; $\theta$ is the polar angle of the tip of the microtubule when it hits the cell surface w.r.t the MTOC.

Therefore, in this case, the capture probability ($P^{\;below}_{\;{glide,\; \tau\Uparrow}}$) would be the product of ${P_{direction}}$ and $P_{no \;cat}^{\;c}$ with an integration of $\theta$ from $\angle AO^{\prime}P$ to $AO^{\prime}U$ (i.e. $\theta_{\text{eff}}$).

Therefore,
\begin{multline} \label{eq:p_capture_c}
P^{\;below}_{\;{glide,\; \tau\Uparrow}}=\frac{1}{2\pi.\theta_{\text{eff}}}\;\cos^{-1}\left[\frac{\cos(2R_{\tau}/R_C)-\cos^2\theta_{\tau} \; \cos^2(R_{\tau}/R_C)}{1-\cos^2\theta_{\tau}\; \cos^2(R_{\tau}/R_C)}\right] \\
\;\times \;\int_{\angle AO^{\prime}P}^{\theta_{\text{eff}}}\exp\left[-\left(\sqrt{R_C^2+h_{MTOC}^2-2R_C\;h_{MTOC}\cos\left(\theta-\sin^{-1}(h_{MTOC}\sin\theta/R_C)\right)} \right. \right.\\
+ \left. \left.2 \pi R_C-R_C\left(\theta-\sin^{-1}(h_{MTOC}\sin\theta/R_C)-\theta_{\tau}\right)-R_{\tau}\right).\frac{f_c}{v_g}\right]d\theta . 
\end{multline}
where $\theta_{\text{eff}}$ and $\angle AO^{\prime}P$ are given by eq.~\ref{eq:AO'U} and eq.~\ref{eq:AO'P} respectively.

(d) Final possibility is that if the MT nucleates in a direction so that its tip falls in the lune $AMLNA$ (FIG.~\ref{fig:figureS1}b): Here, the capture probability would be the same as $P^{\;reverse\;}_{\;glide}$ as calculated in eq.~\ref{eq:p_capture_doubleprime} since the lune $AMLNA$ (FIG.~\ref{fig:figureS1}b) is identical in shape and area to that of lune $AELFA$ (FIG.~\ref{fig:figureS1}b).

Therefore the total probability of a successful search ($P_c$) for a target located above the $TUVW$ plane, would be 
\begin{equation} \label{eq:p_capture1}
P_{c}=P^{\;above}_{\;{glide,\; \tau\Uparrow}}+P_{\;{direct,\; \tau\Uparrow}}+P^{\;below}_{\;{glide,\; \tau\Uparrow}}+P^{\;reverse\;}_{\;glide}
\end{equation}
where $P^{\;above}_{\;{glide,\; \tau\Uparrow}}$, $P_{\;{direct,\; \tau\Uparrow}}$, $P^{\;below}_{\;{glide,\; \tau\Uparrow}}$ and $P^{\;reverse\;}_{\;glide}$ are given by eq.~\ref{eq:p_capture_a}, eq.~\ref{eq:p_capture_b}, eq.~\ref{eq:p_capture_c} and eq.~\ref{eq:p_capture_doubleprime} respectively.

Following eq.~\ref{eq:T_avg}, the approximate average search time is given by
\begin{align}
T_{avg.} = \left[\frac{v_g+v_s}{v_s.f_{c}}\right] \times \left[P^{\;above}_{\;{glide,\; \tau\Uparrow}}+P_{\;{direct,\; \tau\Uparrow}}+P^{\;below}_{\;{glide,\; \tau\Uparrow}}+P^{\;reverse\;}_{\;glide}\right]^{-1} .\label{eq:T_avg_final_above_WXUY}
\end{align}
where $P^{\;above}_{\;{glide,\; \tau\Uparrow}}$, $P_{\;{direct,\; \tau\Uparrow}}$, $P^{\;below}_{\;{glide,\; \tau\Uparrow}}$ and $P^{\;reverse\;}_{\;glide}$ are given by eq.~\ref{eq:p_capture_a}, eq.~\ref{eq:p_capture_b}, eq.~\ref{eq:p_capture_c} and eq.~\ref{eq:p_capture_doubleprime} respectively.

\subsubsection*{\textbf {(iii) The target is in contact with the $TUVW$ plane $\left(\;\left(\frac{\pi}{2}-\sin^{-1}(\frac{R_{N}}{h_{MTOC}})+\cos^{-1}(\frac{R_{N}}{R_{C}}) - \frac{R_{\tau}}{R_C}\right) \leq \theta_{\tau} \leq\;\left(\frac{\pi}{2}-\sin^{-1}(\frac{R_{N}}{h_{MTOC}})+\cos^{-1}(\frac{R_{N}}{R_{C}}) + \frac{R_{\tau}}{R_C}\right)\;\right)$:}}

For this positioning of the target, the nucleated MT can capture the target in three possible ways:

(a) If the MT tip falls in the lune $AELFA$ (FIG.~\ref{fig:figureS1}) but above the target position, then the microtubule will move along the cell periphery and capture the target with a probability same as $P^{\;{above}}_{\;{glide,\; \tau\Uparrow}}$ obtained in eq.~\ref{eq:p_capture_a};

(b) If the MT nucleates with a polar angle between $\angle AO^{\prime}K$ to $\angle AO^{\prime}U$, it will directly capture the target with a probability $P_{\;{direct,\; \tau\eqcirc}}$\;, similar to the description as in eq.~\ref{eq:p_capture_b}; however, the altered limit of the integration will be from $\angle AO^{\prime}K$ to $\theta_{\text{eff}}$. MTs emanating with a polar angle greater than $\theta_{\text{eff}}$ will interact with the nuclear surface and undergo catastrophe. Therefore, 

\begin{multline} \label{eq:p_capture_b'}
P_{\;{direct,\; \tau\eqcirc}}=\frac{1}{2\pi.\theta_{\text{eff}}}\;\cos^{-1}\left[\frac{\cos(2R_{\tau}/R_C)-\cos^2\theta_{\tau} \; \cos^2(R_{\tau}/R_C)}{1-\cos^2\theta_{\tau}\; \cos^2(R_{\tau}/R_C)}\right] \\
\;
\times \;\int_{\angle AO^{\prime}K}^{\theta_{\text{eff}}}\exp\left[-\left(\sqrt{R_C^2+h_{MTOC}^2-2R_C\;h_{MTOC}\cos\left(\theta-\sin^{-1}(h_{MTOC}\sin\theta/R_C)\right)} 
\right).\frac{f_c}{v_g}\right]d\theta .
\end{multline}
where $\theta_{\text{eff}}$ and $\angle AO^{\prime}K$ are given by eq.~\ref{eq:AO'U} and eq.~\ref{eq:AO'K} respectively and here the symbol `$\tau\eqcirc$' is used to account the altered direct capture probability when the IS is adjacent to the plane $TUVW$.

(c) The final possibility is that if the MT nucleates in a direction so that the MT tip falls in the lune $AMLNA$ (FIG.~\ref{fig:figureS1}). Here, the capture probability would be the same as $P^{\;reverse\;}_{\;glide}$, represented by eq.~\ref{eq:p_capture_doubleprime}.

Therefore, the total probability of a successful search ($P_c$) would be 
\begin{equation} \label{eq:p_c_at_WXUY}
P_{c}=P^{\;{above}}_{\;{glide,\; \tau\Uparrow}}+P_{\;{direct,\; \tau\eqcirc}}+P^{\;reverse\;}_{\;glide}.
\end{equation}
where $P^{\;{above}}_{\;{glide,\; \tau\Uparrow}}$, $P_{\;{direct,\; \tau\eqcirc}}$ and $P^{\;reverse\;}_{\;glide}$ are given by eq.~\ref{eq:p_capture_a}, eq.~\ref{eq:p_capture_b'} and eq.~\ref{eq:p_capture_doubleprime} respectively.

The corresponding approximate average search time is given by eq.~\ref{eq:T_avg}
\begin{align}
T_{avg.} = \left[\frac{v_g+v_s}{v_s.f_{c}}\right] \times \left[P^{\;{above}}_{\;{glide,\; \tau\Uparrow}}+P_{\;{direct,\; \tau\eqcirc}}+P^{\;reverse\;}_{\;glide}.\right]^{-1} .\label{eq:T_avg_final_at_WXUY}
\end{align}
%where $P_{c}^{\;a}$, $P_{c}^{\;b^{\prime}}$and $P_{c}^{{\prime}{\prime}}$ are given by eq.~\ref{eq:p_capture_a}, eq.~\ref{eq:p_capture_b'} and eq.~\ref{eq:p_capture_doubleprime} respectively.

\subsubsection*{{{\textbf{Target is at the north polar region of the cell}} $(0\leq \theta_{\tau} \leq \frac{R_{\tau}}{R_C}):$}}

Consider the target is located at the north pole of the cell (i.e. the polar angle of the target with respect to the cell center is $\sim 0$). The target can be captured in two possible ways: 
%Since all the longitudinal line of a sphere passes through the pole, the mt nucleates with any random azimuthal angle $\phi$ will hit the cell surface and move along the corresponding longitudinal line passing through the hitting point on cell surface and reach the target if the MT is in growing state.

(a) MT can capture the target directly, if the nucleation polar angle $\theta$ falls between $0$ and $\angle AO^{\prime}P^{\prime}$, and this process is independent of the azimuthal angle $\phi$ of the MT (refer to FIG.~\ref{fig:figureS1}a).

Therefore, the probability that the target will be captured ($P^{\;NP}_{direct}$) is the product of the probability that the MT will grow in the direction of the target ($P_{direction}^{\;NP}$) and the probability that it will not face any catastrophic event before reaching the target ($P_{no \; cat}^{\;NP}$):

\begin{align}
P^{\;NP}_{direct} &=P_{no \; cat}^{\;NP} \times P_{direction}^{\;NP} .\nonumber\\
&= \int_0^{\angle AO^{\prime}P^{\prime}}\exp\left[-\left(\sqrt{R_C^2+h_{MTOC}^2-2R_C\;h_{MTOC}\cos\left(\theta-\sin^{-1}(h_{MTOC}\sin\theta/R_C)\right)}\right).\frac{f_c}{v_g}\right]\times \frac{d\theta}{\theta_{\text{eff}}} .\nonumber\\
&=\frac{1}{\theta_{\text{eff}}}\int_0^{\angle AO^{\prime}P^{\prime}}\exp\left[-\left(\sqrt{R_C^2+h_{MTOC}^2-2R_C\;h_{MTOC}\cos\left(\theta-\sin^{-1}(h_{MTOC}\sin\theta/R_C)\right)}\right).\frac{f_c}{v_g}\right]d\theta .\label{eq:p_capture_alpha}
\end{align}

where we have considered $\sqrt{R_C^2+h_{MTOC}^2-2R_C\;h_{MTOC}\cos\left(\theta-\sin^{-1}(h_{MTOC}\sin\theta/R_C)\right)}$ to be the distance from the MTOC to the target up to which MT must survive for a successful capture, measured in terms of the polar angle $\theta$ of the MT and $\theta_{\text{eff}}$ given by eq.~\ref{eq:AO'U}. $\angle AO^{\prime}P^{\prime}$ is given by eq.~\ref{eq:AO'P} and substituting the target polar angle $\theta_{\tau}=0$ (since the target is at the north pole):
\begin{align}
\angle AO^{\prime}P^{\prime}&= \cot^{-1}\left(\cot\left(\frac{R_{\tau}}{R_C}\right)-\frac{h_{MTOC}}{R_C}\;\mathrm{cosec}\;\left(\frac{R_{\tau}}{R_C}\right)\right) .\label{eq:AO'P'}
\end{align}

(b) Direct capture of the target will fail if the MT grows with a polar angle $\theta$ exceeding $\angle AO^{\prime}P^{\prime}$ but remains within $\theta_{\text{eff}}$. In this case the MT, with any azimuthal angle $\phi$, will grow along the cell surface and pass through the distal pole. Therefore the directional probability is irrespective of the value of the azimuthal angle $\phi$ with which the MT is nucleated. Consider that the MT hits the cell surface at a point $R$ and move along the $RSLUP^{'}$ ($arc \; RP^{\prime}$) path to reach the target at the north pole (see FIG.~\ref{fig:figureS1}). 

Probability that the target will be captured is given by,
\begin{multline} \label{eq:p_capture_beta}
P_{glide}^{\;NP}=\frac{1}{\theta_{\text{eff}}}
\;\int_{\angle AO^{\prime}P^{\prime}}^{\theta_{\text{eff}}} \exp\left[-\left(\sqrt{R_C^2+h_{MTOC}^2-2R_C\;h_{MTOC}\cos\left(\theta-\sin^{-1}(h_{MTOC}\sin\theta/R_C)\right)} \right. \right.\\
+ \left. \left. 2 \pi R_C-R_C\left(\theta-\sin^{-1}(h_{MTOC}\sin\theta/R_C)\right)-R_{\tau}\right).\frac{f_c}{v_g}\right]d\theta . 
\end{multline}

where $O^{\prime}R=\sqrt{R_C^2+h_{MTOC}^2-2R_C\;h_{MTOC}\cos\left(\theta-\sin^{-1}(h_{MTOC}\sin\theta/R_C)\right)}$, and

$arc \; RSLUP^{\prime}=2 \pi R_C-R_C\left(\theta-\sin^{-1}(h_{MTOC}\sin\theta/R_C)\right)-R_{\tau}$. Here, $\theta$ is the polar angle of the MT with respect to the MTOC before hitting the cell surface, while $\theta_{\text{eff}}$ and $\angle AO^{\prime}P^{\prime}$ are given by eq.~\ref{eq:AO'U} and eq.~\ref{eq:AO'P'} respectively.

Therefore the total probability of a successful search ($P_c$) for a target at the north pole is 
\begin{equation} \label{eq:p_capture2}
P_{c}=P^{\;NP}_{direct}+P^{\;NP}_{glide} 
\end{equation}
where $P^{\;NP}_{direct}$ and $P^{\;NP}_{glide}$ are given by eq.~\ref{eq:p_capture_alpha} and eq.~\ref{eq:p_capture_beta} respectively.

The approximate average search time is given by eq.~\ref{eq:T_avg}
\begin{align}
T_{avg.} = \left[\frac{v_g+v_s}{v_s.f_{c}}\right] \times \left[P^{\;NP}_{direct}+P^{\;NP}_{glide} \right]^{-1} . \label{eq:T_avg_final_north_pole}
\end{align}

Notice that as long as the polar angle of the target $\theta_{\tau}$ is between $0$ and $\frac{R_{\tau}}{R_C}$, the target will be in contact with the south pole (point A) (FIG.~\ref{fig:figureS1}). 

%For this north polar region of the cell average capture time remains more or less %the same.

\subsubsection*{{{\textbf{Target is at the south polar region of the cell}} $\mathbf{(}(\pi-\frac{R_{\tau}}{R_C}) \leq \theta_{\tau} \leq \pi\mathbf{)}$:}}
Consider the target is located at the south pole of the cell (i.e. the polar angle of the target with respect to the cell center is $\pi \;rad$). For this positioning of the target, direct capture is not possible. If the MT nucleates with polar angle $\theta$ between $0$ and $\theta_{\text{eff}}$, it has always a chance to capture the target. Similar to the north-pole scenario, the capture is independent of $\phi$ as the MT with any arbitrary $\phi$ will pass through the south pole.

The capture probability may now be written as:
\begin{multline} \label{eq:p_capture_s}
P^{\;SP}_{glide}=\frac{1}{\theta_{\text{\text{eff}}}}
\;\int_0^{\theta_{\text{eff}}}\exp\left[-\left(\sqrt{R_C^2+h_{MTOC}^2-2R_C\;h_{MTOC}\cos\left(\theta-\sin^{-1}(h_{MTOC}\sin\theta/R_C)\right)}\right. \right. \\
+\left. \left. \pi R_C-R_C\left(\theta-\sin^{-1}(h_{MTOC}\sin\theta/R_C)\right)-R_{\tau}\right).\frac{f_c}{v_g}\right]d\theta . 
\end{multline}

where $O^{\prime}J=\sqrt{R_C^2+h_{MTOC}^2-2R_C\;h_{MTOC}\cos\left(\theta-\sin^{-1}(h_{MTOC}\sin\theta/R_C)\right)}$, and

$arc \;JK^{\prime}=\pi R_C-R_C\left(\theta-\sin^{-1}(h_{MTOC}\sin\theta/R_C)\right)-R_{\tau}$ . Once again, $\theta$ is the polar angle of the MT with respect to the MTOC before hitting the cell surface and $\theta_{\text{eff}}$ is given by eq.~\ref{eq:AO'U}. 

%(\textit(it)the probability that the MT will nucleates in between the polar angle $0$ to ${\pi}{2}$ and the probability that MT will not get any catastrophe before reaching the target 
Therefore, the approximate average search time is given by eq.~\ref{eq:T_avg} 
\begin{align}
T_{avg.} = \left[\frac{v_g+v_s}{v_s.f_{c}}\right].\left[P^{\;SP}_{glide}\right]^{-1} .\label{eq:T_avg_final_south_pole}
\end{align}

where $P^{\;SP}_{glide}$ is given by eq.~\ref{eq:p_capture_s}.

Here, the target remains in contact with the south pole (point $L$) as long as the polar angle of the target falls between ($\pi-\frac{R_{\tau}}{R_C}$) and $\pi$.

\subsubsection*{\textbf{$II$. $N$ searching MTs and a single target :}}
Consider $N$ number of MTs searching independently for a single target.
The probability that a single MT will find the target at a time greater than $T$ is
\begin{align}
Pr(t>T) &= 1-Pr(t\leq T) \nonumber \\
&=e^{-\frac{P_c}{T_u}T} . \label{eq:pr(t>T)}
\end{align}
Since the MTs are independent of each other, the probability that $N$ microtubules will reach the target at a time greater than $T$ may now be expressed as 
\begin{align}
Pr_N \;(t>T) &= (Pr(t>T))^N .\nonumber \\
&=e^{-\frac{P_c.N}{T_u}T} .\label{eq:pr_N(t>T)}
\end{align} 
Therefore, the probability that at least one MT will reach the target at a time less than or equal to $T$ is \citeSup{Wollman} 
\begin{align}
Pr_N \;(t\leq T) &= 1-Pr_N \;(t>T)) \nonumber \\
&=1-e^{-\frac{P_c.N}{T_u}T} . \label{eq:pr_N(t<=T)}
\end{align} 

The corresponding Probability Density Function(PDF) would be 
\begin{equation} \label{eq:f_T}
Pr_N\;(t{=}T) =\dfrac{d}{dT}(Pr_N \;(t\leq T))=\tfrac{P_c.N}{T_u}e^{-\frac{P_c.N}{T_u}T}.
\end{equation}
The mean value of this exponentially distributed random variable $t=T$ is the average capture time of a single target by $N$ searching MTs:
\begin{align}
T^N_{avg.} &=\frac{T_u}{N.P_{c}} .\label{eq:T^N_avg} \\
&=\frac{T_{avg.}}{N} .
\end{align} 
which is $\frac{1}{N}$ times the avg. capture time $T_{avg.}$ for a single MT.

Integrations appeared in the expressions of the average capture time, could be evaluated using standard integration techniques. In the current study, we have employed Simpson's 1/3 rule to numerically integrate the equations. 

Results obtained through mathematical exercise would be able to predict efficiently under the following circumstances:

\begin{enumerate}[(i)]
\item Typical search event requires a large number of unsuccessful searches before a successful capture.
\item For very large number of searching MTs, $P_c.N \simeq 1$ and errors arising from various approximations becomes too great to be ignored. 
\end{enumerate}

\subsection*{\textbf{Inclusion of boundary induced catastrophe of the MTs may regulate the capture process:}}
So far, we have performed the analytical calculation considering MTs that consistently glide along the plasma membrane upon encounter and the growth continue until a spontaneous catastrophe characterized by $f_c$. This hypothesis of MTs' gliding along the cell periphery is based on the observed MTs distribution in T cell~\citeSup{Hammer}. However, it is well known that, in most cell lines, upon hitting the cell surface MTs may undergo catastrophe or continue to grow along the cell periphery determined by the angle of interaction~\citeSup{Picone, Laan, Pavin}. A head-on collision of MT with the  plasma membrane is likely to terminate the MT's growth, whereas, a tangential impingement would favor the MT to glide along the membrane. In order to account for this, we introduce a new parameter $P_{cat}=P_c^b \cos \beta$ as the probability at which a MT undergoes catastrophe upon hitting the plasma membrane. The parameter $\beta$ is regarded as the angle between the MT and the plasma membrane normal vector. MTs that hit the cell wall normally ($\beta=0^{\circ}$), would face a catastrophe with probability $P_{cat}=P_c^b$, and would spontaneously glide along the cell membrane (with $P_{cat}=0$) if hits the wall tangentially ($\beta=90^{\circ}$). From the geometrical argument one can write $\beta$ as $(\theta-\theta_1)$ and the probability that the MT would glide along the plasma membrane is given by $P^{\;cat}_{glide}=[1-P_{cat}]= [1- P_c^b \cos (\theta-\theta_1)]$. Consequently, the probability of a successful capture requires the conjectures of three probabilities: (a) The probability for a successful growth towards the target ($P_{direction}$), (b) The probability that the MT would glide along the cell periphery ($P^{\;cat}_{glide}=1-P_{cat}$), and (c) the probability that it would survive before the target is reached i.e., it would not undergo any catastrophe (due to spontaneous catastrophe frequency $f_c$) before a successful MT-target attachment ($P_{no \; cat}$). 

Therefore,
\begin{equation} \label{eq:P_c_glide}
P_c = P_{direction} \;  . P^{\;cat}_{glide} \; . P_{no \; cat}
\end{equation} 

and the average search time would take the form 
\begin{equation} 
\label{eq:T_u_dash}
T_{avg} = \frac{T_u^{\prime}}{P_c}
\end{equation}

Note that the unsuccessful cycle time $T_u^{\prime}$ is not the same as derived earlier (i.e., $T_u$) without considering the boundary induced catastrophe of the MT. The exact formulation of $T_u^{\prime}$  is non-trivial and beyond the scope of the current exercise.
%; but one could certainly speculate that as the cell boundary acts as an obstacle for the MT growth, it would effectively reduce the average unsuccessful cycle time of the MTs that hit the plasma membrane.
Further, the above expression of the capture probability ($P_c$) is valid only for the MTs that glide along the cell surface aiming to capture the target with $P_{glide}$. However, the direct capture probability ($P_{direct}$) i.e, when MTs captures the target prior to interacting with the cell membrane, would be a product of $P_{direction}$ and $P_{no \; cat}$ only. Therefore, to incorporate the effect of cell boundary on the MT dynamics, we have multiplied the product with additional factor $P^{\;cat}_{glide}$ in the expression of $P_{glide}$ as derived earlier for different target locations.

It is also to be expected that for specific positioning of the target where the direct capture, as well as the indirect capture by the gliding MTs participate in the capturing process, the capture time would increase for a finite catastrophe due to the boundary as it would decrease the net capture probability. Target position dominated by the direct capture (e.g. target located near the north pole), finite catastrophe due to the cell boundary may decrease the capture time by preventing a large number of MTs to glide along the cell wall which would otherwise make futile attempts to capture the target.

\vspace{15mm}

\section{SUPPORTING RESULTS}
%\begin{center}
% \normalsize\bfseries\MakeUppercase{I. SUPPORTING FIGURES}
%\end{center}

\begin{figure}[hbt!]
\centering
\includegraphics[width=1.0\linewidth]{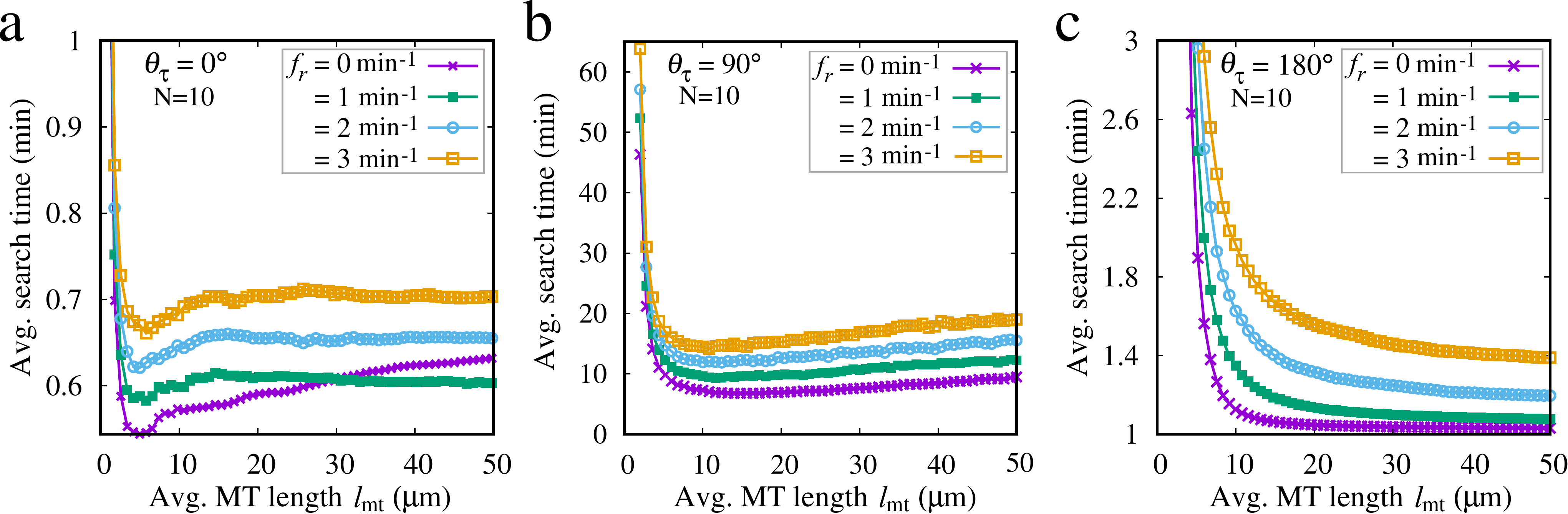}
\caption{{\bf{Average search time is plotted against average MT length for different rescue frequencies}}. The target is placed at $\theta_{\tau}= 0^{\circ}$, $90^{\circ}$ and $180^{\circ}$ in (a), (b) and (c) respectively. Notice, the global minima in the average search time is obtained with zero rescue frequency. To see this figure in color, go online.}
\label{fig:figureS2}
\end{figure}

\vspace{15mm}

\subsection*{MTOC position dependent maneuvering of the capture time analysing through different capture probabilities}

\begin{figure}[hbt!]
\centering
\includegraphics[width=0.5\linewidth]{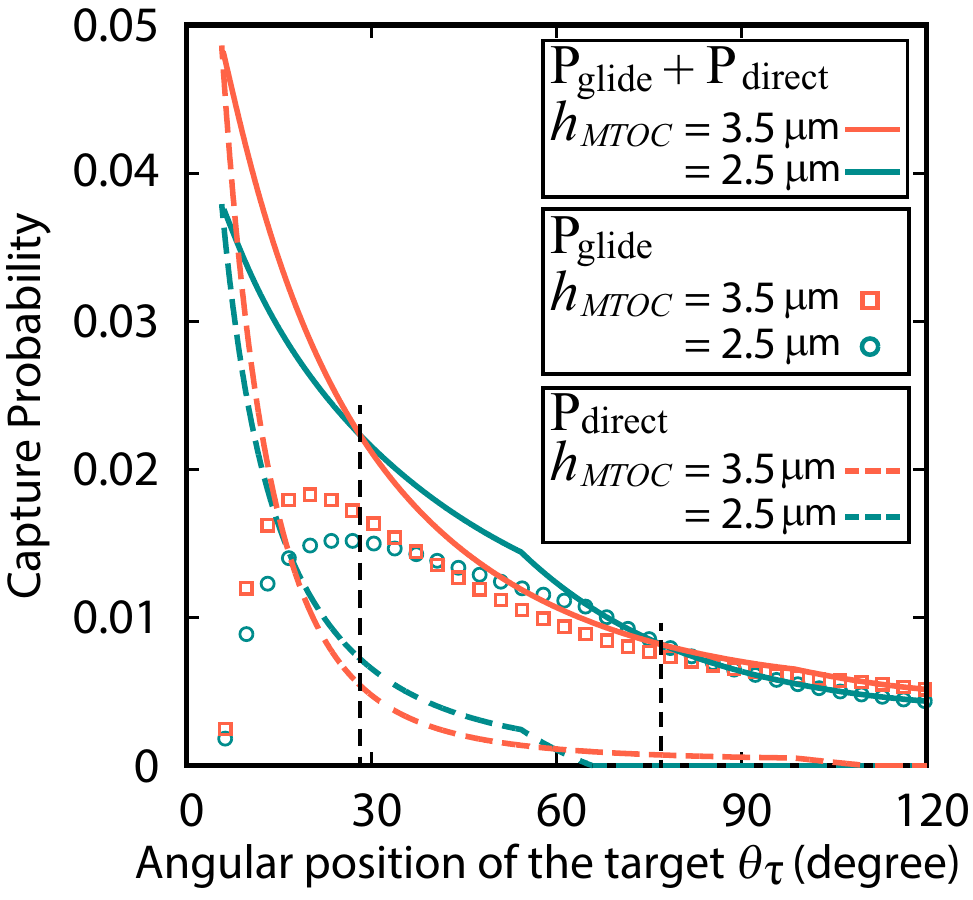}
\caption{{\bf{Different capture probability contribution plotted as a function of the angular position of the target $\theta_{\tau}$}}. We find two crossover in $(P_{\text{glide}}+P_{\text{direct}})$ plot for $h_{MTOC}=2.5\; \SI{}{\micro\metre}$ and $3.5\; \SI{}{\micro\metre}$ at $\theta_{\tau}$ near about $28^{\circ} \; \text{and} \; 78^{\circ}$. Other parameters are as in FIG.~\ref{fig:figure2}c (see the main text). To see this figure in color, go online.}
\label{fig:figureS3}
\end{figure}

\begin{figure}[hbt!]
\centering
\includegraphics[width=1\linewidth]{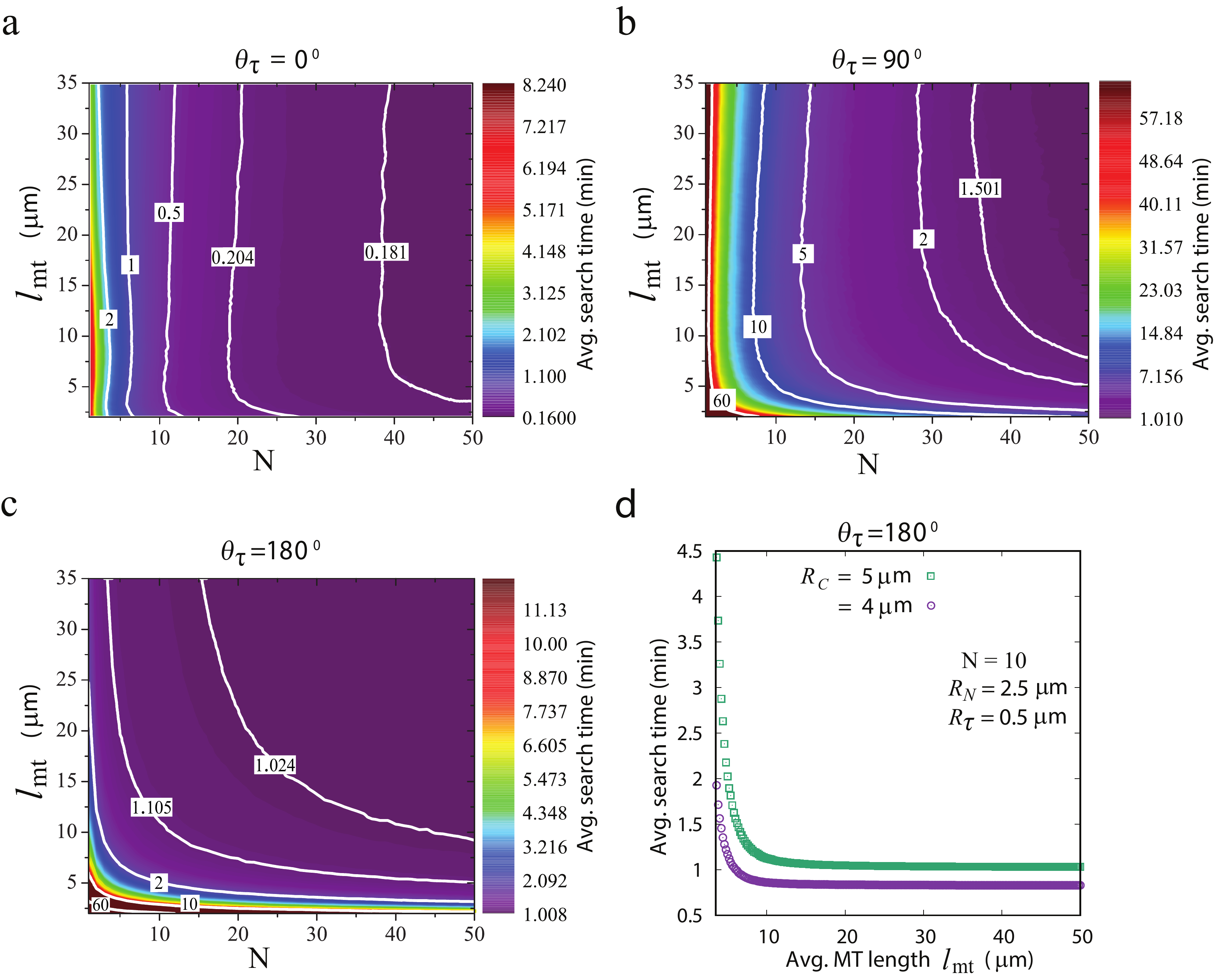}
\caption {{\bf{Average search time vs. average MT length and the number of searching MTs}}. (a) Color plot representing the average search time as a function of average microtubule length $l_{mt}$ and number of searching microtubules $N$ for target size $R_{\tau}=0.5\;\SI{}{\micro\metre}$ and location $\theta_{\tau}=0^{\circ}$. Lines represent the contours corresponding to specific average search time. (b), (c) Similar to (a) but the target is placed at the equatorial plane (i.e $\theta_{\tau}=90^{\circ})$ and at the distal pole $\theta_{\tau}=180^{\circ}$ respectively. (d) Average search time plotted as a function of the average MT length $l_{mt}$ for different cellular radii $R_C=5\;\SI{}{\micro\metre}$ and $4 \;\SI{}{\micro\metre}$ with target located at $\theta_{\tau}=180^{\circ}$ searched by $N=10$ microtubules. We find that the minimum of the average search time $\sim 0.83 \; \text{min}$ for $R_C=4\;\SI{}{\micro\metre}$ is lower than the minimum of average search time ($\sim 1.03 \;\text{min}$) for $R_C=5\;\SI{}{\micro\metre}$. To see this figure in color, go online.}
\label{fig:figureS4}
\end{figure}

\begin{figure}[hbt!]
\centering
\includegraphics[width=1.0\linewidth]{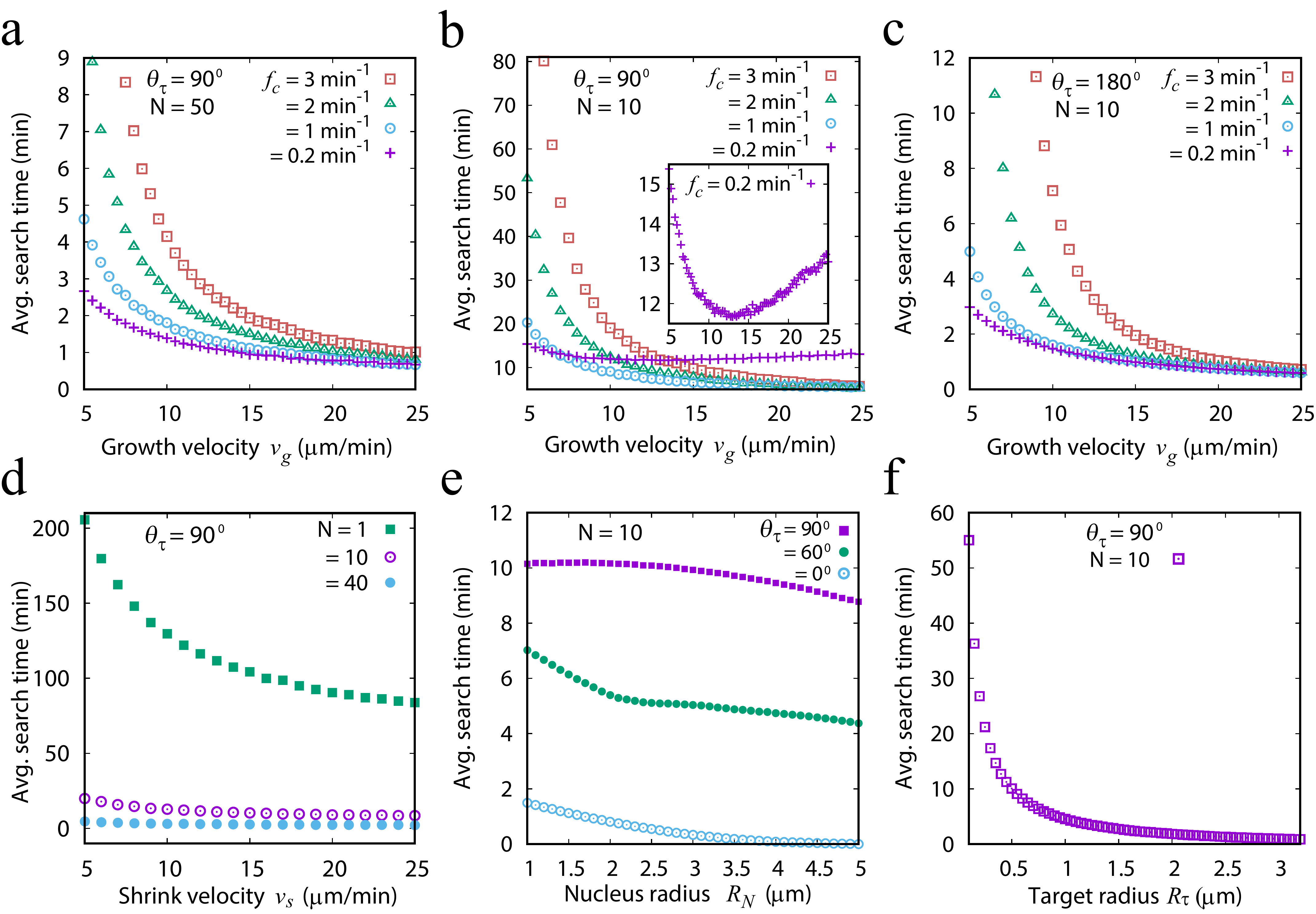}
\caption{{\bf{Variation of the average search time with various system parameters.}} (a) Average search time vs growth velocity for different catastrophe frequencies $f_c$: $0.2\;min^{-1}$, $1\;min^{-1}$, $2\;min^{-1}$, and $3\;min^{-1}$ respectively, and for $f_r=0$. The target with $R_{\tau}=0.5\;\SI{}{\micro\metre} $ is located at $\theta_{\tau}=90^{\circ}$ and the number of searching MT is $50$. Notice, the search time becomes smaller at large growth velocities and small catastrophe frequencies. (b) Similar to (a), but for $N=10$ microtubules. A minimum in the average search time is obtained at small catastrophe frequencies (i.e., $f_c=0.2\;min^{-1}$). (c) Similar to (a), but the target is located at the polar angle $\theta_{\tau}=180^{\circ}$ and number of searching MTs is $N=10$. Note that, no optimization in the average search time as a function of growth velocity is observed. (d) Average search time plotted as a function of MT shrink velocity. The effect of shrink velocity on the average search time is insignificant for large number of MTs, but the effect becomes significant for smaller number of MTs. (e) The search time varies non-monotonously with nuclear radius for $\theta_{\tau}=90^{\circ}$, however, decreases monotonically for $\theta_{\tau}=0^{\circ}$. For $\theta_{\tau}=$ $60^{\circ}$, the search time decreases monotonically up to a certain value of the nucleus radius and becomes non-monotonous for larger nucleus. (f) Larger target reduces the time required for capture. To see this figure in color, go online.}
\label{fig:figureS5}
\end{figure}

\begin{figure}[hbt!]
\centering
\includegraphics[width=1\linewidth]{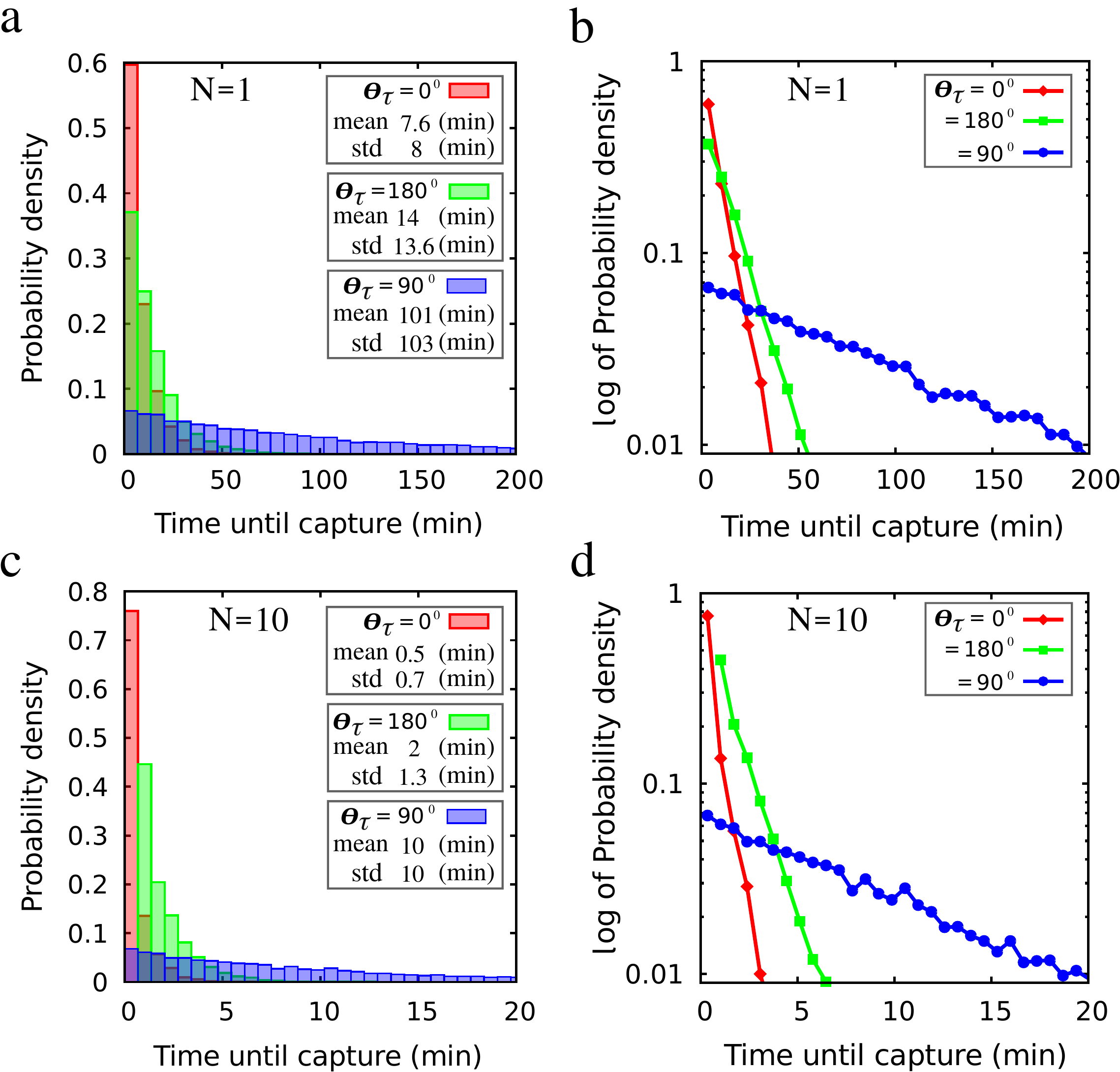}
\caption {{\bf{Distribution of time until capture}}. (left: [a and c]) Simulation results showing the probability density of time until capture with number of searching MTs 1 and 10 for three different target positions: $\theta_{\tau}=0^{\circ}$, $\theta_{\tau}=180^{\circ}$, and $\theta_{\tau}=90^{\circ}$ respectively. Bars are histograms of 10000 simulated values of the time until capture. Notice that the variance of the time until capture (i.e., ${\text{std}}^2$) which measures the spread of the distribution is maximum for $\theta_{\tau}=90^{\circ}$ with respect to $\theta_{\tau}=0^{\circ}$ and $180^{\circ}$. Also, the variance of the time until capture for $\theta_{\tau}=180^{\circ}$ is larger than the variance for $\theta_{\tau}=0^{\circ}$. (right: [b and d]) Logarithm of the probability density demonstrates that the time until capture is approximately exponentially distributed for $N=1$ and $10$, because the log of an exponential function is a linear function. Other parameters are as in FIG.~\ref{fig:figure1}c (see the main text). To see this figure in color, go online.}
\label{fig:figureS6}
\end{figure}

In order to understand the MTOC position dependent capture time presented in FIG.~\ref{fig:figure2}c of the main text, we resort to the mathematical expressions derived in the \ref{s:supplementary}. Essentially, the net capture probability is a sum of the probabilities due to direct and indirect captures by the MTs gliding along the cell surface. Target positioned below the plane $TUVW$ (See FIG.~\ref{fig:figureS1}a) require indirect search by the gliding MTs, since the nuclear envelope prevent the MTs capturing the target directly. The total capture probability is given $P_{c}={P^{\;{above}}_{\;{glide,\; \tau\Downarrow}}}+{P^{\;{reverse}}_{\;{glide}}}$ (see eq.~\ref{eq:p_capture} in \ref{s:supplementary}), where, $P^{\;above}_{\;{glide,\; \tau\Downarrow}}$ is the probability due to the MTs that fall in the target lune $AELFA$ (see FIG.~\ref{fig:figureS1}a) and glide along the cell surface toward the target, and ${P^{\;{reverse}}_{\;{glide}}}$ is the probability that the MTs fall in the lune $AMLNA$ (FIG.~\ref{fig:figureS1}, \ref{s:supplementary}) and capture the target by gliding along the cell surface in a direction opposite to the previous one. The key contribution comes from the probability $P^{\;above}_{\;{glide,\; \tau\Downarrow}}$ since the remaining probability vanishes for MTs that do not grow beyond the cell diameter. On the other hand, the direct capture also comes into play if the target is located above the plane $TUVW$ (See FIG.~\ref{fig:figureS1}b) or stay in touch with the plane. For the target positioned above the plane $TUVW$ (See FIG.~\ref{fig:figureS1}b), the total capture probability is given by $P_{c}=P^{\;above}_{\;{glide,\; \tau\Uparrow}}+P_{\;{direct,\; \tau\Uparrow}}+P^{\;below}_{\;{glide,\; \tau\Uparrow}}+P^{\;reverse\;}_{\;glide}$ (see eq.~\ref{eq:p_capture1} in \ref{s:supplementary}). Here, $P^{\;above}_{\;{glide,\; \tau\Uparrow}}$ is the probability due to the gliding MT that falls in the target lune $AELFA$ (see FIG.~\ref{fig:figureS1}b) but above the position of target and glide along the cell surface for capture. Probability $P_{\;{direct,\; \tau\Uparrow}}$ is the direct capture probability i.e., the probability that the MT capture the target directly. The major contributions come from these two terms as the remaining terms vanish for MTs that do not grow beyond the cell diameter, which is a key factor for these terms to contribute. When the target is in contact with the plane $TUVW$, the direct capture probability becomes $P_{\;{direct,\; \tau\eqcirc}}$ (eq.~\ref{eq:p_capture_b'}, \ref{s:supplementary}). For simplicity, neglecting $P^{\;below}_{\;{glide,\; \tau\Uparrow}}$ and $P^{\;reverse\;}_{\;glide}$ for short MTs and considering only the dominant term $P^{\;above}_{\;{glide}}$, we denote the probability for the gliding MTs as $P_{\text{glide}}$ and the probability for the direct capture as $P_{\text{direct }}$. In FIG.~\ref{fig:figureS3}, we plotted three different probabilities: $P_{\text{glide}}$, $P_{\text{direct}}$ and their sum ($P_{\text{glide}}+P_{\text{direct}}$) as a function of $\theta_{\tau}$ of the target position for $h_{MTOC}=2.5 \;\SI{}{\micro\metre}$ and $3.5\;\SI{}{\micro\metre}$. Clearly, the two crossovers in the plot of ($P_{\text{glide}}+P_{\text{direct}}$) appearing at $\sim 28^{\circ}$ and $\sim 78^{\circ}$, resemble the scenario presented in the Inset of FIG.~\ref{fig:figure2}c in main text. For very small polar angles of the IS, MTs grow relatively less from a distant MTOC (large $h_{MTOC}$ and near the plasma membrane) to capture the target directly compared to the MTOC placed adjacent to the nucleus (small $h_{MTOC}$ and away from the plasma membrane). However, if the polar angle $\theta_{\tau}$ of the IS exceeds a certain limit, length of the direct path between the MTOC and the target becomes longer from the distant MTOC compared to the adjacent one - this gives rise to the first crossing of $P_{\text{direct}}$ at $\theta_{\tau}$ $\sim 16^{\circ}$. The second crossing happens at $\sim 60^{\circ}$ where the target is partially visible to the MTs from MTOC adjacent to the nucleus but fully visible to the MTs nucleated from the distant MTOC. Now $P_{\text{glide}}$, being a function of the path length between the target and the MTOC, also plays a crucial rule in the capture process. Here, we find two crossings in the plot of $P_{\text{glide}}$ for $\theta_{\tau}$ $\sim 40^{\circ}$ and $\sim 90^{\circ}$. Finally, combining the two probabilities we find the crossings of the capture probabilities at $\theta_{\tau}$ $\sim 28^{\circ}$ and $\sim 78 ^{\circ}$ which conform the results presented in the Inset of FIG.~\ref{fig:figure2}c (see the main text). 

\vspace{10mm}

\clearpage

\clearpage

\clearpage
%%%%%%%%%%%%%%%% TABLE %%%%%%%%%%%%%%%%%%
\begin{table}[hbt!]

\centering
\footnotesize
\caption{\footnotesize \textbf{List of variables used in this work}}
\label{tab:table}

\begin{threeparttable}
\newcolumntype Y{X[c]{S[per-mode=symbol,
table-align-text-pre=false,
table-align-text-post=false,
input-symbols=,
input-open-uncertainty=,
input-close-uncertainty=,
detect-all
]}}

\bgroup
\def\arraystretch{2.0}% 1 is the default, change whatever you need
\setlength{\tabcolsep}{1.5em} % for the horizontal padding

\begin{tabular}{p{1.7cm} p{7.0cm} p{8.5cm}} 
%\begin{tabular}{c l r}
\hline
\textbf{Abbreviations } & \textbf{Meaning} & \textbf{Equation / Value | Range | Reference}\\
\hline

$Pr(t\leq T)$ &Probability that the target is eventually captured at a time ($t$) less than or equal to $T$ for a single MT & Equation~\ref{eq:Pr} \\

$P_c$ & Probability of a successful search for a single MT & 
Equations~\ref{eq:p_capture},~\ref{eq:p_capture1},
~\ref{eq:p_c_at_WXUY},~\ref{eq:p_capture2} and~\ref{eq:p_capture_s} \\

$T_u$ & Average unsuccessful cycle time & Equation~\ref{eq:T_u} \\

$P_{direction}$ & Probability to nucleate in a direction so that MT can capture the target & Equations~\ref{eq:p_direction} and~\ref{eq:p_direction_final} \\

$P_{no \; cat}$ & Probability of not undergoing catastrophe before the target is reached & Equations \ref{eq:p_no_cat_final},~\ref{eq:p_no_cat_final'}, and~\ref{eq:p_no_cat_final^c}\\

$P_{direct}$ & Probability that MT will capture the target directly without attacking the cell surface & Equations \ref{eq:p_capture_b},~\ref{eq:p_capture_b'}, and~\ref{eq:p_capture_alpha} \\

$P_{glide}$ & Probability of capture the target by gliding along the cell surface & Equations~\ref{eq:p_capture_prime},~\ref{eq:p_capture_doubleprime},
~\ref{eq:p_capture_a},~\ref{eq:p_capture_c},~\ref{eq:p_capture_beta}, and~\ref{eq:p_capture_s} \\

$ T_{avg.}$ & Average time for a single MT to capture the target & Equations~\ref{eq:T^1_avg},~\ref{eq:T_avg},~\ref{eq:T_avg_final},
~\ref{eq:T_avg_final_above_WXUY},~\ref{eq:T_avg_final_at_WXUY},~\ref{eq:T_avg_final_north_pole}, and~\ref{eq:T_avg_final_south_pole} \\

$T_{avg.}^N$ & Average time to capture the target by N number of MTs & Equation~\ref{eq:T^N_avg} \\

$R_C$ & Cell radius & $5\;\SI{}{\micro\metre}$ | $4-10\;\SI{}{\micro\metre}$ | \citetSup{Hammer}\\

$R_{N}$ & Nucleus radius & $R_C/2$ | $1-5\;\SI{}{\micro\metre}$ | \citetSup{Peglow, Maccari} \\

$R_{\tau}$ & Target radius & $0.5\;\SI{}{\micro\metre}$ | $0.05-3.2\;\SI{}{\micro\metre}$\\

$h_{MTOC}$ & variable distance between the cell center and the MTOC & $2.5-3.5\;\SI{}{\micro\metre}$\\

$N$ & Number of dynamic MTs & $1-100$ | \citetSup{Hammer}\\

$l_{mt}$ & Average MT-length & $2-35 \;\;\SI{}{\micro\metre}$\\

$v_g$ & MT growth velocity & $14.3 \;\; \;\SI{}{\micro\metre}/\text{min}$ | $5-25 \;\; \;\SI{}{\micro\metre}/\text{min}$ | \citetSup{Leibler}\\

$v_s$ & MT shrink velocity &$16\;\; \;\SI{}{\micro\metre}/\text{min}$ | $5-25 \;\; \;\SI{}{\micro\metre}/\text{min}$ | \citetSup{Leibler}\\

$f_c$ & MTs catastrophe frequency &${v_g}/{l_{mt}}$ | $0.2-14.3 \;\; \text{min}^{-1}$ | \citetSup{Leibler, Verde}\\

$f_r$ & MTs rescue frequency & $0$ | \citetSup{Leibler, Wollman} \\

\hline

\end{tabular}
\egroup

\end{threeparttable}
\end{table}

\clearpage

\bibliographystyleSup{biophysj}
\bibliographySup{SuppBibTex}

\end{document}